\documentclass[useAMS,usenatbib,article]{mn2e}

\usepackage{longtable}

\usepackage[pdftex]{graphicx}
\usepackage{epstopdf}

\def \lbul{\ensuremath{L_\mathrm{bul}}}
\def \lbs{\ensuremath{L_\mathrm{3.6,bul}}}

\def \msun{\ensuremath{M_\odot}}
\def \mbh{\ensuremath{M_\mathrm{BH}}}
\def \mstar{\ensuremath{M_\star}}
\def \mdyn{\ensuremath{M_\mathrm{dyn}}}
\def \re{\ensuremath{R_\mathrm{e}}}

\def  \wav{3.6 $\mu$m}

\def \spitzer      {{\it Spitzer}}

\title[\mbh-bulge scaling relations at \wav]
{The \spitzer/IRAC view of black hole - bulge scaling relations}

\author[E. Sani et al.]
{E.~Sani,$^{1,2}$\thanks{E-mail: sani@arcetri.astro.it}
A.~Marconi,$^1$ L.~K.~Hunt,$^2$ G.~Risaliti$^{2,3}$
\newauthor 
           \\
$^1$Dipartimento di Fisica e Astronomia, Universit\`a di Firenze, Largo
E. Fermi 2, I-50125 Firenze, Italy\\
$^2$INAF - Osservatorio Astrofisico di Arcetri, Largo
E. Fermi 5, I-50125 Firenze, Italy\\
$^3$Harvard-Smithsonian Center for Astrophysics, 60
Garden Street, Cambridge, MA 02138}

\date{Released 2010 Xxxxx XX}

\pagerange{\pageref{firstpage}--\pageref{lastpage}} \pubyear{2010}

\def\LaTeX{L\kern-.36em\raise.3ex\hbox{a}\kern-.15em
  T\kern-.1667em\lower.7ex\hbox{E}\kern-.125emX}

\begin{document}

\label{firstpage}

\maketitle

\begin{abstract}
We present a mid-infrared investigation of the scaling relations 
between supermassive black hole 
masses (\mbh) and the structural parameters of the host spheroids in local galaxies.
This work is based on two-dimensional bulge-disk decompositions of 
\spitzer/IRAC \wav\ images of 57 galaxies with \mbh\ estimates.
We first verify the accuracy of our decomposition by examining the fundamental plane (FP) of spheroids
at \wav. Our estimates of effective radii (\re) and average surface brightnesses, combined with velocity
dispersions from the literature, define a FP relation consistent with previous determinations but doubling the observed range in \re. None of our galaxies is an outlier of the FP, demonstrating the accuracy of our bulge-disk decomposition which also allows us to independently identify pseudobulges in our sample.
We calibrate M/L at \wav\ by using the tight \mdyn-\lbul\ relation ($\sim 0.1$ dex intrinsic dispersion) 
and find that no color corrections are required to estimate the stellar mass. 
The \wav\ luminosity is thus the best tracer of stellar mass yet studied.
We then explore the connection between \mbh\ and bulge structural parameters (luminosity, mass, effective radius).
We find tight correlations of \mbh\ with both \wav\ bulge luminosity and dynamical mass ($\mbh/\mdyn\sim 1/1000$), with intrinsic
dispersions of $\sim0.35$ dex, similar to the \mbh$-\sigma$ relation. Our results are
consistent with previous determinations at shorter wavelengths. 
By using our calibrated M/L,  we rescale \mbh-\lbul\ to obtain the \mbh-\mstar\ relation, which can be used as the local reference for high-z studies which probe the cosmic evolution of \mbh-galaxy relations and where the stellar mass is inferred directly from luminosity measurements.
The analysis of pseudobulges shows that 4 out of 9 lie on the scaling relations within the observed scatter, 
while those with small \mbh\ are significantly displaced. 
We explore the different origins for such behavior, while considering the possibility 
of nuclear morphological components not reproduced by our two-dimensional decomposition.
\end{abstract}


\section{Introduction}
Supermassive black holes (BHs) are believed to dwell in almost all galaxy spheroids (hereafter bulges) as a result of past quasar activity (Soltan~1982, 
Marconi et al.~2004, Shankar~2009 and references therein).
A direct link between bulge formation and the growth of the central BHs has been inferred from tight relations between 
BH mass (\mbh) and the bulge structural parameters, such as velocity dispersion 
($\sigma$, Ferrarese and Merrit~2000, Gebhardt et al.~2000), luminosity 
(\lbul, Kormendy \& Richstone~1995, Marconi \& Hunt~2003 - hereafter, MH03) and mass (Magorrian et al.~1998, MH03, H{\"a}ring \& Rix~2004, hereafter HR04).\\
The underlying physics is encoded not only in the slopes of the
correlations, but also in the intrinsic or cosmic scatter ($rms$), i.e.~the
scatter not accounted for by measurement uncertainties. 
Particular care has been taken with its measurement and, for instance, 
over the last 15 years the intrinsic scatter of \mbh- \lbul\ decreased from 
an initial value of $rms\sim0.5$~dex (in log~\mbh, Kormendy \& Richstone 1995) to 
$\sim 0.3-0.4$~dex (MH03, G\"ultekin et al.~2009, Hu~2009 hereafter G09 and H09 respectively). 
\mbh-$\sigma$ is considered one of the tightest relations
($rms\sim0.25-0.3$~dex, Tremaine et al.~2002, T02), suggesting that  
bulge dynamics or mass rather than luminosity are driving the correlations. 
Indeed the \mbh-\mdyn\ relation has a 
scatter comparable with \mbh-$\sigma$ ($\sim0.3$~dex, MH03, HR04, H09 and references therein). 
Nonetheless, slopes and $rms$ of scaling relations are still a subject of debate; 
an accurate assessment of the actual $rms$ is indeed crucial to draw conclusions 
on the physical origin of the correlations and turns out to be fundamental 
to both constrain BH-galaxy evolutionary models and compute the space density of BHs (e.g.~G09, Hopkins et al.~2007).

Since a successful theoretical model should be able to account for \emph{all} empirical 
BH-bulge relations, discerning which are the truly fundamental ones,
there has also been a growing interest in searching for all possible connections of the BH with other structural 
parameters. These include the effective radius \re\ (MH03) and its various combinations with $\sigma$ such as virial mass, gravitational binding energy of the host spheroid (e.g.~MH03, Aller \& Richstone 2007, H09, Feoli et al.~2010), thus leading to a possible fundamental plane for BHs (Barway \& Kembhavi~2007, Hopkins et al.~2007). 

To investigate the interplay between \mbh, bulge dynamics and luminosity, it is crucial (\emph{i}) to perform a proper photometric decomposition of the galaxy components (e.g.~disk, bulge, bar etc.) and (\emph{ii}) to obtain good estimators
of stellar mass (\mstar), dynamical mass (\mdyn) and the galaxy mass-to-light ratio (M/L).

An accurate bulge-disk decomposition can be performed by simultaneously fitting 
multiple bi-dimensional components to galaxy images. 
To compute the bulge mass, one can apply the virial theorem which turns out to be quite accurate (Cappellari et al.~2006) but requires
a spectroscopical estimate of the velocity dispersion in addition to the bulge effective radius \re. 
Alternatively, one can estimate \mstar\ by combining \lbul\ with a M/L ratio 
based on calibrated relations available in the literature (see, e.g., Bell \& de Jong~ 2001, Bell et al.~2003, Cappellari et al.~2006). 
In this case, \mstar\ can be affected by dust extinction, depending on wavebands, 
and by the assumption of a constant M/L over the entire galaxy.
In any case, it is believed that \mstar\ is more easily estimated than \mdyn\ especially at higher redshifts where
it is the only host property not affected 
by strong biases (Merloni et al.~2010, Trakhtenbrot \& Netzer~2010, Lamastra et al.~2010).

Recently, attention has been drawn to the coexistence of pseudobulges and BHs 
(Graham 2008b, Greene, Ho \& Barth~2008, H09, Nowak et al.~2010, Erwin~2010). 
Pseudobulges are bulges which are photometrically and morphologically disk-like 
(possibly containing typical disk features such as bars, rings or ovals) and which present 
'cool' kinematics, dominated by rotation (Kormendy \& Kennicutt~2004).
Apparently, pseudobulges follow their own relation with BHs, 
hosting less massive BHs than classical bulges (Greene, Ho, \& Barth~2008, H09). 
Indeed, in these kind of structures, \mbh\ seems to better correlate with the small classical 
bulge component only (Nowak et al.~2010, Erwin~2010). 
However, these results are only based on a few pseudobulges ($\sim5$) which have been analyzed as a separate class, and 
the poor statistics preclude a firm conclusion.

In this paper we present a mid-infrared (MIR) view of the \mbh-bulge scaling relations based on a 2D photometric decomposition of 
\spitzer/IRAC images at \wav. The superb performance of \spitzer/IRAC (Fazio et al.~2004) provides 
images of unprecedented quality in the MIR for the 57 galaxies analyzed here, and IRAC sensitivity permits the 
clear identification of morphological features.
Therefore we identify classical bulges and pseudobulges with our analysis and compare their properties with \mbh\ for one 
of the largest samples ever considered.
\wav\ observations are an ideal tracer of \mstar, and are less affected by extinction 
than shorter wavelengths.

Our aims are to:\\
- determine the bulge structural parameters with high accuracy;\\
- investigate the \mbh-bulge scaling relations and their intrinsic scatter, taking 
into account the nature of the bulge (classical vs pseudobulges);\\
- calibrate the M/L ratio thus supplying the reference \mbh-\mstar\ in the local Universe for studies of the \mbh-galaxy scaling relations at higher $z$, based on \mstar\ derived from luminosity measurements (by, e.g., Merloni et al.~2010, Trakhtenbrot \& Netzer~2010).

In Section~2 we describe sample selection, data reduction and the grid-method adopted
to decompose the IRAC \wav\ images. We then present  the fundamental plane (FP) for spheroids defined by the objects in our sample.
\mbh-bulge scaling relations and the linear regression methods adopted to estimate slopes and intrinsic dispersion of the correlations are described in Section~3.
Section~4 is dedicated to a discussion about (a) the utility of \mbh-\mdyn\ as 
a benchmark to probe BH vs host galaxy co-evolution, and (b) 
pseudobulges location in the scaling relations.

Throughout this paper we assume the standard cosmology with 
H$_0=70$~km~s$^{-1}$~Mpc$^{-1}$, $\Omega_M=0.3$, $\Omega_\Lambda=0.7$.
\begin{table*}{OBSERVATIONS}
\begin{center}
\centerline{\begin{tabular}{lccccc}
\hline
Source Name  & PID  & Date   & Frame Time & Frames & Position\\
(1)         & (2)  & (3)    & (4)        & (5)    & (6) \\
\hline
Circinus & 40936& 2007-09-09 & 12         & 1      & 4  \\
IC1459   & 20371& 2005-11-28 & 2          & 1      & 22 \\
IC2560   & 40936& 2007-07-05 & 12         & 1      & 4  \\
IC4296   & 20371& 2006-02-13 & 2          & 1      & 22  \\
NGC221   & 00069& 2004-07-19 & 12         & 1      & 5  \\
NGC524   & 50630& 2008-09-18 & 30         & 1      & 5  \\
NGC821   & 20371& 2005-08-21 & 2          & 1      & 22 \\
NGC1023  & 00069& 2004-02-11 & 30         & 1      & 5  \\
NGC1068  & 00032& 2004-01-16 & 12         & 2      & 5  \\
NGC1300  & 61065& 2009-09-06 & 30         & 1      & 11 \\
NGC1316  & 00159& 2004-02-23 & 30         & 1      & 9  \\
NGC2549  & 50630& 2008-12-23 & 30         & 1      & 5  \\
NGC2748  & 61063& 2009-12-02 & 30         & 1      & 4  \\
NGC2778  & 30318& 2006-11-24 & 100        & 1      & 5  \\
NGC2787  & 03674& 2004-10-30 & 30         & 1      & 3  \\
NGC2974  & 30318& 2006-12-28 & 100        & 1      & 5  \\
NGC3031  & 00159& 2004-05-01 & 30         & 1      & 12 \\
NGC3079  & 00059& 2004-04-05 & 12         & 1      & 4  \\
NGC3115  & 00069& 2004-04-29 & 12       & 1           & 4 \\
NGC3227  & 03269& 2004-12-21 & 12       & 1           & 1 \\
NGC3245  & 03674& 2004-11-28 & 30       & 1           & 3 \\
NGC3368  & 00069& 2004-05-19 & 30       & 1           & 5 \\
NGC3377  & 00069& 2004-05-27 & 30       & 1           & 5 \\
NGC3379  & 00069& 2004-12-15 & 12       & 1           & 5 \\
NGC3384  & 30318& 2006-12-27 & 100      & 1           & 5 \\
NGC3414  & 50630& 2009-01-29 & 30       & 1           & 5 \\
NGC3489  & 00069& 2004-05-19 & 12       & 1           & 5 \\
NGC3585  & 30318& 2006-07-09 & 100      & 1           & 5 \\
NGC3607  & 00069& 2004-05-19 & 12       & 1           & 5 \\
NGC3608  & 30218& 2006-12-27 & 100       &1            & 5 \\
NGC3998  & 00069& 2004-04-21 & 12       & 1           & 5 \\
NGC4026  & 30318& 2006-12-26 & 100       &1            &5 \\
NGC4151  & 03269& 2004-12-17 & 12       & 1           & 5 \\
NGC4258  & 20801& 2005-12-25 & 30       & 2           & 7 \\
NGC4261  & 00069& 2004-05-27 & 12       & 1           & 5 \\
NGC4374  & 00069& 2004-05-27 & 12       & 1           & 5 \\
NGC4459  & 03649& 2005-01-22  & 12        & 1           &5 \\
NGC4473  & 03649& 2005-01-22  & 12       & 1           & 5 \\
M87$^a$ & 03228 & 2005-06-11 & 30      & 1            & 5 \\
             & 03228 & 2005-06-11 & 30      & 1           & 5 \\
NGC4486A & 03228 & 2005-06-11 & 30      & 1            & 5 \\
NGC4552  & 00159 & 2004-05-27 & 30      & 1           & 7 \\
NGC4564  & 20371 & 2006-02-09 & 2       & 1           & 22 \\
NGC4594  & 00159 & 2004-06-10 & 30      & 1           & 6 \\
NGC4596  & 03674 & 2005-06-10 & 30      & 1           & 3 \\
NGC4621  & 03649 & 2005-06-10 & 12      & 1           & 5 \\
NGC4649  & 00069 & 2004-06-10 & 12      & 1           & 5 \\
NGC4697  & 03403 & 2005-06-17 & 30      & 1           & 5 \\
NGC5077  & 00069 & 2005-05-11 & 12      & 1           & 5 \\
CenA     & 00101 & 2004-02-11 & 12      & 1           & 5 \\
NGC5576  & 03403 & 2005-07-15 & 30      & 1           & 5 \\
NGC5813  & 00069 & 2004-02-17 & 12      & 1           & 5 \\
NGC5845  & 20371 & 2005-08-23 & 2       & 1           & 22 \\
NGC5846  & 00069 & 2004-03-09 & 12      & 1           & 5 \\
NGC6251  & 02418 & 2004-12-16 & 30      & 2           & 6 \\
NGC7052  & 30877 & 2006-11-26 & 30      & 1           & 5 \\
NGC7457  & 30318 & 2006-07-12 & 100     & 1           & 5 \\
NGC7582  & 03269 & 2004-11-27 & 12      & 1           & 1 \\
\hline
\end{tabular}}
\end{center}
\caption{Observations Log. Columns: (1) Source name. $^a$ mean of 
2 observations (see Section~2.2 for details). 
(2) Proposal identification number (PID). 
(3) Observing date. (4) Exposure time for each frame. (5) Numbers of frames. (6) Number of positions 
to realize the source maps.}
\label{tb:obs}
\end{table*}

\section{Data selection and analysis}
In this section we describe the criteria adopted to select our sources (Section~2.1), 
the techniques adopted to co-add the basic calibrated data 
(BCD) provided by version 18.7 of the \spitzer\ pipeline (Section~2.2), and 
the 2D fitting procedure followed to decompose IRAC \wav\ images (Section~2.3). 
Finally Section~2.4 contains the analysis of the  FP at \wav.

\subsection{Sample selection}
We consider the samples compiled by G09 and H09 for a total of 64 galaxies. 
These authors have selected only galaxies with reliable \mbh\ estimates 
(i.e. by stellar or gas dynamics and masers).
This is one of the largest samples used so far 
to investigate BH-galaxy scaling relations which also allows 
a consistent comparison of our results with previous works.

G09 and Batcheldor (2010) have recently shown that discarding sources 
whose sphere of influence is not spatially resolved can bias or even simulate 
the observed linear relations. Therefore, we do not apply any cut based on that criterion.
To increase the number of objects with low \mbh, we add to the sample NGC3368 and NGC3489 whose \mbh\ 
is estimated in Nowak et al.~(2010). 
These objects have also a pseudobulge and their inclusion allows us to better investigate at \wav\ the discrepancies 
between pseudo and classical bulge properties found by H09 and Nowak et al.~(2010).

We thus obtain a sample of 66 galaxies, 7 of which
(CygnusA, NGC1399, NGC3393, NGC4291, NGC4342, NGC4742, NGC5252) have no \wav\ 
public \spitzer/IRAC observations. 
In addition, we exclude M31 because the available maps do not cover the entire galaxy and
we exclude the Milky Way as there are no secure measurements of its \wav\ bulge properties.
The final sample thus consists of 57 galaxies. 
Tables~\ref{tb:obs} and \ref{tb:properties} provide the basic parameters of the observations 
together with distances, \mbh\ and $\sigma$ measurements\footnote{We assume that the 
effective velocity dispersion $\sigma_e$ is consistent, within the errors, with the central 
velocity dispersion $\sigma_c$ and we refer to it as $\sigma$. This choice is supported by G09, 
whose Fig.~2 shows that no bias is implied by this choice.}. 
To identify pseudobulges in \wav\ images we adopt a criterion based on the S\'{e}rsic 
index and examine \emph{a posteriori}
the sources location in the \wav\ FP of elliptical galaxies.
\begin{table*}{SAMPLE PROPERTIES.}
\begin{scriptsize}
\begin{center}
\centerline{\begin{tabular}{lcccccc}
\hline
Source Name & Type  &Distance& \mbh\             & $\sigma$    & $\log$L$_V$             & $\log$L$_K$ \\
 (1)   &  (2)&      (3)    &  (4)                    & (5)        & (6)               & (7)         \\
\hline
Circinus & Sb  &     4.0     &   0.017$^{+0.004}_{-0.003}$   & 158$\pm$18 &  -                & 10.03       \\ 
IC1459   & E4  &       30.9  &    28$^{+11}_{-12.0}$         & 340$\pm$17 & 10.96$^{+0.06}_{-0.06}$ & 11.59       \\
IC2560   & SBb &     40.7    &  0.044$^{+0.044}_{-0.022}$    & 144$\pm$7  &  -                & 10.46       \\
IC4296   & E   &     50.8    &  13.5$^{+2}_{-2}$             & 226$\pm$10 &  -                & 12.36       \\
NGC221   & E2  &     0.86    &  0.031$^{+0.006}_{-0.006}$    & 75 $\pm$3  & $8.66^{+0.02}_{-0.02}$  & 8.83        \\ 
NGC524   & S0  &     33.6    &  8.32$^{+0.60}_{-0.37}$       & 235$\pm$12 & -                 & 11.46       \\
NGC821   & E4  &     25.5    &  0.42$^{+0.28}_{-0.08}$       & 209$\pm$10 & $10.42^{+0.05}_{-0.06}$ & 10.8        \\
NGC1023  & SB0 &     12.1    &  0.46$^{+0.05}_{-0.05}$       & 205$\pm$10 & $10.17^{+0.10}_{-0.13}$ & 10.54       \\
NGC1068  & Sb  &      15.4   &  0.086$^{+0.003}_{-0.003}$    & 151$\pm$7  & $10.8^{+0.07}_{-0.09}$  & 10.82       \\
NGC1300  &SB(rs)bc&  20.1    &  0.71$^{+0.69}_{-0.35}$       & 218$\pm$10 & -                 & -           \\
NGC1316  & SB0 &     19      &  1.62$^{+0.28}_{-0.27}$       & 226$\pm$11 & -                 & 11.2        \\
NGC2549  & S0  &     12.3    &  0.14$^{+0.10}_{-0.12}$       & 145$\pm$7  & -                 &  10.18      \\
NGC2748  & Sc  &     24.9    &  0.47$^{+0.38}_{-0.38}$       & 115$\pm$5  & -                 & -           \\
NGC2778  & E2  &     24.2    &  0.16$^{+0.90}_{-0.10}$       & 175$\pm$8  & $9.78^{+0.05}_{-0.06}$  & -           \\
NGC2787  & SB0 &     7.9     &  0.43$^{+0.04}_{-0.05}$       & 189$\pm$9  & -                 & 9.86        \\
NGC2974  & E4  &     21.5    &  1.7$^{+0.3}_{-0.3}$          & 227$\pm$11 & -                 & 10.95       \\
NGC3031  & Sb  &     4.1     &  0.80$^{+0.2}_{-0.11}$        & 143$\pm$7  & -                 & 10.44       \\
NGC3079  & SBcd&     15.9    &   0.025$^{+0.002}_{-0.002}$   & 146$\pm$7  & -                 & 10.26       \\
NGC3115  & S0  &     10.2    &   9.6$^{+5.4}_{-2.9}$         & 230$\pm$11 & $10.4^{+0.02}_{-0.02}$  & 10.46       \\
NGC3227  & SBa &     17.0    &  0.20$^{+0.37}_{-0.08}$       & 133$\pm$6  & -                 & 9.92        \\
NGC3245  & S0  &     22.1    &  2.2$^{+0.5}_{-0.5}$          & 205$\pm$10 & -                 & 10.69       \\
NGC3368  & SABa&    10.4     &  0.075$^{+0.015}_{-0.015}$    &98.5$\pm$5  & -                 &  -          \\
NGC3377  & E6  &     11.7    &  1.1$^{+1.1}_{-0.1}$          & 145$\pm$7  & $9.97^{+0.04}_{-0.04}$  & 10.22       \\
NGC3379  & E0  &     11.7    &  1.2$^{+0.8}_{-0.58}$         & 206$\pm$10 & $10.37^{+0.01}_{-0.01}$ & 10.96       \\
NGC3384  & SB0 &     11.7    &  0.18$^{+0.01}_{-0.03}$       & 143$\pm$7  & $9.90^{+0.08}_{-0.10}$   & 10.44       \\
NGC3414  & S0  &     25.2    &  2.51$^{+0.30}_{-0.31}$       & 205$\pm$10 & -                 & 10.71       \\
NGC3489  & SB0 &    12.1     & 0.06$^{+0.0012}_{-0.0012}$    & 91 $\pm$5  & -                 &  -          \\
NGC3585  & S0  &     21.2    &  3.4$^{+1.5}_{-0.8}$          & 213$\pm$10 & $10.65^{+0.07}_{-0.09}$ & 11.18       \\
NGC3607  & E1  &     19.9    &  1.2$^{+0.4}_{-0.41}$         & 229$\pm$11 & $10.58^{+0.04}_{+0.04}$ & 11.09       \\
NGC3608  & E1  &     23      &  2.1$^{+1.1}_{-0.7}$          & 182$\pm$9  & $10.35^{+0.04}_{-0.04}$ & 10.81       \\
NGC3998  & S0  &     14.9    &  2.4$^{+2.1}_{-1.8}$          & 305$\pm$15 & -                 & 10.63       \\
NGC4026  & S0  &     15.6    &  2.1$^{+0.7}_{-0.4}$          & 180$\pm$9  & $9.86^{+0.07}_{-0.09}$  & 10.57       \\
NGC4151  & Sa  &     20.0    &  0.65$^{+0.07}_{-0.07}$       & 156$\pm$8  & -                 & 10.27       \\
NGC4258  &SABbc&     7.2     &  0.378$^{+0.001}_{-0.001}$    & 115$\pm$10 & -                 & 9.93        \\
NGC4261  & E2  &     33.4    &  5.5$^{+1.1}_{-1.2}$          & 315$\pm$15 & $11.02^{+0.02}_{-0.02}$ & 11.42       \\
NGC4374  & E1  &     17      &  15$^{+11}_{-6}$              & 296$\pm$14 & $10.91^{+0.02}_{-0.02}$ & -           \\
NGC4459  & E2  &     17      &  0.74$^{+0.14}_{-0.14}$       & 167$\pm$8  & $10.35^{+0.02}_{-0.02}$ & 10.58       \\
NGC4473  & E5  &     17      &  1.3$^{+0.5}_{-0.94}$         & 190$\pm$9  & $10.38^{+0.02}_{-0.02}$ & 10.92       \\
M87      & E1  &     17      &  36$^{+10}_{-10}$             & 375$\pm$18 & $11.1^{+0.02}_{-0.02}$  & 11.46       \\
NGC4486A & E2  &     17      &  0.13$^{+0.05}_{-0.04}$       & 111$\pm$5  & $9.41^{+0.02}_{-0.02}$  & 10.19       \\
NGC4552  & E   &     15.3    &  5.0$^{+0.6}_{-0.4}$          & 252$\pm$12 & -                 & 11.04       \\
NGC4564  & S0  &     17.0    &  0.69$^{+0.04}_{-0.10}$       & 162$\pm$8  & $9.77^{+0.11}_{-0.15}$  & 10.35       \\
NGC4594  & Sa  &     10.3    &  5.7$^{+4.3}_{-4.0}$          & 240$\pm$12 & $10.9^{+0.06}_{-0.06}$  & -           \\
NGC4596  & SB0 &     18.0    &  0.84$^{+0.36}_{-0.25}$       & 136$\pm$6  & -                 & 10.44       \\
NGC4621  & E5  &     18.3    &  4.0$^{+0.5}_{-0.4}$          & 225$\pm$11 & -                 & 10.77       \\
NGC4649  & E2  &     16.5    &  21$^{+5}_{-6}$               & 385$\pm$19 & $10.99^{+0.02}_{-0.02}$ & 11.46       \\
NGC4697  & E6  &     12.4    &  2.0$^{+0.2}_{-0.2}$          & 177$\pm$8  & $10.44^{+0.04}_{-0.05}$ & 10.52       \\
NGC5077  & E3  &     44.9    &  8.0$^{+5.0}_{-3.3}$          & 222$\pm$11 & $10.74^{+0.05}_{-0.06}$ & 11.35       \\
CenA     & S0/E&     4.4     &  3.0$^{+0.4}_{-0.2}$          & 150$\pm$7  & $10.66^{+0.03}_{-0.03}$ & 10.49       \\
NGC5576  & E3  &     27.1    &  1.8$^{+0.3}_{-0.4}$         & 183$\pm$9  & $10.43^{+0.05}_{-0.06}$ & 10.5        \\
NGC5813  & E1  &     32.2    &  7.1$^{+0.9}_{-0.6}$          & 230$\pm$11 & -                 & 11.06       \\
NGC5845  & E3  &     28.7    &  2.9$^{+0.5}_{-1.7}$          & 234$\pm$11 & $9.84^{+0.05}_{-0.06}$  & 10.62       \\
NGC5846  & E0  &     24.9    &  11.0$^{+1.1}_{-1.0}$         & 237$\pm$12 & -                 & 11.33       \\
NGC6251  & E1  &     106     &  6.0$^{+2.0}_{-2.0}$          & 290$\pm$14 & -                 & 11.81       \\
NGC7052  & E3  &     70.9    &  4.0$^{+2.4}_{-1.6}$          & 266$\pm$13 & -                 & 11.39       \\
NGC7457  & S0  &     14.0    &  0.041$^{+0.012}_{-0.017}$    & 67 $\pm$3  & $9.42^{+0.04}_{-0.05}$  & 9.69        \\
NGC7582  & SBab&     22.0    &  0.55$^{+0.26}_{-0.19}$       & 156$\pm$19 & -                 & -           \\
\hline
\end{tabular}}
\end{center}
\linespread{0.4}
\vspace{-.6cm}
\caption{Columns: (1) galaxy name. (2) Hubble type taken from the Hyper-Leda catalogue. (3) Distance in Mpc.  
(4) \mbh\ in $10^8$\msun, see G09, H09 and Nowak et al.~(2010) for references on BH mass measurements. (5) Velocity dispersion in km/s. 
(6)-(7) logarithmic V- and K-band luminosities in solar units. The $1\sigma$ error of L$_K$ is $10\%$ (H09). 
All values in columns (4)-(7) are from G09 and H09 if not included in V-band sample, 
and are scaled from the original publications to our preferred distances (column 3) assuming the standard cosmology.
}
\label{tb:properties}
\end{scriptsize}
\end{table*}
\subsection{IRAC images}
Given the wavelength dependence of the emission from 
stellar photospheres, the \wav\ IRAC band 
is the best suited for our study. 
In fact, the STARBURST99 models (Leitherer et al.~1999) show that only $\sim10\%$ and even less of the 
emission in the 4.5~$\mu$m band is stellar, and that at longer wavelengths the fraction is even lower (Helou et al.~2004). 
Hence we used only channel 1 (centered at $3.55~\mu$m) basic calibrated data (BCD).

Individual BCD frames were processed with  version 18.7 of the SSC pipeline. 
The image mosaicking and source extraction package (MOPEX, Makovoz \& Marleau 2005) was 
used to co-add BCD frames for each source. Bad pixels, i.e.~pixels masked in the 
Data Collection Event (DCE) status files and in the static masks (\textit{pmasks}), were 
ignored. Additional inconsistent pixels were removed by means of the MOPEX outlier 
rejection algorithms. We relied on the dual outlier technique, 
together with the multi-frame algorithm. 
The frames were corrected for geometrical distortion and projected onto 
a North-East coordinate system with pixel sizes of 1.20~arcsec, equivalent to 
the original pixels. Mosaics for data images were realized with standard linear interpolation. 
The same was done for the uncertainty images, i.e.~the maps for the
standard deviations of the data frames.
The signal-to-noise ratio in our post-pipeline MOPEX mosaics are comparable to or better than 
those in the SSC products.
M87 has two maps with different spatial extension, thus 
we average them with a weighted mean where the weights are given by uncertainty images. 
\begin{table*}{DECOMPOSITION PARAMETERS}
\begin{center}
\centerline{\begin{tabular}{lcccccccccccc}
\hline
Source Name& $n$  &     $m_{bulge}$   &     $R_e$       &  $(b/a)_b$ & $C_0$ & $m_{disk}$ & $R_0$ &$(b/a)_d$ & $m_{bar}$& FWHM & $(b/a)_{bar}$ & $m_{AGN}$ \\
 (1)   &  (2)   &        (3)       &     (4)          &  (5)       &    (6)   &    (7)     & (8)   & (9)      &   (10)   & (11) &    (12)     & (13)      \\
\hline
Circinus &   2    &  5.48$\pm$ 0.04  &   0.21$\pm$0.01	& 0.68       &   0.2    & 4.4        &1.58   & 0.39     & -     &   -     & -           & - \\
IC1459   &    6   &  6.15$\pm$ 0.3   &   9.15$\pm$3.71	& 0.76       &   -0.14  & -          &-      & -        & -     &   -     & -           & 12        \\
IC2560   &   2    &  9.1 $\pm$ 0.7  &   5.43$\pm$2.84	& 0.37       &   0.62   & 9.0        &6.18   & 0.57     & -     &   -     & -           & 10.8      \\
IC4296   &    4   &  7.09$\pm$ 0.5   &   8.28$\pm$5.91	& 0.92       &   0.66   & -          &-      & -        & -     &   -     & -           & 11.6      \\
NGC221   &    4   &  5.2 $\pm$ 0.63  &   0.12$\pm$0.07	& 0.7        &   0.08   & 6.5        &0.09   & 0.89     & -     &   -     & -           & -         \\
NGC524   &    3   &  7.01$\pm$ 0.19  &   4.37$\pm$2.54	& 0.95       &   -0.07  & 8.7        &4.65   & -        & -     &   -     & -           & -         \\
NGC821   &    7   &  6.9 $\pm$ 0.26  &   7.86$\pm$4.15	& 0.62       &   -0.19  & -          &-      & -        & -     &   -     & -           & -         \\
NGC1023  &    3   &  6.7 $\pm$ 0.18  &   1.41$\pm$0.80	& 0.61       &   -0.55  & 6.67       &48.51  & 0.3      & 9.01  &   1.13  & 0.55        & -         \\
NGC1068  &    1   &  6.15$\pm$ 0.05  &   0.73$\pm$0.02	& 0.73       &   0.19   & 6.1        &1.76   & 0.81     & -     &   -     & -           & 7.37      \\
NGC1300  &    3   &  8.40$\pm$ 0.33  &  8.32 $\pm$1.30  & 0.29      &   0.0    & 7.56       &7.21   & 0.74     & -     &   -     & -           & -         \\
NGC1316 &    5   &  4.7 $\pm$ 0.8   &   8.57$\pm$0.55	& 0.71       &   -0.15  & 6.91       &11.52  & 0.63     & -     &   -     & -           & 11.13     \\
NGC2549  &    7   &  8.32$\pm$ 0.13  &  0.69 $\pm$0.15  & 0.5        &   -0.42  & 9          &1.81   & 0.24     & -     &   -     & -           & -         \\
NGC2748  &    4   &  9.64$\pm$ 0.09  &  1.87 $\pm$0.54  & 0.95       &   -0.31  & 8.72       &1.9    & 0.24     & -     &   -     & -           & -         \\
NGC2778  &    2.5 &  10.44$\pm$0.15  &   0.29$\pm$0.05	& 0.73       &   0.09   & 9.74       &1.22   & 0.79     & -     &   -     & -           & -         \\
NGC2787  &    3   &  7.57$\pm$ 0.57  &   0.60$\pm$0.19	& 0.62       &   0.0    & 8.07       &1.12   & 0.52     & 11.5   &   0.17     & 0.11           & 12.3         \\
NGC2974  &    3   &  7.28$\pm$ 0.14  &   2.83$\pm$0.61	& 0.67       &   -0.06  & -          &-      & -        & -     &   -     & -           & 11.8      \\
NGC3031  &    3   &  4   $\pm$ 0.31  &   2.53$\pm$0.87	& 0.62       &   -0.03  & 4.51       &3.35   & 0.49     & -     &   -     & -           & 10.2      \\
NGC3079  &   2    &  6.87$\pm$ 0.08  &   5.71$\pm$0.37	& 0.17       &   0.33   & 9.48       &5.31   & 0.86     & 10    &   0.27  & 0.13        & 9.64      \\
NGC3115  &    3   &  5.99$\pm$ 0.1   &   1.35$\pm$0.15	& 0.33       &   -0.62  & 6.69       &3.4    & 0.41     & -     &         & -           & -         \\
NGC3227 &    4   &  7.27$\pm$ 0.5   &   6.83$\pm$2.27	& 0.49       &   0.18   & 9.44       &2.36   & 0.44     & -     &         & -           & -         \\
NGC3245  &    2.5 &  8.73$\pm$ 0.2   &   0.49$\pm$0.04	& 0.58       &   -0.13  & 8.14       &2.3    & 0.52     & -     &         & -           & -         \\
NGC3368  &   1    &  6.57$\pm$ 0.1   &   2.3 $\pm$0.12  & 0.63       &   -0.13  & 6.83       &6.17   & 0.58     & 7.25  &   0.7   & 0.55        & -         \\
NGC3377  &    6   &  6.93$\pm$ 0.18  &   3.13$\pm$0.88	& 0.55       &   0.03   & -          &-      & -        & -     &   -     & -           & -         \\
NGC3379  &    5   &  5.85$\pm$ 0.16  &   2.61$\pm$0.64	& 0.88       &   0.03   & -          &-      & -        & -     &   -     & -           & -         \\
NGC3384  &    2.5 &  8.11$\pm$ 0.07  &   0.25$\pm$0.01	& 0.76       &   -0.04  & 7.06       &2.5    & 0.49     & 8.73  &   1.13  & 0.79        & -         \\
NGC3414  &    5   &  7.87$\pm$ 0.5   &  2.67 $\pm$1.29  & 0.75       &   -0.44  & 9.28       &3.04   & 0.87     & 12.4  &   0.29  & 0.36        & -         \\
NGC3489  &   1.5  &  8.3 $\pm$ 0.45  &   0.27$\pm$0.21  & 0.82       &   0.15   & 7.7        &1.11   & 0.54     & -     &   -     & -           & -         \\
NGC3585  &    2.5 &  7.21$\pm$ 0.63  &   1.59$\pm$1.97	& 0.55       &   -0.28  & 7.08       &5.25   & 0.61     & -     &   -     & -           & -         \\
NGC3607  &    5   &  6.44$\pm$ 0.25  &   4.3 $\pm$2.02  & 0.83       &   0.0    & -          &-      & -        & -     &   -     & -           & 12.9      \\
NGC3608  &    6   &  7.19$\pm$ 0.23  &   6.29$\pm$2.51	& 0.79       &   0.12   & -          &-      & -        & -     &   -     & -           & -         \\
NGC3998  &   1.5  &  8.14$\pm$ 0.05  &   0.34$\pm$0.36  & 0.82       &   0.07   & 7.75       &1.51   & 0.79     & -     &   -     & -           & 10.3      \\
NGC4026  &    3.5 &  8.08$\pm$ 0.18  &   0.86$\pm$0.30	& 0.60       &   0.0     & 8.36       &2.34   & 0.16     & -     &   -     & -           & -         \\
NGC4151  &    3.5 &  8.13$\pm$ 0.41  &   0.52$\pm$0.42	& 0.87       &   0.07   & 8.17       &2.99   & 0.61     & -     &   -     & -           & 14.6      \\
NGC4258  &    2   &  5.27$\pm$ 0.08  &   3.9 $\pm$0.42  & 0.46       &   0.18   & 8.73       &6.84   & 0.66     & -     &   -     & -           & 9.6       \\
NGC4261  &    4   &  7.23$\pm$ 0.15  &   3.66$\pm$2.80	& 0.80       &   0.0    & 8.81        &6.65   & 0.99     & -     &   -     & -           & 13.09     \\
NGC4374  &    7   &  5.06$\pm$ 0.04  &   8.73$\pm$1.00	& 0.90       &   0.0  & -          &-      & -        & -     &   -     & -           & 11.7       \\
NGC4459  &    2.5 &  7.7 $\pm$ 0.1   &   0.85$\pm$0.10	& 0.83       &   0.02   & 7.7        &2.66   & 0.76     & -     &   -     & -           & -         \\
NGC4473  &    7   &  6.59$\pm$ 0.05  &   4.06$\pm$0.87	& 0.56       &   -0.07  & -          &-      & -        & -     &   -     & -           & -         \\
M87   &    4   &  5.10$\pm$ 0.05  &   8.20$\pm$0.60	& 0.82       &   0.0   & -          &-      & -        & -     &   -     & -           & 12.26     \\
NGC4486A &    2.5 &  9.04$\pm$ 0.08  &   0.59$\pm$0.07	& 0.8        &   0.03   & -          &-      & -        & -     &   -     & -           & -         \\
NGC4552  &    4   &  6.65$\pm$ 0.08  &   1.8 $\pm$0.81  & 0.92       &   -0.05  & 7.78       &7.17   & 0.8      & -     &   -     & -           & 11.5      \\
NGC4564  &    7   &  7.76$\pm$ 0.13  &   2.06$\pm$0.03	& 0.62       &   0.33   & 9.46       &1.39   & 0.2      & -     &   -     & -           & -         \\
NGC4594  &    1.5 &  5.51$\pm$ 0.14  &   3.3 $\pm$0.35  & 0.26       &   -0.93  & 5.79       &2.74   & 0.61     & 6.32  &   7.79  & 0.71        & 9.64      \\
NGC4596  &    3   &  7.6 $\pm$ 0.05  &   2.44$\pm$0.12	& 0.53       &   -0.23  & 7.66       &4.91   & 0.83     & 11.76 &   0.05  & 0.99        & -         \\
NGC4621  &    5   &  6.22$\pm$ 0.1   &   5.46$\pm$1.09	& 0.63       &   -0.19  & -          &-      & -        & -     &   -     & -           & -         \\
NGC4649  &    3 &  5.55$\pm$ 0.10  &   3.77$\pm$0.34	& 0.82       &   0.0   & 8.3          &3.3      & 0.93        & -     &   -     & -           & -         \\
NGC4697  &    5   &  5.69$\pm$ 0.09  &   6.04$\pm$1.39	& 0.65       &   -0.1   & -          &-      & -        & -     &   -     & -           & -         \\
NGC5077  &    6   &  7.61$\pm$ 0.26  &   6.35$\pm$2.34	& 0.71       &   0.11   & -          &-      & -        & -     &   -     & -           & 15.53     \\
CenA     &    3.5 &  3.49$\pm$ 0.27  &   2.21$\pm$1.2	& 0.83       &   0.0  & 4.16       &3.92   & 5.20     & 0.50     &   -     & -           & 10.2      \\
NGC5576  &    7   &  7.26$\pm$ 0.15  &   4.51$\pm$1.29	& 0.71       &   0.03   & -          &-      & -        & -     &   -     & -           & -         \\
NGC5813  &    6   &  6.52$\pm$ 0.18  &   17.4$\pm$3.80	& 0.78       &   0.1    & -          &-      & -        & -     &   -     & -           & -         \\
NGC5845  &    3   &  8.96$\pm$ 0.02  &   0.51$\pm$0.02	& 0.67       &   -0.06  & -          &-      & -        & -     &   -     & -           & -         \\
NGC5846  &    3   &  6.70$\pm$ 0.06  &   4.40$\pm$0.90	& 1          &   0.0    & -          &-      & -        & -     &   -     & -           & -         \\
NGC6251  &    7   &  8.2 $\pm$ 0.19  &   21.8$\pm$7.71  & 0.84       &   -0.09  & -          &-      & -        & -     &   -     & -           & 12.8      \\
NGC7052  &    5   &  7.96$\pm$ 0.16  &   13.5$\pm$3.81  & 0.49       &   0.46   & -          &-      & -        & -     &   -     & -           & -         \\
NGC7457  &    7   &  9.54$\pm$ 0.90  &   0.94$\pm$0.81	& 0.52       &   0.19   & 8.37       &1.61   & 0.56     & -     &   -     & -           & -         \\
NGC7582  &    4   &  7.33$\pm$ 0.44  &   9.22$\pm$4.39	& 0.5        &   0.23   & 8.14       &3.18   & 0.22     & -     &   -     & -           & 8.62      \\
\hline
\end{tabular}}
\end{center}
\linespread{0.5}
\caption{Columns: (1) galaxy name. 
(2)-(6) bulge structural parameters: S\'{e}rsic index, \wav\ magnitude, effective radius 
in kpc, axis ratio, diskyness-boxiness ($C_0<0$ for disky ellipsoids, $C_0>0$ for the boxy ones). 
(7)-(9) disk properties: \wav\ magnitude, scale length in kpc, axis ratio. 
(10-12) bar properties: \wav\ magnitude, FWHM in kpc, axis ratio. (13) AGN \wav\ magnitude.
}
\label{tb:spitzer}
\end{table*}
\subsection{Image analysis}
We use the public program GALFIT V.3 (Peng, et al.~2002, 2007) 
to perform a two-dimensional decomposition of \wav\ images in bulge/disk components.
Besides the standard inputs (e.g. data, point spread function\footnote{We use the 
instrumental PSF image provided by the SSC, after checking that its full width 
at half maximum (FWHM) is consistent with what observed for the 
foreground stars in our images.} (PSF) images etc.), 
GALFIT requires  a standard-deviation image, used to give 
relative weights to the pixels during the fit, and a bad pixel mask. 
We therefore use the uncertainty data obtained from MOPEX as sigma images and construct 
a bad pixel frame masking out foreground stars, background galaxies and possible 
irregularly shaped regions such as dust lanes across the galaxy.

We choose the number and kind of model components on the basis of the Hubble morphological 
types (listed in Table~\ref{tb:obs}) and after a visual inspection of the images. 
Thus, for elliptical galaxies we use a pure S\'{e}rsic profile, while 
for lenticular (S0) and spiral galaxies we add an exponential disk. In the case of barred galaxies (SB), we consider an 
additional Gaussian component, equivalent to a S\'{e}rsic profile with index $n=0.5$. 
In some cases, even if the source is classified as a barred galaxy, the bar cannot be 
identified in the MIR (e.g.~NGC2778). In these cases we do not add any Gaussian component to the model. 
In the case of active galaxies, we also include a nuclear point source to account for the emission of the active galactic nucleus 
(AGN). 

The contribution of thermal dust to the observed emission at \wav\ can be powered  
by an active galactic nucleus (AGN) and/or star formation. 
The former can be accounted for by including a nuclear PSF in the model, while the latter can be masked out. 
Indeed intense star formation occurs in well-localized knots of emission, and does not affect
the entire galaxy disk. 
Overall we estimate that hot dust can contribute
no more than a few percent to the bulge (or disk) total luminosity and, given
its localized nature, should not significantly affect the structural parameters
derived from our global two-dimensional fits.

We fix the background in the fits, estimating it 
as the mean surface brightness (with the relative standard deviation) 
over an annular region surrounding the galaxy between 2-3 times the optical radius.
Foreground sources such as stars or galaxies are not considered in the background calculation
by means of a 2.5 sigma rejection criterion.

Some aspects of the photometric decomposition are potentially problematic. 
Due to the functional form of the S\'{e}rsic profile, the effective radius \re\ 
is coupled to the S\'{e}rsic index $n$. To control this degeneracy we follow 
Hunt, Pierini \& Giovanardi (2004, hereafter HPG04) and fit every 
galaxy with S\'{e}rsic indexes $n$ fixed to values 
$n=~1,1.5,2,2.5,3,3.5,4,5,6,7$, which reasonably span all galaxy morphologies. 
We then assume $0.5$ as the error on the S\'{e}rsic index retrieved from the best fit (see below for
the criteria defining the best model).

In addition, background measurements can be affected by galaxy extended emission; 
indeed, in several cases the faint tail of source emission covers most of the 
field-of-view (NGC3115, NGC4621, NGC4374, NCG5846, NGC7052 are some extreme examples).
Thus for each galaxy, we fix three different 
background values: $s, s-\sigma_s$ and $s+\sigma_s$, where $s$ and $\sigma_s$ are the sky 
flux and its standard deviation estimated as described above. This allows to account for  
possible large variations of the background. 

Once the number of components is defined, we run 
30 different models; for each of the 10 fixed $n$ values we use the 3 different background estimates.  
The free parameters are the bulge and disk brightness\footnote{To compute the 
magnitudes at \wav\ we adopt a zero point 
of 17.25 in Vega magnitudes, according to the IRAC photometric system (Reach et al.~2005).}, their scale lengths,
ellipticities, position angles and the bulge diskyness/boxyness. Additional free 
parameters can be the bar brightness, FWHM, ellipticity and position angle and/or the 
AGN point-source \wav\ magnitude. 
As starting guesses we use the axis 
ratio, the major axis position angle, 
provided by the HyperLeda\footnote{http://leda.univ-lyon1.fr/} database.
The initial disk and bulge scale lengths ($R_0$ and $R_e$ respectively), are proportional 
to the optical radius ($R_{opt}$ listed in HyperLeda): $R_0\sim0.25R_{opt}$; 
the effective radius follows the bulge shape according to the S\'{ersic} index 
$R_e\sim\frac{n}{10}R_{opt}$ for $n<4$ and $R_e\sim0.4R_{opt}$ for $n\geq4$ 
(Giovanardi \& Hunt~1988, Moriondo, Giovanardi \& Hunt~1998). 

As shown by HPG04, the bulge component influences the $\chi^2$  value only in the inner region of the 
galaxy; indeed, the reduced $\chi^2$ depends on $n$ only at small distances (radii) from the center, while the large area 
of the disk strongly dilutes the influence of the bulge on the global $\chi^2$. 
Appendix~A (available online) shows examples of the trend reduced $\chi^2$ with radii for our galaxies outlining that fits with acceptable $\chi^2$ values can differ significantly in the inner region where the bulge is important.
Therefore, following HPG04, we selected the fits with the best $n$ and sky values considering  the lowest reduced $\chi^2$ 
within half optical radius $R_{opt,0.5}$. The fits are then considered reliable on the basis of 
circumnuclear flux conservation; 
the ratio between the inner counts in the residual and source images must 
lower than 0.05. This means that less than $5\%$ of the central flux can be lost or introduced by the model. 
As described in the following section, the recovery of a tight FP at \wav\  shows the accuracy of our analysis in estimating bulge structural parameters. 

The absolute errors on the free parameters are derived as the variations between the 
best fit values and the ones obtained from the model having the closest 
$s,~s+\sigma_s$ or $s-\sigma_s$ value and the lowest possible 
reduced $\chi^2$ within $R_{opt,0.5}$ according to the criterion described above. 
With this choice we overestimate the statistical fit errors, but 
carefully constrain the effects due to an uncertain estimate of the background.
The fit parameters and their uncertainties are listed in Table~\ref{tb:spitzer}. 
See also Appendix~A (available online) for further details on the error estimates. 
\subsection{The fundamental plane at \wav\ and the identification of pseudobulges}

Early type galaxies and the bulges of disk galaxies follow the well known relations connecting their effective radius \re,
mean surface brightness $<I_e>$ within \re, and velocity dispersion
$\sigma$ (e.g. Djorgovsky \& Davis~1987, Prugniel \& Simien~1996, Jorgensen, Franx \& Kjaergaard~1996, 
Burstein et al.~1997, Cappellari et al.2006). 
In particular, Jun \& Im (2008, hereafter JI08) studied the FP at \wav\ considering a sample of elliptical galaxies spanning an order of magnitude in effective radii. 
Here, to analyze the bulge properties and verify the reliability of our analysis, we adopt the 
relation found by JI08: 
\begin{equation}
\log R_e=1.55\times\log\sigma-0.89\times\log<I_e>-9.89,
\label{eq:fp}
\end{equation}
where $\sigma$ in km/s, \re\ is in kpc and $<I_e>$ in mag~arcsec$^{-2}$. 
\begin{figure*}
\begin{center}
\includegraphics[scale=0.9]{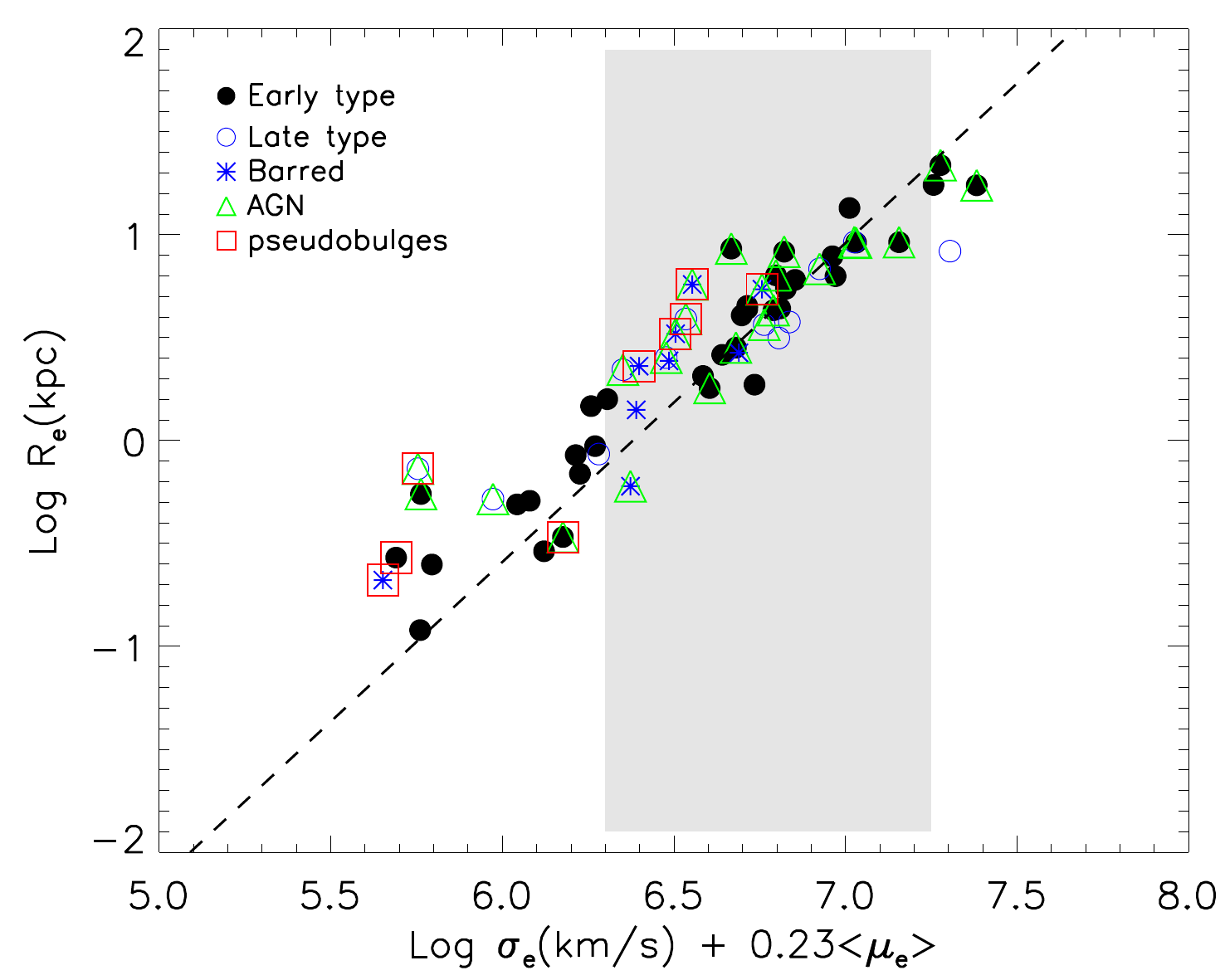}
\end{center}
\linespread{1.0}
\caption{
MIR fundamental plane for our decomposed bulges. 
The \wav\ FP coordinate axes are given by equation~(\ref{eq:fp}) (JI08), represented by the black dashed line. 
The bulge parameters are the
effective radius \re\ in kpc (Tab.~\ref{tb:spitzer}), the velocity dispersion $\sigma$ 
in km/s (Tab.~\ref{tb:properties}) and the mean surface brightness within \re, 
$<\mu_e>=m_{bulge}+2.5\log[2\pi (b/a)_b R_e^2]$, where $(b/a)_b$ is the bulge axis ratio.
$<\mu_e>$ is expressed in mag/arcsec$^2$ (for a consistent comparison with JI08, 
Vega magnitudes in Tab.\ref{tb:spitzer} are converted 
to the AB photometric system as $m$(AB)=$m$(Vega)+2.79). 
The color codes reflect the adopted 
fitting model, i.e.~black points correspond to pure S{\'e}rsic profiles, 
blue symbols refer to  
S{\'e}rsic and an exponential disk (open circles), possibly with an additional Gaussian 
component (asterisks). The empty green triangles correspond to active galaxies, whose AGN emission 
is reproduced adding a point-like source to the models listed above. Open 
red squares highlight bulges with $n\leq2$ which we thus classify as pseudobulges 
according to the FD10 criterion; no additional components are added to model these galaxies. 
The gray-shaded region highlights the range of velocity dispersion and surface brightness of the JI08 sample.
} 
\label{fg:fp}
\end{figure*}
%

Figure~\ref{fg:fp} shows the \wav\ FP for all our decomposed bulges obtained by 
using equation~(\ref{eq:fp}) (dashed black line).  
From Fig.~\ref{fg:fp} we can draw two main conclusions: (\emph{i}) our bulges follow the \wav\ 
FP of elliptical galaxies with low dispersion, and (\emph{ii}) pseudobulges (open red squares) 
lie on the FP within the observed dispersion. 
Indeed, (\emph{i}) if we follow JI08 and apply a maximum likelihood method with a bootstrap 
resampling to account for errors, the intrinsic dispersion is: $0.20\pm0.04$~dex for the entire sample, 
and $0.11\pm0.04$~dex for non-active galaxies. The last result is consistent within the errors 
with JI08 ($rms=0.09\pm0.04$, private communication). Thus the presence of an AGN can 
slightly worsen the 2D decomposition, introducing some degeneracy with the bulge component. 
Nevertheless, the low dispersion of the \wav\ FP for our bulges and the absence of strong outliers 
suggests that our 2D decomposition is reliable. 
A tight FP and its extension to reliably compact bulges then permit us to calibrate 
M/L (Fig.~\ref{fg:ml} and see Section~3.2). 

A detailed analysis of the FP is beyond the scope of this paper.
However, it should be noted that our sample spans an order of magnitude larger 
range in galaxy sizes, extending the FP to smaller galaxy radii than in JI08.  
This could be the origin of the possible small disagreement between the FP defined by our 
sources and the determination by JI08 (see Fig.~\ref{fg:fp}).

Classical bulges and pseudobulges differ in morphology and dynamics: the first 
behave like early-type galaxies, scaled to small sizes, surrounded by a disk component. 
Classical bulges thus are either prolate or oblate spheroids and are 
kinematically\footnote{The bulge kinematics are defined according to Erwin~(2010) where V is the 
stellar rotational velocity deprojected to its in-plane value and $\sigma$ is the 
stellar velocity dispersion.}
hot, (tending to be pressure dominated with a 
stellar $V/\sigma<1$). On the contrary pseudobulges are disk-like components with 
$V/\sigma>1$ cold kinematics dominated by rotation and possibly containing disk 
features such as nuclear bars 
and rings as well as young stellar populations 
(Andredakis, Peletier \& Balcells~1995, Fisher \& Drory~2010, hereafter FD10, and references therein). 
A correlation between the bulge morphology and the S\'{e}rsic index has been 
proposed by considering the morphological 
properties of pseudobulges as shown in FD10. 
We thus follow the FD10 S\'{e}rsic index criterion to classify those bulges in 
spirals and lenticulars with $n<2$ as pseudobulges. 
By means of the S\'{e}rsic index, we identify nine disk galaxies as hosting pseudobulges: 
Circinus, IC2560, NGC1068, NGC3079, NGC3368, NGC3489, NGC3998, NGC4258, NGC4594. 
The pseudobulges (\emph{ii}) which we have identified are marked with open red squares in Fig.~\ref{fg:fp} and are not outliers of the  
\wav\ FP. This is not surprising since the physical meaning of the fundamental 
plane is that galaxies are virialized systems (Burstein et al.~1997) and this applies 
also to rotationally supported pseudobulges. 
Our sample partly overlaps the FD10 one, and for the 8 sources in common the classical versus 
pseudo classification of bulges is in perfect agreement.
\section{Results}
In the previous section we have verified the validity of our 2D decomposition by exploring 
the \wav\ FP for ellipticals, which turns out to be as tight as that 
observed by JI08 (private communication).
Here we study the relations between  \mbh\ and the bulge structural parameters  listed in Tab.~\ref{tb:properties}  and \ref{tb:spitzer}), respectively.
To analyze the \mbh-bulge scaling relations, we adopt three different fitting 
methods:

1. a bisector linear regression (Akritas \& Bershady~1996), which uses the bivariate 
correlated errors and intrinsic scatter (BCES) method. 
Whereas this method takes into account the intrinsic scatter, it does not allow any 
determination of it. Hence the intrinsic $rms$ has been estimated with 
a maximum likelihood method assuming normally distributed values.

2. the linear regression FITEXY method as modified by T02, that accounts for the intrinsic scatter 
by adding, in quadrature, a constant value to the error of the dependent variable 
in order to obtain a reduced $\chi^2$ of 1. 

3. a Bayesian approach to linear regression, LINMIX\_ERR (Kelly~2007), which accounts 
for measurement errors, non-detections and intrinsic scatter to compute the posterior 
probability distribution of parameters.

The \mbh-bulge relations we fit are in the following form:
\begin{equation}
\log M_\bullet/M_\odot=\alpha+\beta\times(x-x_0),
\label{eq:rel}
\end{equation}
where $x$ is the logarithm of a measured bulge structural parameter expressed in solar 
units and $x_0$ is its mean value, used to reduce the covariance between the fit parameters.
As described below, the three methods provide consistent results for the \mbh-bulge relations. 

Since the LINMIX\_ERR is argued to be among the most robust regression methods 
for reliable estimates of  the intrinsic dispersion (Kelly~2007), 
we use it to obtain the final results on the  \mbh-bulge scaling relations. 
We still use the BCES and FITEXY methods for comparison with previous works. 

The fitting results are plotted in Fig.~\ref{fg:mbh}, \ref{fg:mass}, and \ref{fg:sigma}, 
where we present the \mbh-\lbul, \mbh-\mdyn\ and \mbh-$\sigma$ relations respectively.
The \mbh-$\sigma$ relation is analyzed only to provide a consistent comparison 
with \mbh-\lbul\ and \mbh-\mdyn\ in terms of sample and fitting methods.

We exclude from the linear regressions the nine sources 
classified as pseudobulges. This allows us to verify, and possibly 
quantify, whether pseudobulges 
follow the same scaling relations as classical bulges. 
From a visual inspection of Fig.~\ref{fg:mbh}-\ref{fg:sigma}, pseudobulges with 
large BH masses (\mbh$>10^7$\msun) follow the relations while those with 
low BH masses deviate significantly from the scaling relations defined by classical bulges. 
Deviant pseudobulges are analyzed in the following sections and discussed in Section~4. 

In the following, after describing in details the \mbh-bulge relations at \wav, we 
compare them with the published results obtained at shorter wavelengths. 
Finally in Section~3.4 we explore the possible \mbh\ correlation with $\sigma$ and 
$R_e$ \emph{separately} as suggested by MH03, Hopkins et al.~2007, Graham~2008a. 
\subsection{\mbh\ versus \wav\ bulge luminosity}
\begin{figure*}
\begin{center}
\includegraphics[scale=0.7]{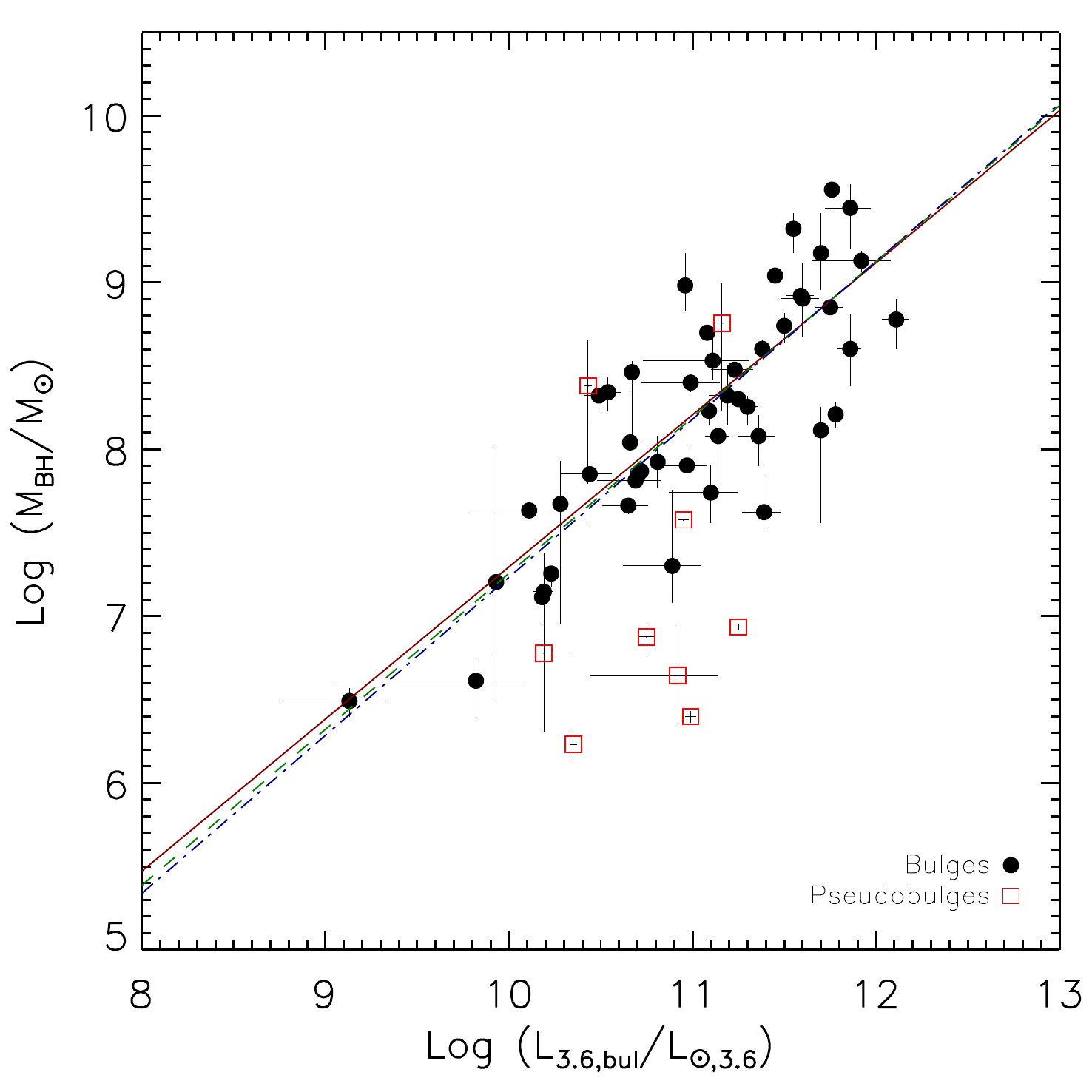}
\end{center}
\linespread{1.0}
\caption{Scaling relations. The \mbh\ as a function of \wav\ luminosity. 
The linear regressions are shown as dot dashed blue, dashed green and red continuous lines 
respectively for the BCES, FITEXY and LINMIX\_ERR methods and are obtained from classical bulges only (47 sources). Pseudobulges are open red squares.}
\label{fg:mbh}
\end{figure*}

To construct the \mbh-\lbs\ correlation, we compute the bulge \wav\ luminosity using
the magnitudes obtained with the two-dimensional decomposition (see Section 2.3 for details)
and reported in Tab. 3. To derive \wav\ luminosities in solar units we use the \wav\ Solar 
absolute magnitude obtained with a K-band value of 3.3 mag and a K-[3.6]=0.05~mag color
correction (Allen 1976, see also Bessell \& Brett 1988, Bell \& De Jong 2001).
The data for classical bulges are fitted with the three linear regression methods described
at the beginning of this section. We obtain the following scaling relation of \mbh\ with
\wav\ bulge luminosity (see Tab. 4 for the results from all fitting methods):
\begin{eqnarray}
\log\mbh/\msun & = & (8.19\pm0.06)+\nonumber\\
               &  & (0.93\pm0.10)\times[\log(\lbs/L_{3.6,\odot})-11]\nonumber\\
               &  & [rms=0.38\pm0.05]\nonumber \\
\label{eq:l}
\end{eqnarray}
%
\begin{table*}{LINEAR REGRESSION COEFFICIENTS}
\begin{center}
\centerline{\begin{tabular}{c c l c c c c}
\hline
$\log(\mbh)-\log(x)$ & $x_0$ & Method & $\alpha$ & $\beta$ & $rms$ & Ref. \\
\hline
                   & 11.0  & BCES   & $8.18\pm0.06$ & $0.95\pm0.10$ & $0.35\pm0.02$ & - \\
\mbh-\lbs\           & 11.0  & FITEXY & $8.21\pm0.06$ & $0.91\pm0.10$ & $0.35\pm0.05$ & - \\
                   & 11.0  &LINMIX\_ERR & $8.19\pm0.06$ & $0.93\pm0.10$ & $0.38\pm0.05$ & - \\
\hline
                   & 10.9  & BCES   & $8.21\pm0.07$ & $1.13\pm0.12$     & $0.31$ & MH03$^a$ \\
$L_{bul,K}$         & 10.9  & FITEXY & $8.29\pm0.08$ & $0.93\pm0.10$     & $0.33$ & G07 \\  
                   & 10.9  & BCES   & $8.38\pm0.05$ & $0.97\pm0.08$     & $0.36$ & H09 \\
\hline
$L_{bul,V}$         & 11.0  & MAX. LIKE.$\dagger$ & $8.95\pm0.11$ & $1.11\pm0.18$ & $0.38$ & G09$^b$ \\  
                   & 10.3  & BCES   & $8.41\pm0.11$ & $1.40\pm0.17$     & - & L07$\ddagger$ \\ 
\hline
                   & 11.0  & BCES   & $8.18\pm0.06$ & $0.83\pm0.08$ & $0.33\pm0.02$ & - \\
                   & 11.0  & FITEXY & $8.20\pm0.06$ & $0.80\pm0.08$ & $0.35\pm0.05$ & - \\
\mbh-\mdyn\        & 11.0  & LINMIX\_ERR & $8.20\pm0.06$ & $0.79\pm0.09$ & $0.37\pm0.05$ & - \\
                   & 10.9  & BCES   & $8.28\pm0.06$ & $0.96\pm0.07$ & $0.25$ & MH03$^a$ \\
                   & 11.0  & BCES   & $8.20\pm0.10$ & $1.12\pm0.06$ & $0.30$ & HR04 \\ 
                   & 10.9  & BCES   & $8.61\pm0.05$ & $0.88\pm0.06$ & $0.27$ & H09 \\
\hline      
                   & 11.0   & BCES   & $8.15\pm0.05$ & $0.80\pm0.08$ & $0.35\pm0.03$ & - \\
\mbh-\mstar\         & 11.0   & FITEXY & $8.17\pm0.06$ & $0.77\pm0.08$ & $0.35\pm0.05$ & - \\
                   & 11.0   & LINMIX\_ERR & $8.16\pm0.06$ & $0.79\pm0.08$ & $0.38\pm0.05$ & - \\
                   & 11.0   & BCES   & $8.24\pm0.05$ & $1.07\pm0.09$ & $0.32$ & H09 \\
\hline                       
                   & 200  & BCES   & $8.27\pm0.05$ & $4.0\pm0.3$ & $0.32\pm0.04$ & - \\
\mbh-$\sigma$        & 200  & FITEXY & $8.30\pm0.05$ & $4.0\pm0.3$ & $0.30\pm0.04$ & - \\
                   & 200  & LINMIX\_ERR & $8.29\pm0.05$ & $4.0\pm0.4$ & $0.33\pm0.04$ & - \\ 
\mbh-$\sigma$        & 200  & FITEXY & $8.13\pm0.06$ & $4.0\pm0.3$ & $0.3$ & T02 \\
\mbh-$\sigma$        & 200  & MAX. LIKE. & $8.23\pm0.08$ & $4.0\pm0.4$ & $0.31\pm0.06$ & G09$^b$ \\
\hline                                             
\end{tabular}}
\end{center}
\caption{Linear regression coefficients of \mbh\ as a function of the bulge property $x$, 
starting from the top of column (1): bulge \wav\ luminosity (equation~\ref{eq:l}) compared with 
previous works and with K and V band results; bulge dynamical (equation~\ref{eq:my}) and stellar
(equation~\ref{eq:mst}) masses; bulge velocity dispersion (equation~\ref{eq:s}) and effective radius 
(equation~\ref{eq:re}). 
The coefficients are obtained using the normalizations
in column (2) and with the methods listed in column (3). 
$\dagger$ G09 use a generalized maximum likelihood method to account for upper limits. 
Columns (4)-(6) list the intercept, slope and rms (in dex) respectively. 
We compare our results with the works listed in column (7). $\ddagger$ Lauer et al.~(2007). 
$^a$ Coefficients obtained cutting
the sample on the basis of the sphere of influence argument for the \mbh\ estimates. 
$^b$ Coefficients obtained only for elliptical galaxies.}
\label{tb:rel}
\end{table*} 
All the fit parameters obtained with the three different methods 
are consistent within the errors (see Tab.~\ref{tb:rel}, and Fig.~\ref{fg:mbh}). 
Our results are in excellent agreement with the \mbh-L$_{bul,\lambda}$ 
obtained at shorter wavelengths. 
More specifically using the BCES and FITEXY results for the comparison, 
the intercept, slope and $rms$ are fully consistent with the 
K-band measurements of MH03, Graham~(2007), and H09. 
We find an intrinsic dispersion for the \mbh-\lbs\ scaling relation with $rms=0.35\pm0.05$dex (BCES, FITEXY in Tab.\ref{tb:rel}), consistent 
with the correlation in the K-band: $rms=0.31$ and $0.36$ dex in MH03 and H09 respectively. 
Optical observations in the V-band (Lauer et al.~2007, G09) produce slightly 
higher values for $\alpha$ and $\beta$. 
Different linear correlation coefficients can be due to a combination of factors: 
most obviously color corrections, but also incomplete sample overlap and the range 
of Hubble types (G09 fit only early types). 
Furthermore our \mbh-\lbs\ is tighter than the V-band correlation 
whose $rms$ is $0.38\pm0.09$ for early-type galaxies only. 
Moreover extinction can 
affect the luminosity, especially the V-band. 

Figure~\ref{fg:mbh} illustrates that 4 out of 9 pseudobulges (NGC3489, NGC3998, NGC4258, NGC4594) 
are consistent with the correlation for classical bulges, while the others (CIRCINUS, IC2560, NGC1068, NGC3079, NGC3368) are outliers, 
at more than $4\sigma$ below the LINMIX\_ERR 
linear regression. 
The poor statistics for pseudobulges prevent us from drawing firm conclusion on the physical nature of 
this behavior. Nevertheless we examine the possible disagreement also for the other 
\mbh\ linear regressions in Section~4.

\subsection{\mbh\ versus bulge mass}
In this section we first describe two methods to derive the bulge mass, and then 
relate \mbh\ with \mdyn\ and \mstar. 
Indeed one of our aims is to verify whether the \wav\ waveband is suitable 
to investigate the \mbh-bulge mass relations. 
We determine (\emph{i}) the bulge dynamical mass \mdyn\ by applying the virial theorem (as in MH03, H09), 
and (\emph{ii}) the bulge stellar mass by calibrating the M/L as a function of the \wav\ 
luminosity both with and without V-[3.6] and K-[3.6] colors.
Thus (\emph{i}) the bulge \mdyn\ is computed as follows:
\begin{equation}
M_{dyn}=\kappa R_e \sigma^2/G,
\label{eq:md}
\end{equation}
where G is the gravitational constant, $\sigma$ and R$_e$ the velocity dispersion (in Tab.\ref{tb:properties})
and the bulge effective radius (in Tab.~\ref{tb:spitzer}) respectively. It is generally 
assumed that bulges behave as isothermal spheres and the factor $\kappa$ is usually fixed at a value 
of 3 (Gebhardt et al.~2003) or 8/3 (MH03). 
However, since bulges are well reproduced by S\'{e}rsic profiles, $\kappa$ is in general a
function of the S\'{e}rsic index $n$. Cappellari et al. (2006), using full dynamical models, provide
the $\kappa(n)$ relation for classical bulges and show that even the use of a constant average value
of $\bar\kappa = 5$ can provide an accurate estimate of \mdyn\ to within 0.10 dex. \
We have therefore adopted $\bar\kappa = 5$.

\begin{figure*}
\begin{center}
\includegraphics[scale=0.55,angle=-90]{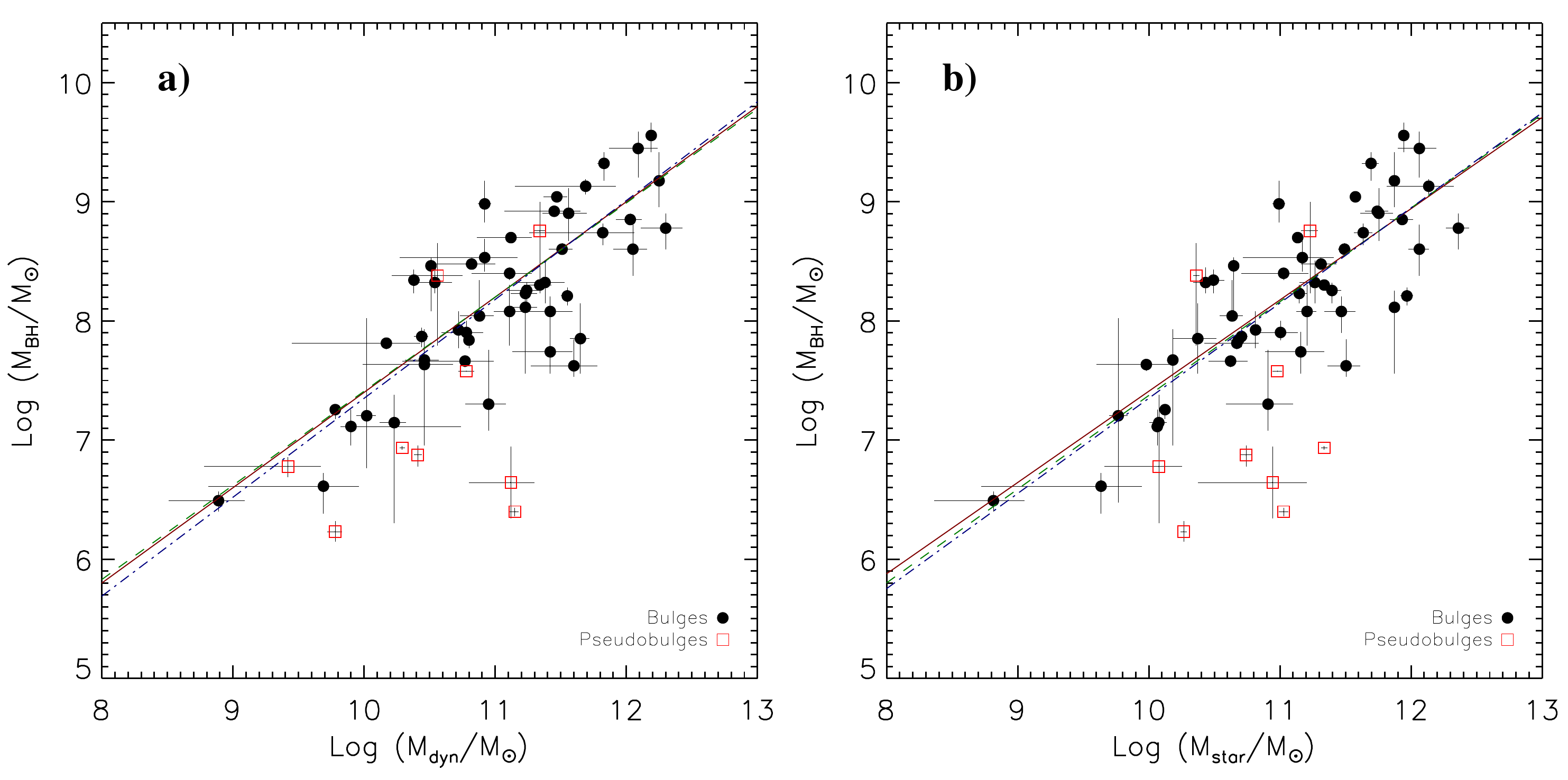}
\end{center}
\linespread{1.0}
\caption{Scaling relations. The \mbh\ as a function of bulge 
dynamical mass (panel a) and bulge stellar mass (panel b).
Pseudobulges are highlighted as red open squares. 
In panel (b) the M/L is from equation~\ref{eq:mll}. 
See Section~3.2 and Fig.~\ref{fg:ml} for details on the M/L calibrations.}
\label{fg:mass}
\end{figure*}

The \mbh-\mdyn\ bulge relation shown in Fig.~\ref{fg:mass}A together with the three 
linear regressions.
The \mdyn\ dependence for the 47 classical bulges is:
\begin{eqnarray}
\log\mbh/\msun& =& (8.20\pm0.06)+\nonumber\\
              &  & (0.79\pm0.09)\times[\log(\mdyn/\msun)-11]\nonumber \\
              &  & [rms=0.37\pm0.06]\nonumber\\
\label{eq:my}
\end{eqnarray}
The three linear regressions are consistent within the errors 
(See Tab.~\ref{tb:rel}, Fig.~\ref{fg:mass}A). 
The \mbh-\mdyn\ relation is rather  
tight with an $rms\sim0.35$~dex from LINMIX\_ERR, comparable to the scatter in 
the \mbh-$\sigma$ function (see Section~3.3). 
The  \mbh-\mdyn\ relations are in excellent agreement with the similar K-band analysis in MH03 and H09.
The BCES regression is also consistent with a sample of 30 galaxies by HR04 
although their slope is slightly superlinear.

The position of pseudobulges is the same observed in Fig.~\ref{fg:mbh} and described 
in Section~3.1, where about half of the pseudobulges lie more than $4\sigma$ below the regression. 
The result for dynamical bulge mass can be compared with those for stellar mass to assess wether 
there are discrepancies between the two relations with \mbh.

To estimate (\emph{ii}) the bulge stellar mass \mstar\ we have calibrated the M/L 
ratio at \wav\ through color corrections. 
It is known that the M/L ratio is wavelength dependent 
and not constant for a spiral galaxy in any waveband. 
M/L trends have been found to be minimized in the K-band (Bell \& de Jong~2001, Bell et al.~2003). 
Published works assume a constant 
M/L with radius, while we are interested in the M/L for the bulge alone. 
The \wav\ wavelength is in principle a good tracer of stellar light for both young and old stellar 
populations as well, 
thus our data are expected to trace M/L with a low scatter. 
Moreover the extent of our FP to compact sizes, its tight dispersion valid 
for all Hubble types, and the agreement with the \wav\ FP observed for pure elliptical galaxies  
should guarantee a reliable estimate for the M/L of our bulges. 
\begin{figure*}
\begin{center}
\includegraphics[scale=0.9]{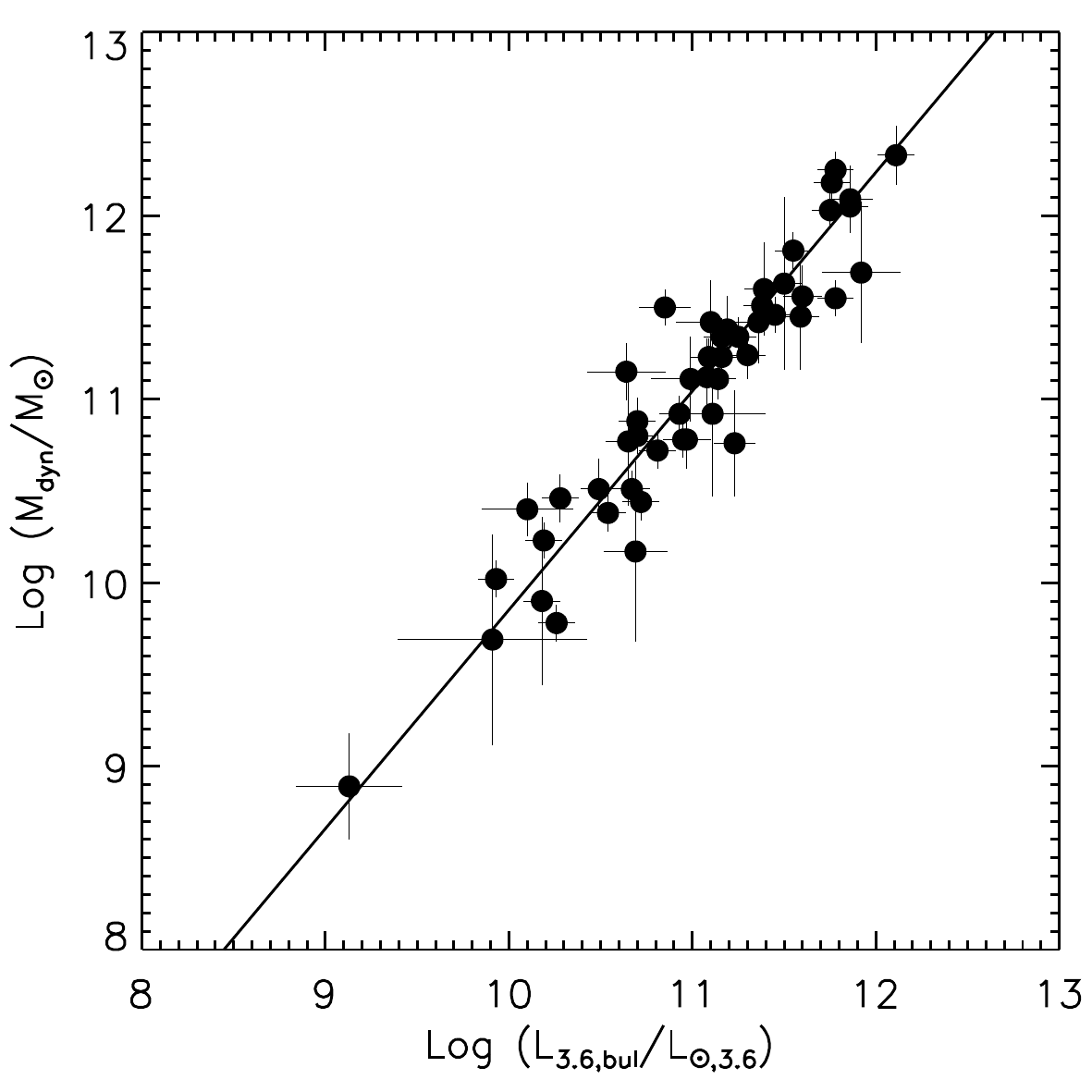}
\end{center}
\linespread{1.0}
\caption{\mdyn\ versus \lbs. The black line represents the linear regression of 
equation~(\ref{eq:mll}).
}
\label{fg:ml}
\end{figure*}

We assume that \mdyn\ in bulges is dominated by the stellar mass, 
with a negligible contribution of dark matter and gas (Drory, Bender
\& Hopp 2004; Padmanabhan et al. 2004). 
We then consider our sample of spheroids (49 sources, 
see Tab.\ref{tb:properties}) and fit a linear relation to \mdyn, computed 
as in equation~(\ref{eq:md}), vs \lbs\ as: 
\begin{eqnarray}
\log\mdyn &=&(11.04\pm0.03)+\nonumber\\
         & &(1.18\pm0.07)\times[\log(\lbs/L_{3.6,\odot})-11]\nonumber\\
         & &[rms=0.13\pm0.04]\nonumber\\
\label{eq:mll}
\end{eqnarray}
This is determined using LINMIX\_ERR and is very tight with $rms\sim0.10$. 
A tight \mdyn-\lbs\ relation is not unexpected as it is just a different way (i.e. $\sigma^2$\re\ vs \lbs) to express the FP.
The \mdyn-\lbs\  relation is shown in Fig.~\ref{fg:ml}. 

We now include a color correction to test how it affects the M/L ratio. 
Only 25 sources are included in this exercise, i.e. those whose classical bulges having V- and K- 
magnitudes available in the literature (listed in Tab.~\ref{tb:properties}). 
This allows a comparison with the above calibration and the following multi-linear regression:
\begin{footnotesize}
\begin{eqnarray}
\log\mdyn& = & (11.00\pm0.05)+\nonumber\\
& & (1.35\pm0.11)\times[\log(\lbs/L_{3.6,\odot})-11]\nonumber\\
& & +(0.07\pm0.10)\times[(V-K)-2.54]\nonumber\\
& & [rms=0.11\pm0.06]\nonumber\\
\label{eq:multi}
\end{eqnarray}
\end{footnotesize}
where we add V-K color correction relative to the sample mean V-K color. 
The fit was performed using the multivariate extension of LINMIX\_ERR (MLINMIX\_ERR, Kelly~2007). 
The zeropoints and slopes of \lbs\  in equation~(\ref{eq:mll}) and (\ref{eq:multi}) 
are the same, as well as intrinsic dispersions, while the slope of $(V-K)$ is consistent 
with zero within the errors. 
This, clearly indicates that M/L for bulges is not affected by the V-K color and the \wav\ 
waveband is confirmed to be a better tracer of the stellar light than the K-band for both old 
and young stellar populations. 
%
With the \mdyn-\lbs\ relation we have (\emph{i}) calibrated the M/L ratio for the 
bulge component independently of the Hubble type of our galaxies, (\emph{ii}) demonstrated how 
the \wav\ passband traces the stellar bulge mass better than optical or near-infrared colors 
for both young and old stellar populations, and finally (\emph{iii}) found that 
V- and K- colors do not strongly influence \mstar\ determinations based on \lbs.

Figure~\ref{fg:mass}B shows the \mbh-\mstar\ linear regressions:
\begin{eqnarray}
\log\mbh/\msun&=&(8.16\pm0.06)+\nonumber\\
              & & (0.79\pm0.08)\times[\log(\mstar/\msun)-11]\nonumber \\
              & & [rms=0.38\pm0.05]\nonumber \\
\label{eq:mst}
\end{eqnarray}
where we use equation~(\ref{eq:mll}) to calibrate \mstar. Equation~(\ref{eq:mst}) is just the 
\mbh-\lbs\ relation rescaled using equation~(\ref{eq:mll}) to determine \mstar,  
in order to obtain a relation where \mstar\ is
derived by the host luminosity as done in studies at high redshifts (e.g.~Merloni et al.~2010).
Thus the black hole vs bulge mass correlation can be used as the reference scaling 
relation in the local Universe to probe a possible evolution of \mbh-\mstar. This
issue is discussed in Section 4. The coefficients and rms of equation~(\ref{eq:mst}) are consistent
within the errors with the \mbh-\mdyn\ scaling relation (for our choice fitting method, see equation~(\ref{eq:my}), and Tab.~\ref{tb:rel}). H09
find a steeper slope, but the use of an extinction corrected B-V color to calibrate
M/L in the K band possibly affect the resulting \mbh-\mstar. The M/L calibration
presented here provide a robust estimate for \mstar; 
this issue is pursued further in Section~4. 

\subsection{A comparison with the \mbh-$\sigma$ relation}
In the previous sections we have analyzed all the \mbh-bulge scaling relations which can
be obtained using results from our bi-dimensional decomposition. We found (\emph{i}) the same
intrinsic dispersion for all the scaling relations $rms\sim0.35$~dex, and (\emph{ii}) that pseudobulges
have a bimodal distribution with half of them lying on the relations within the observed
scatter, while those harboring smaller BHs (\mbh$\leq10^7$\msun) significantly deviate from the
regression.
Here we compare our \mbh- bulge relations with \mbh-$\sigma$. A detailed analysis
of the \mbh-$\sigma$ relation is beyond the scope of this paper, our aim is just to obtain a consistent
comparison of \mbh-$\sigma$ with the other relations in terms of samples and fitting method. 
\begin{figure*}
\begin{center}
\includegraphics[scale=0.7]{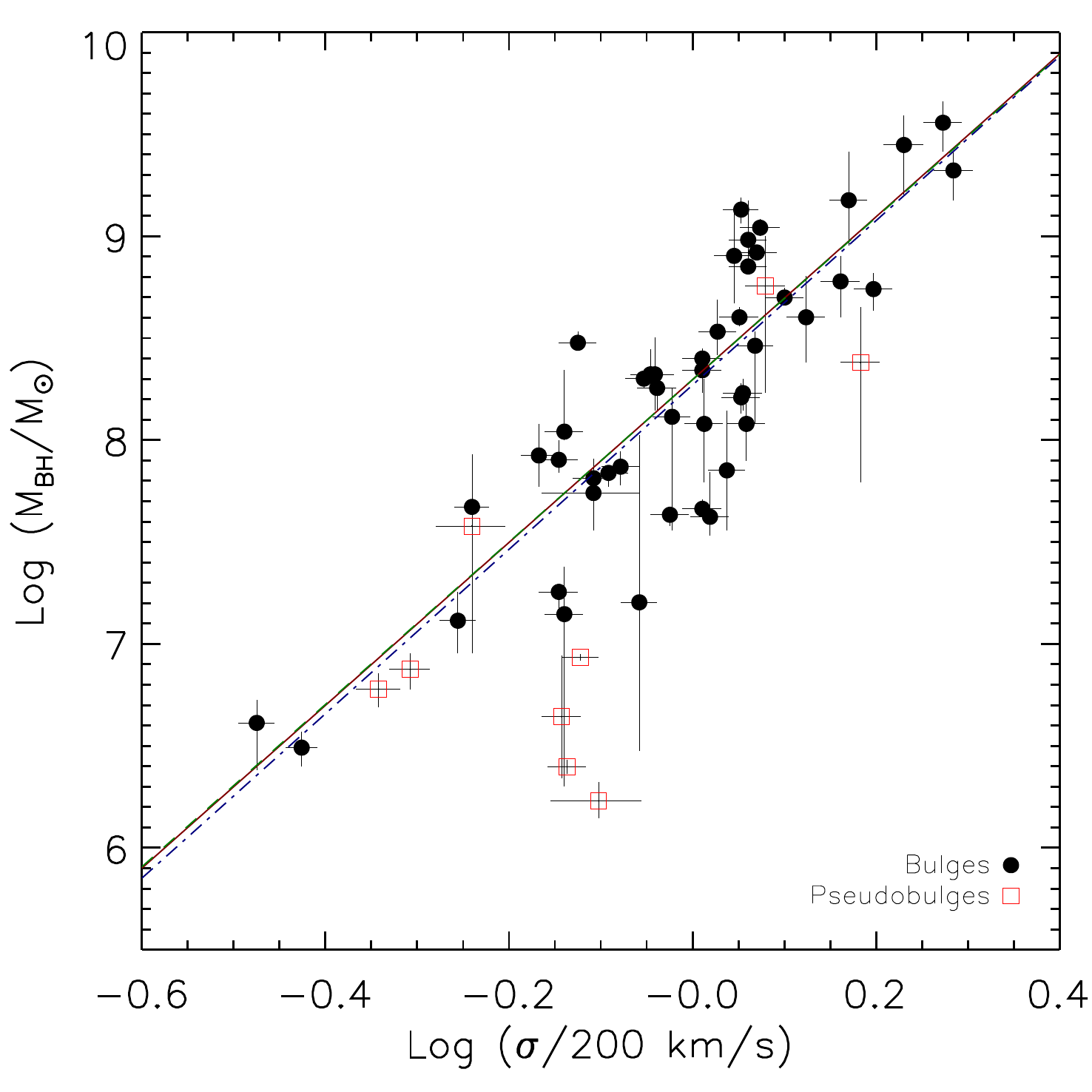}
\end{center}
\caption{Scaling relations. The \mbh\ is plotted as a function of velocity dispersion. 
The linear regressions and symbols are as in Fig.~\ref{fg:mbh}.}
\label{fg:sigma}
\end{figure*}
The classical bulges lie on 
the following correlation:
\begin{eqnarray}
\log\mbh&=&(8.29\pm0.05)+\nonumber\\
        & & (4.0\pm0.4)\times\log(\sigma/200~km\,s^{-1})\nonumber\\
        & &[rms=0.33\pm0.04]\nonumber\\
\label{eq:s}
\end{eqnarray}
In equation~(\ref{eq:s}) the velocity dispersion is normalized at 200~km/s according to T02, G09. 
Figure~\ref{fg:sigma} displays \mbh\ as a function of $\sigma$ and the three linear regressions, 
with the parameters in Tab.~\ref{tb:rel}.

Although the samples do not fully overlap, our BCES and FITEXY results are in 
good agreement with G09 and T02. 
More interesting is the correspondence, within the errors, 
between the \mbh-\mdyn\ (\mstar, \lbs)  and \mbh-$\sigma$ 
intrinsic scatters ($rms\sim0.3-0.35$~dex). 
Thus at \wav\ is not possible to assess whether the bulge dynamics or luminosity drive the 
observed scaling relations.

As expected, some, but not all, the pseudobulges 
are displaced from the inferred regressions. 
These are CIRCINUS, IC2560, NGC1068, NGC3079 and NGC3998; therefore there is a good overlap 
(4/6 where the exceptions are NGC3368 and NGC3998) 
with the deviant pseudobulges found in the previous sections. 
This discrepancy, also found in H09 and Nowak et al.~2010, 
could arise from several different causes; there could be an 
additional nuclear component not identified with our image 2D decomposition; 
or e. g. pseudobulges could have a different evolutionary history than 
classical bulges. We defer a detailed discussion to the following section. 
\begin{figure*}
\begin{center}
\includegraphics[scale=0.7]{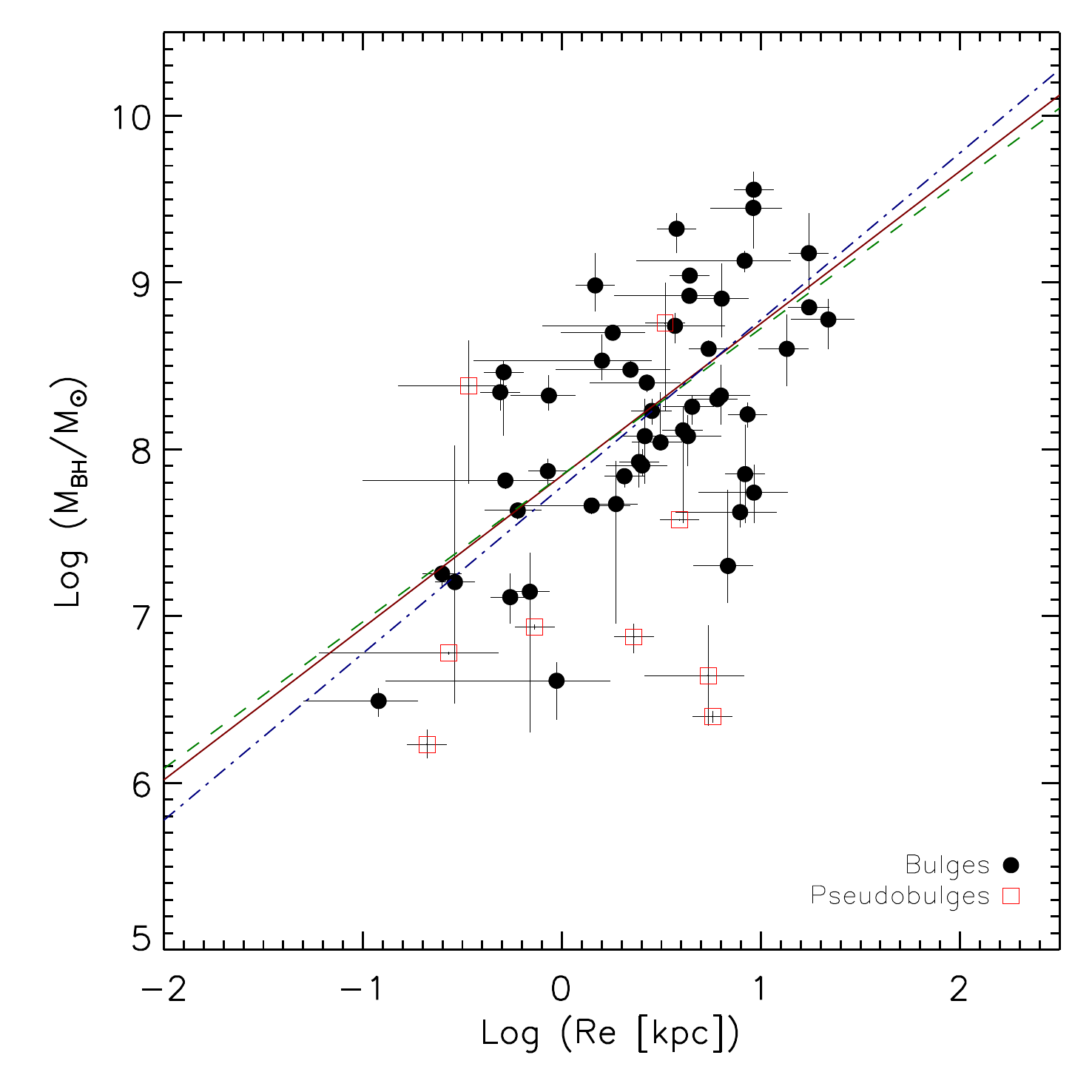}
\end{center}
\caption{Scaling relations. The \mbh\ is plotted as a function of the bulge effective radius \re. 
The linear regressions and symbols are as in Fig.~\ref{fg:mbh}.}
\label{fg:re}
\end{figure*}

\subsection{\mbh\ versus \re, the effective radius at \wav}
\begin{figure*}
\begin{center}
\includegraphics[scale=0.7]{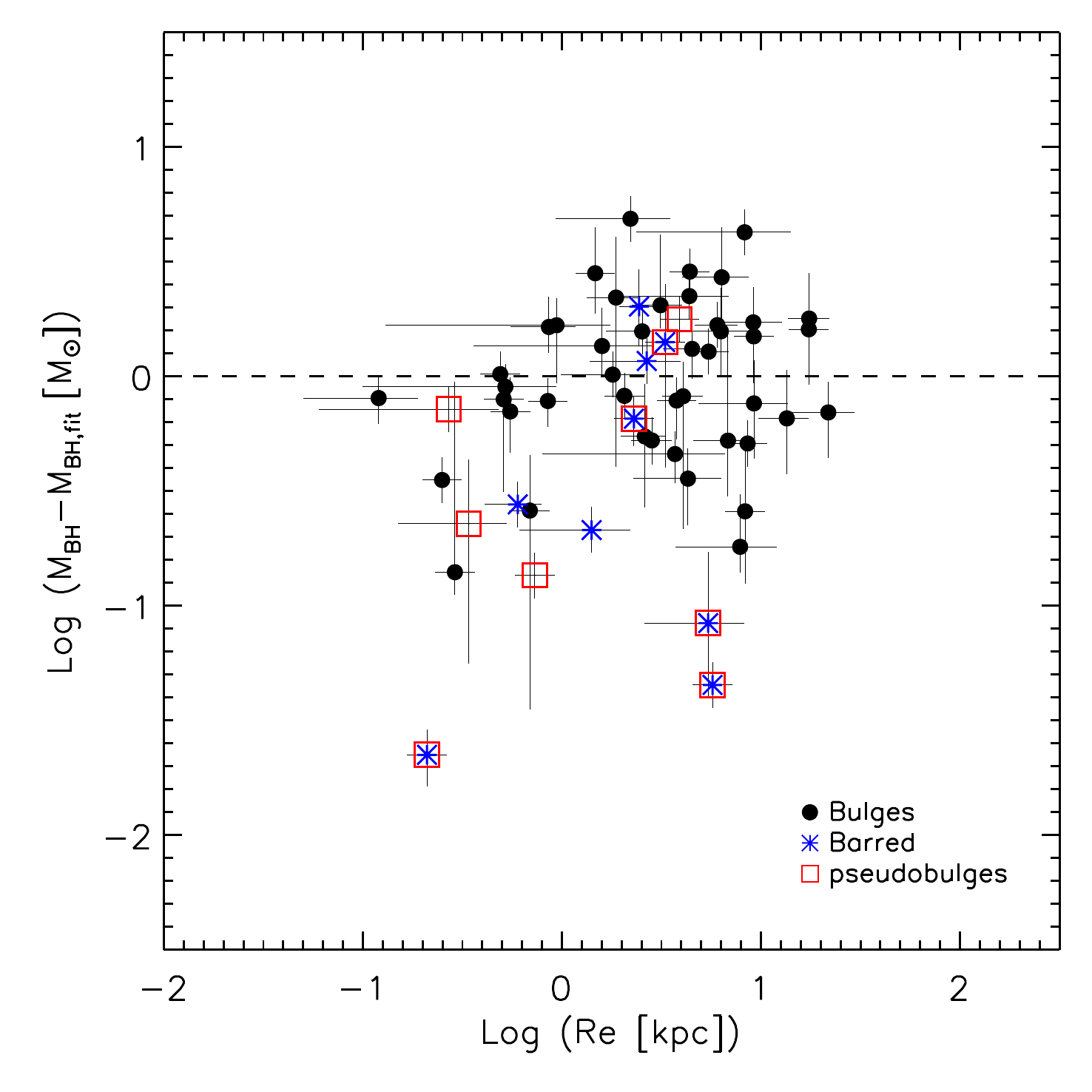}
\end{center}
\caption{\mbh\ (in Tab.\ref{tb:properties}) residuals about their \mbh-$\sigma$ 
relation (Eq~\ref{eq:s}) versus \re. The dashed line represents null residuals. 
Black points are early- and late- type galaxies, 
blue asterisks are barred galaxies and open red 
squares are pseudobulges.}
\label{fg:res}
\end{figure*}

\mbh\ has been observed to be separately related to the bulge stellar velocity dispersion $\sigma$
and its effective radius \re. This has been first observed in MH03 by a partial correlation
analysis. Figure~\ref{fg:re} shows the \mbh-\re\ relation obtained with our \wav\ 2D decomposition,
where the LINMIX ERR method gives:
\begin{eqnarray}
\log\mbh &=& (8.22\pm0.08)+\nonumber\\
         & & (0.88\pm0.17)\times[\log(\re/kpc)-0.4]\nonumber\\
         & & [rms=0.54\pm0.07]\nonumber\\
\label{eq:re}
\end{eqnarray}
The three linear regression in Tab.~\ref{tb:rel} are in good agreement within the errors. 
The intrinsic dispersion is significantly larger than in other relations. 
Nevertheless the possible existence of \mbh-\re\ is 
intriguing and prompts investigation of a possible fundamental 
plane for BHs, in analogy with the galaxies FP. 

It is still unclear whether {\it one} \mbh-bulge relation is significantly tighter 
than the others. 
Hopkins et al.~(2007) explained the lack of a dominant \mbh-bulge scaling relation because each
of them is the projection of the same fundamental plane relating \mbh\ with two or more bulge
properties. However, the issue of a BH fundamental plane is still matter of debate: Graham~2008a
showed that the need for a \mbh\ fundamental plane originates on a displacement of barred
galaxies with respect to the residuals of the \mbh-$\sigma$ relation. 

If a \mbh\ FP actually exists, it should be easily observed at \wav. Thus here we 
follow MH03 and Graham~(2008a) and compare the residuals of \mbh-$\sigma$ for 
our bulges (expressed in equation~\ref{eq:s}) with the effective radius 
(column 4 in Tab.~\ref{tb:spitzer}). Figure~\ref{fg:res} shows this comparison where we 
highlight the position of barred galaxies and pseudobulges. Hence, we check for a possible 
significant correlation performing both the Pearson and Spearman rank correlation test. 
As shown in Tab.\ref{tb:res}, there is \emph{no} significant correlation either for the entire sample 
or excluding barred galaxies  and/or pseudobulges. 
\begin{table*}{STATISTICAL TESTS FOR \mbh\ RESIDUALS}
\begin{center}
\centerline{\begin{tabular}{llcc}
\hline
Sample & Test & $\rho$ & significance \\
\hline
all galaxies & Spearman & 0.22 & 0.10   \\
           & Pearson  & 0.29 & - \\
\hline
no barred  & Spearman & 0.21 & 0.16   \\
           & Pearson  & 0.29 & - \\
\hline
no pseudobulges & Spearman & 0.16 & 0.26   \\
           & Pearson  & 0.24 & - \\
\hline            
no pseudobulges, & Spearman & 0.11 & 0.48   \\
no bar             & Pearson  & 0.20 & - \\
\hline         
\end{tabular}}
\end{center}
\caption{Statistical test for the \mbh\ residuals about their \mbh-$\sigma$ correlation with $R_e$ in Fig.~\ref{fg:res}.  Column 1: tests are performed for the total sample and sub-samples excluding barred galaxies and pseudobulges. 
Column 2: the kind of statistic applied to data (Spearman rank correlation and Pearson correlation). 
Column 3: correlation coefficient $\rho$. 
Column 4: and the two-sided significance of $\rho$ deviation from zero in the Spearman rank correlation. 
The significance is a value in the interval [0.0, 1.0]; a value consistent with zero indicates a significant correlation.
}
\label{tb:res}
\end{table*}  
Thus, our analysis at \wav\ does not confirm the existence of BH FP. 
We refer to a forthcoming paper (Marconi et al.~2011 in prep.) for a detailed study 
of the origin of the FP for supermassive BHs. 

\section{Discussion}

The high signal-to-noise images provided by \emph{Spitzer}/IRAC and our improved 2D analysis have allowed us to obtain accurate estimates of bulge luminosities and effective radii at \wav, as demonstrated by the tight fundamental plane and \mdyn-\lbul\ relations.
We have been able to (\emph{i}) calibrate the \mstar-\lbul\ relation, accurate to $0.10\pm0.05$ rms without any color correction and (\emph{ii}) securely define pseudobulges through the S\'{ersic} index value ($n\leq2$) and study their location with respect to the scaling relations for classical bulges. 
We have then obtained the \mbh-galaxy relations for classical bulges using our photometry at  \wav. They
are expressed in Eqs.~\ref{eq:l}, \ref{eq:my}, and \ref{eq:mst}, and shown 
in Figs.~\ref{fg:mbh}, \ref{fg:mass} and~\ref{fg:sigma}. 

All the above scaling relations have an intrinsic dispersion of 0.35~dex (from a Bayesian approach 
to linear regression, see Tab.\ref{tb:rel}) and are in perfect agreement with the \mbh-$\sigma$ 
relation for our galaxies in equation~(\ref{eq:s}). Considering the methods adopted in literature to perform 
linear regressions that account for the intrinsic scatter (BCES and FITEXY) our \wav\ correlations 
are as tight as the \mbh- bulge K-band luminosity (MH03, H09) and tighter than in V-band (G09). 
Moreover the \mbh-\mdyn\ and \mbh-\mstar\ agree very well with MH03, HR04 and H09, with the 
new advantage of an accurate M/L relation for bulges at \wav\ calibrated without any color correction.

Figure \ref{fg:mass} present the \mbh-\mstar\ relation, where \mstar\ is  obtained by combining \lbul\ with our calibrated  \mstar-\lbul\ relation. This relation represents the reference \mbh-\mstar\ in the 
local Universe for studies of the \mbh-galaxy scaling relations at higher z, based on 
\mstar\ derived from luminosity measurements (Merloni et al.~2010, Trakhtenbrot \& Netzer~2010).
The $z$ evolution of \mbh-galaxy scaling relations has been studied in several papers but remains controversial. 
For instance, Woo et al.~(2008) report a significant evolution of the BH-bulge relations at $z=0.3-0.6$, with larger BH masses for given bulge $\sigma$/luminosity with respect to local values. This evolution is however not found by Labita et al.~(2009) or Shen et al.~(2008). 
Shields et al.~(2003) find no evolution of \mbh-$\sigma$ up to z$\sim2$ but using [OIII] line width as a surrogate for $\sigma$.
Indeed, at high redshift a dynamical measurement of the bulge mass based on $\sigma$  is difficult due to the limited 
signal-to-noise and spatial resolution. On the contrary the total 
stellar mass of the host galaxy is relatively easier to estimate from host galaxy luminosities and colors (e.g. Peng et al. 2006), or by deconvolving AGN and galaxy spectral energy distributions (Merloni et al.~2010) or by applying the correlation between the AGN bolometric luminosity and \mbh\ together with the relation between the star formation rate and \mstar\ in star forming galaxies (Trakhtenbrot \& Netzer~2010). 
The above observational works suggest that the \mbh/\mstar\ at high redshift is 
significantly higher than the local value. In any case, observations at high redshifts are fundamental 
to constrain models of BH-galaxy coevolution and to probe different paths followed by galaxies during their evolution and the building up
of the \mbh-bulge relations (Lamastra et al.~2010 and references therein). 

In Section~3.2, the tight \mdyn-\lbs\ relation with negligible color corrections 
justifies the validity of 2D decomposition to measure M/L at \wav. 
This (\emph{i}) unveils the capability of our \mbh-\mdyn\ (and thus \mstar) to be a suitable 
benchmark to probe the \mbh-bulge co-evolution and to constrain theoretical models. 
In fact, in the next decade the instruments onboard the \emph{James Webb Space Telescope} (JWST) will increase the possibility to further explore with unprecedented accuracy this waveband.  

The other (\emph{ii}) intriguing result found here is related to the behavior of pseudobulges. 
The displacement of pseudobulges, located below the scaling relations, is reported in 
H09 and Nowak et al.~(2010). 
In \S2.4 we followed FD10 and identified 9 pseudobulges with S\'{e}rsic index 
$n\leq2$. Of these, CIRCINUS, IC2560, NGC1068, and NGC3398 lie below \emph{all} 
the \mbh\ scaling relations; NGC3368 deviates from \mbh-\lbs, -\mdyn but not from 
\mbh-$\sigma$, while the contrary happens for NGC3998. The other pseudobulges NGC3489, 
NGC4258, NGC4374, NGC4594 are within the observed scatter. 
As mentioned in \S3.3 this discrepancy can arise from several causes. 
We first note that the $n\leq2$ is a statistical property of pseudobulges; indeed 
the tails of the S\'{ersic} index distributions for classical and pseudobulges overlaps 
around $n\sim2$ (see Fig.~9 in Fisher \& Drory~2008 and FD10). Thus, not all bulges 
with a low $n$ would be actual pseudobulges from the dynamical point of view 
(Gadotti~2009).

Anyway, it is interesting to explore the possible explanations once 
we assume that all the nine galaxies actually harbor a pseudobulge.

(1) Additional nuclear component on small scales ($\simeq10$~arcsec), such as bars, 
ovals and lenses are observed not only in 
disks but also in S0 galaxies (Laurikainen et al.~2009). 
With the grid method applied 
for 2D fitting (see \S2.3), an additional component on small image scales of 10-15 pixels 
can not be taken into account, because (a) it is difficult to constrain input parameters 
while automatically running the fits, and (b) the model does not require any additional 
component to match our quality criteria (see \S2.3).

(2) Nowak et al.~(2010) define NGC3368 and NGC3489 as pseudobulges by resolving the 
nuclear dynamics, and find the coexistence of a classical bulge with a pseudobulge, 
where the \mbh\ seems to better correlate with the classical bulge alone. 
For the same reasons outlined above, it is beyond the possibilities of our analysis to deconvolve the classical bulge 
from the pseudobulge component.

(3) The nine pseudobulges could be in different evolutionary stages. 
Montecarlo simulations in Lamastra et al.~(2010) 
show how in the buildup of the local \mbh-\mstar\ relation, galaxies can follow 
various evolutionary paths that proceed from below the correlation 
and reach their final position passing above it. Thus, the position of each source could just be the result of the different evolutionary paths being followed.

(4) Finally, nuclear star clusters (NSCs) may be common in those spheroid with 
\mstar$\sim10^8-10^{11}$~\msun\ harboring a \mbh$<10^7$~\msun, and
may contribute significantly at the mass of the central object increasing its value of 
a factor of $\sim2$ (Graham \& Spitler~2009 and references therein). 
A larger value of \mbh\ can prevent the coexistence of a NSC, 
because the very central stars should be either disrupted by the BH gravity or kicked away by tidal forces. 
Moreover Graham \& Spitler~(2009) derive a linear regression between 
the nuclear mass (BH plus NSC) and the stellar mass of the host spheroid. 
Thus,  the proper quantity to be plotted on the y-axis seems to be \mbh\ plus the NSC 
mass, at least for \mbh$<10^7$~\msun. The identification of NSCs requires the highest spatial resolutions available 
with the current instrumentation and can not be reached with \emph{Spitzer} even 
for the closest of our sources. 

We conclude that the observed displacement of pseudobulges could arise from observational 
limitations, true evolution, or a combination of the two. 
\section{Conclusions}
In this work we have investigated the scaling relations observed in the local Universe between 
\mbh\ and the structural parameters of the host bulges. The analysis is based on a 2D 
decomposition of \wav\ Spitzer/IRAC images of 57 
early- and late- type galaxies with \mbh\ measurements. Given the well known degeneracy
between the bulge S\'{e}rsic index $n$ and effective radius \re, we have adopted a grid-method to fit 
IRAC images and determine $n$ rather than let it freely vary. 
The accuracy of our analysis is verified by the agreement of our bulges with the mid-infrared fundamental
plane determination by JI08.
Our galaxies extend their \wav\ FP by doubling the \re\ range towards 
smaller radii and none of our galaxies deviate significantly from the JI08 FP. 
This allows us 
to reliably (\emph{i}) calibrate M/L at \wav, (\emph{ii}) study the \mbh- scaling relation and (\emph{iii}) identify 
pseudobulges in our sample. 

We obtained a tight ($rms=0.10$~dex) \mdyn-\lbs\ relation: 
\[
\log\mdyn/\msun= 11.04+1.18\times[\log(\lbs/L_{3.6,\odot})-11], 
\]
which allows us to estimate stellar masses based on luminosities at \wav\ with high accuracy. 
The \wav\ luminosity appears to be the   best tracer of \mstar\ yet found. 

The relations between \mbh, luminosity, masses, and effective radius, fitted with a Bayesian 
approach to linear regression are:
\[\log\mbh/\msun=8.19+0.93\times[\log(\lbs/L_{3.6,\odot})-11],\] 
\[\log\mbh/\msun=8.20+0.79\times[\log(\mdyn/\msun)-11],\] 
\[\log\mbh/\msun=8.16+0.79\times[\log(\mstar/\msun)-11],\] 
\[\log\mbh/\msun=8.22+0.9\times[\log(\re/kpc)-0.4],\]
\emph{all} with the same intrinsic dispersion of 
$rms\sim0.35$~dex except for \mbh-\re\ which has $rms\sim0.5$~dex.
Our \mbh-\mstar\ relation can 
be used as the local reference for high redshift studies which probe the cosmic evolution of \mbh-galaxy scaling scaling relations. 

These \wav\ \mbh-\lbs, \mbh-\mdyn, \mbh-\mstar\ relations turn out to be as tight as \mbh-$\sigma$ and, moreover,
they are consistent with previous determinations from the literature at shorter wavelengths.

We independently identified as pseudobulges those galaxies with S\'{e}rsic index lower than 2 
and found 9 sources that satisfy this criterion. Of these, 4 pseudobulges lie on scaling relations 
within the observed scatter, while those with \mbh\ lower than $10^7$~\msun\ are significantly 
displaced. We discussed the 
different physical and evolutionary origins for such behavior, while considering the presence 
of nuclear morphological components not reproduced by our two-dimensional decomposition. 

Finally, we verified the existence of a possible FP for supermassive BHs, relating \mbh\ with two 
or more bulge properties. We did not find any correlation between the residuals of \mbh-$\sigma$ and the effective radius, showing that our 
data do not require the existence of any BH fundamental plane.

\section{Acknoledgments}
The authors are grateful to the anonymous referee for the constructive comments and suggestions. 
We acknowledge B.~M.~Peterson for discussing the background effects and analysis method.
E. Sani thanks  R.~I.~Davies, A.~W.~Graham, and N.~Neumayer for precious discussions on the pseudobulges 
and barred galaxies behaviors. 
This work has been partially supported by the NASA grants Spitzer/1343503 and GO-11735.01. 
We acknowledge the usage of the HyperLeda database (http://leda.univ-lyon1.fr). 
This research has made use of the NASA/IPAC extragalactic database (NED).


\newpage
\appendix
\section{details of the 2D fitting method}
An important issue for bulge decomposition is related to the fitting method. 
In the case of one dimensional fitting, radially averaged surface brightness profiles 
are fit with one or more components. Consequently, information related to, e.g., 
isophotal twists or changes in ellipticity is lost and the galaxy profile is poorly defined. 
Moreover, in one-dimensional fitting different morphological components may appear to merge smoothly,
generating degeneracy and non-uniqueness in the profile decomposition. A two-dimensional 
analysis breaks this degeneracy (see, e.g., Peng et al.~2002, de Souza, Gadotti \& dos Anjos~2004). 

In Section~2.3 we describe in detail our grid method for a two-dimensional decomposition 
of \emph{Spitzer}/IRAC 3.6~$\mu$m images. 
This is an improved version of the two-dimensional method commonly 
adpted in the literature, where all galaxy parameters 
are left free to vary. 
In fact, degeneracies also  affect two dimensional methods: one is related to the functional 
form of the bulge profile, ''S\'{e}rsic'' profile, in which the $n$ and \re\ parameters are strongly coupled. Moreover, the fainter outer parts of 
galaxies may be affected by errors on background estimates.
In Section 2.3 we also justify our choice of the best fit 
by a quantitative inspection of the $\chi^2$ as a function of the radius. We note that the 
errors on the bulges parameters in Tab.~3 are \emph{not} one standard deviation 
errors, rather are given by the discrepancy between the 
best fit values and the ones obtained from the model having the closest 
$s,~s+\sigma_s$ or $s-\sigma_s$ value and the lowest possible 
reduced $\chi^2$ within $R^{opt}_{0.5}$ according to the criteria described in Section~2.3. 
For each galaxy, Fig.~\ref{fg:ap} shows the image, 2D model, and residuals. 
It also shows the trend of $\chi^2$ vs the source radius for: the best model 
(i.e. best $n$ in red), the fit for the nearest $n$ with 
low $\chi^2$ within $R^{opt}_{0.5}$ maintaining the same sky values 
(in dark blue, is plotted to show the dependence on $n$ at small radii), and the model 
from which we compute the absolute error on free parameters (in sky blue).
Even with the above technique when real 
galaxies are fitted in two dimensions, 
their surface brightness in the circumnuclear regions is rarely well reproduced only by axisymmetric components. 
In fact, it is not uncommon to find inner disks and non-axisymmetric features
in elliptical galaxies 
(e.g. Moriondo et al. 1998, Laurikainen et al. 2005). 
Substructure in the circumnuclear regions 
of both elliptical and spiral galaxies (such ovals, disks, lenses) 
occurs frequently, and is usually impossible to fit them with only axisymmetric 
components, with or without the addition of a bar. 
This is the origin of nuclear structures in the residual images shown in Fig.~\ref{fg:ap}. 
Finally, we also note that our models (red curves in Fig.~\ref{fg:ap}) do not require 
any additional nuclear component to respect the quality criteria of Section~2.3, 
for which, no more than $5\%$ of the nuclear flux should be lost.
\begin{figure*}
\centerline{
\hbox{
\includegraphics[width=0.25\linewidth,angle=90]{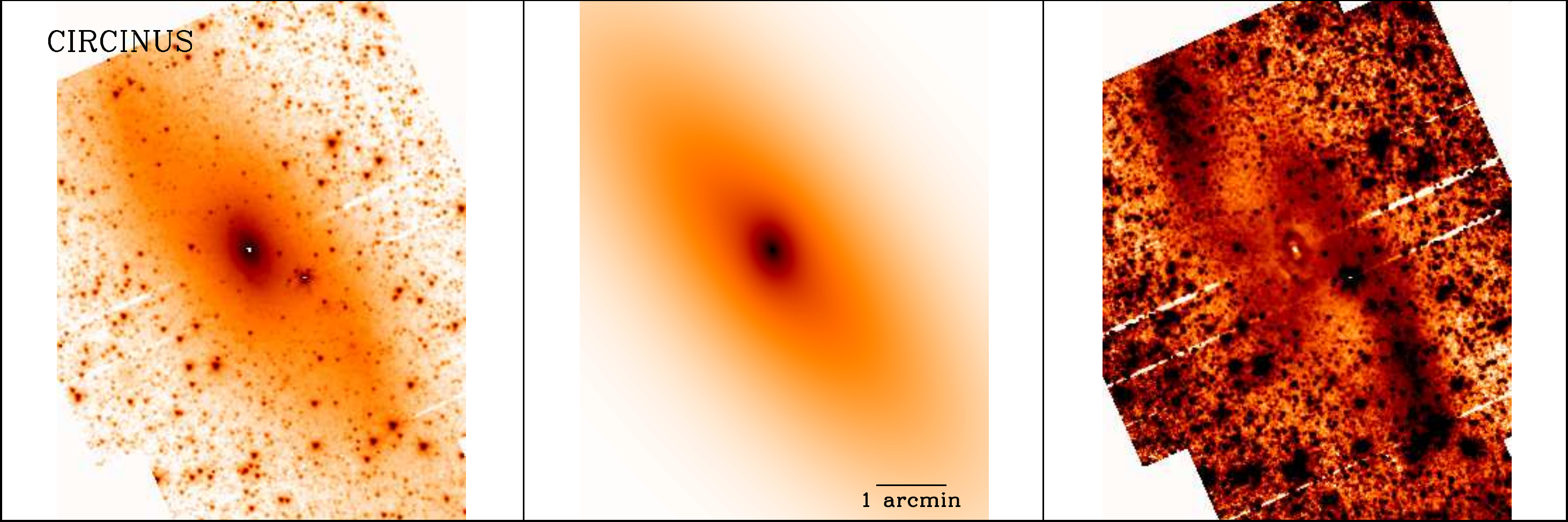}
\includegraphics[height=0.25\linewidth]{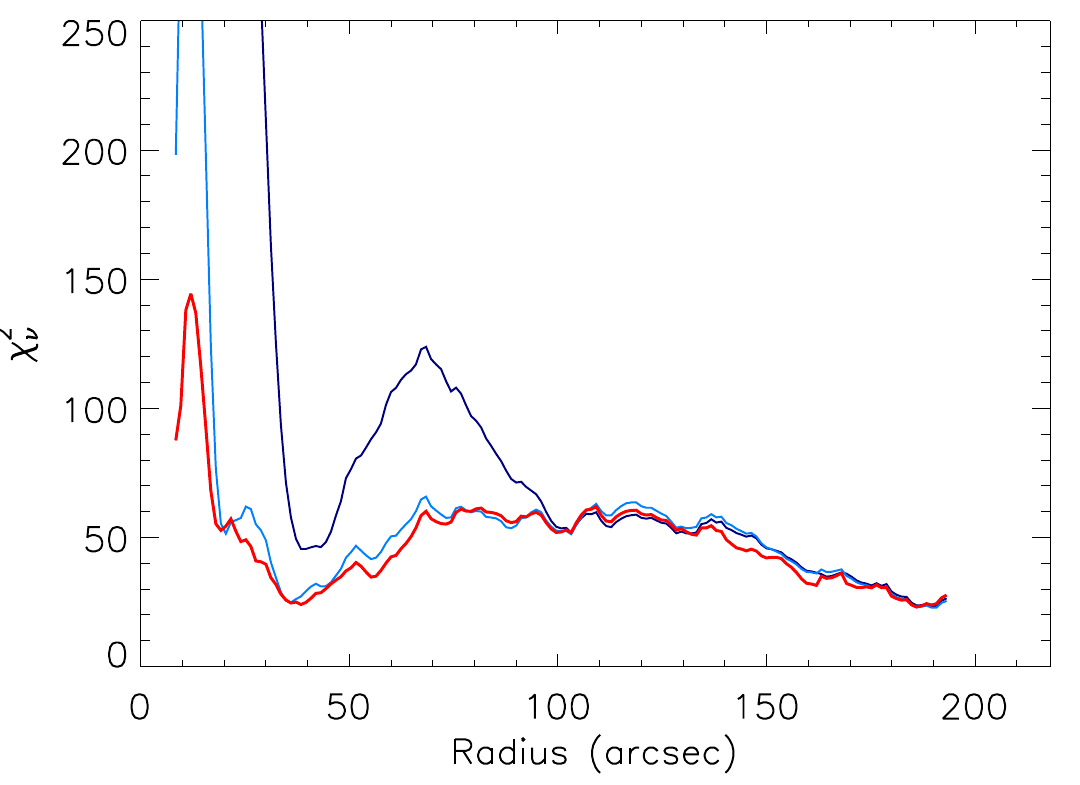}
}
}
\centerline{
\hbox{
\includegraphics[width=0.25\linewidth,angle=90]{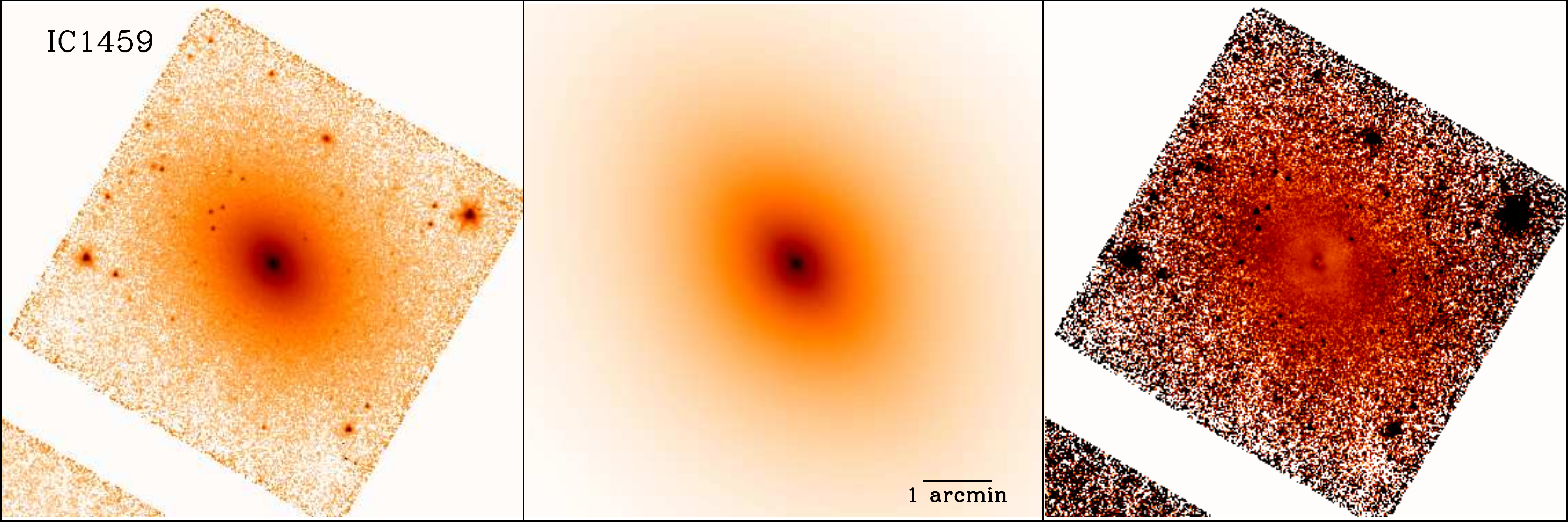}
\includegraphics[height=0.25\linewidth]{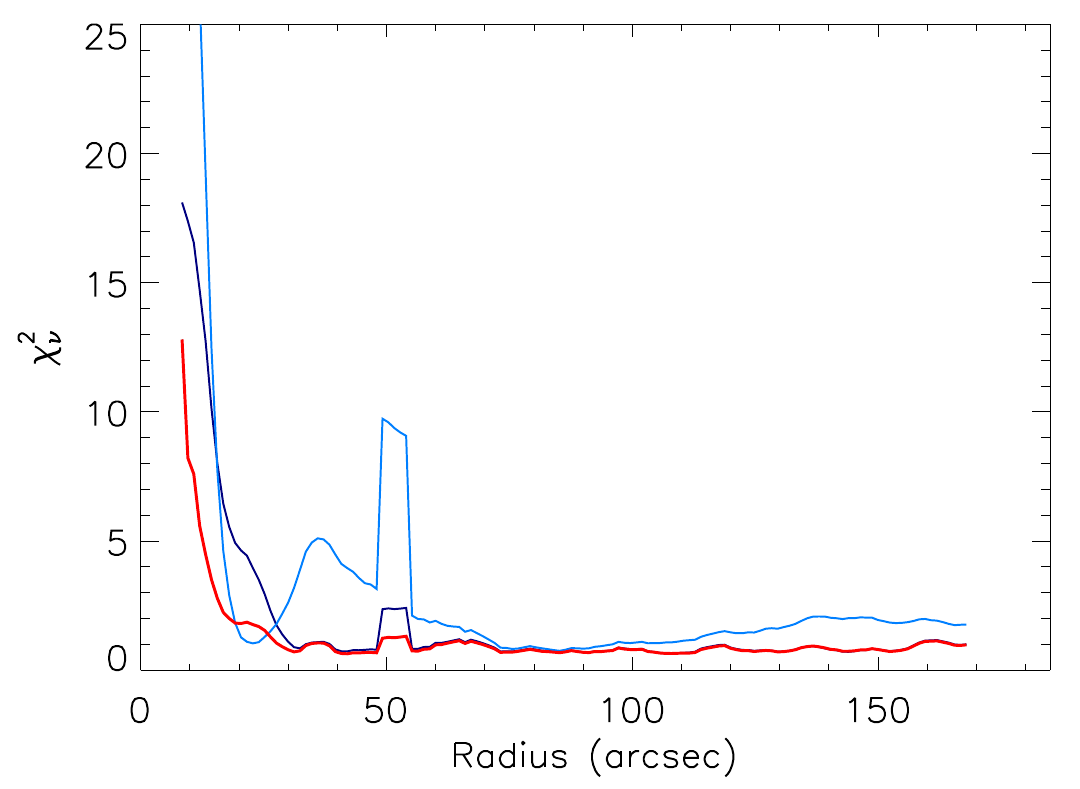}
}
}
\centerline{
\hbox{
\includegraphics[width=0.25\linewidth,angle=90]{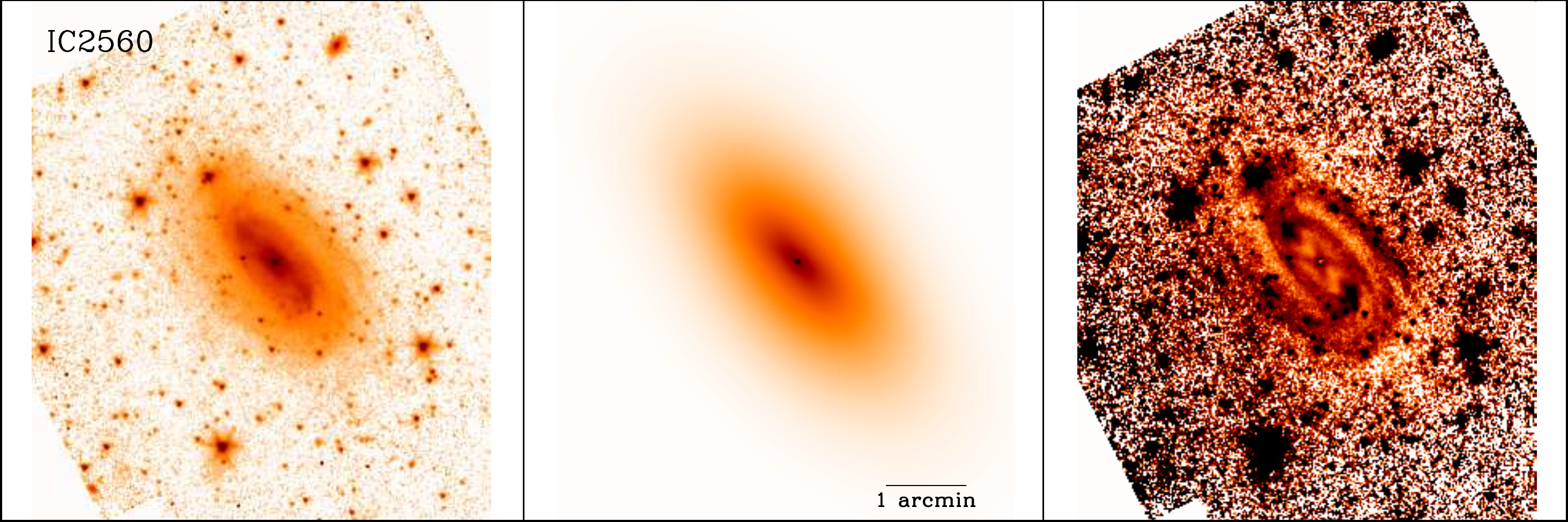}
\includegraphics[height=0.25\linewidth]{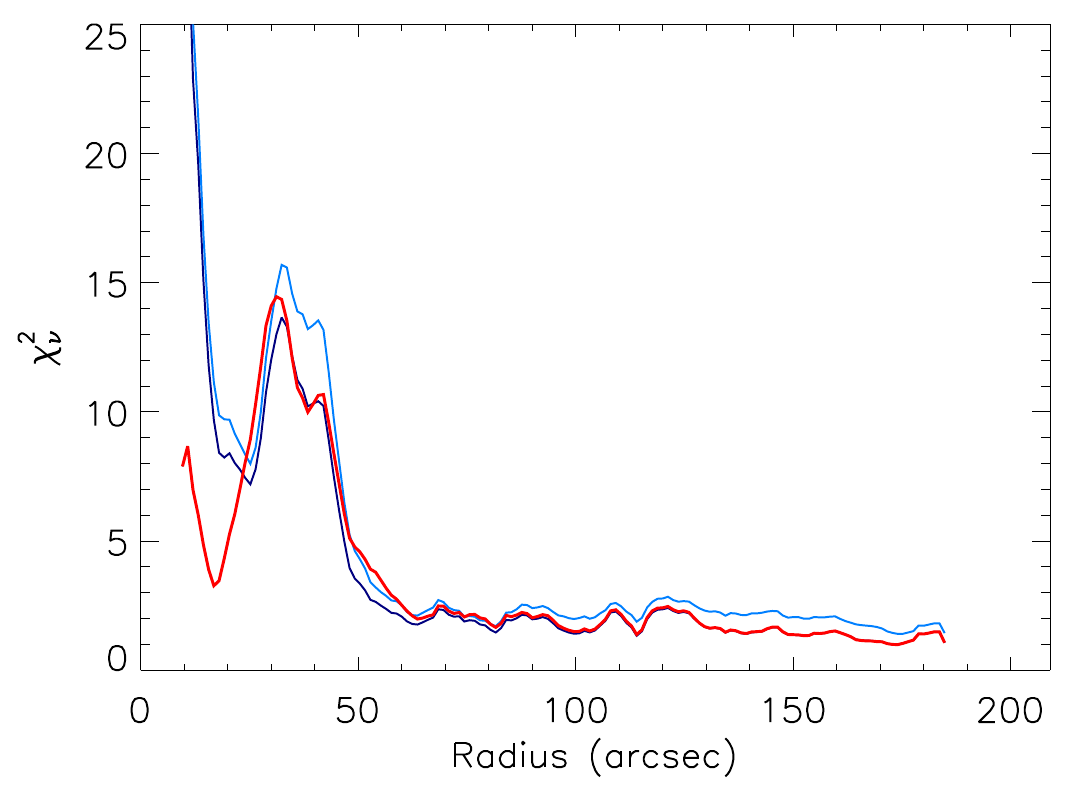}
}
}
\centerline{
\hbox{
\includegraphics[width=0.25\linewidth,angle=90]{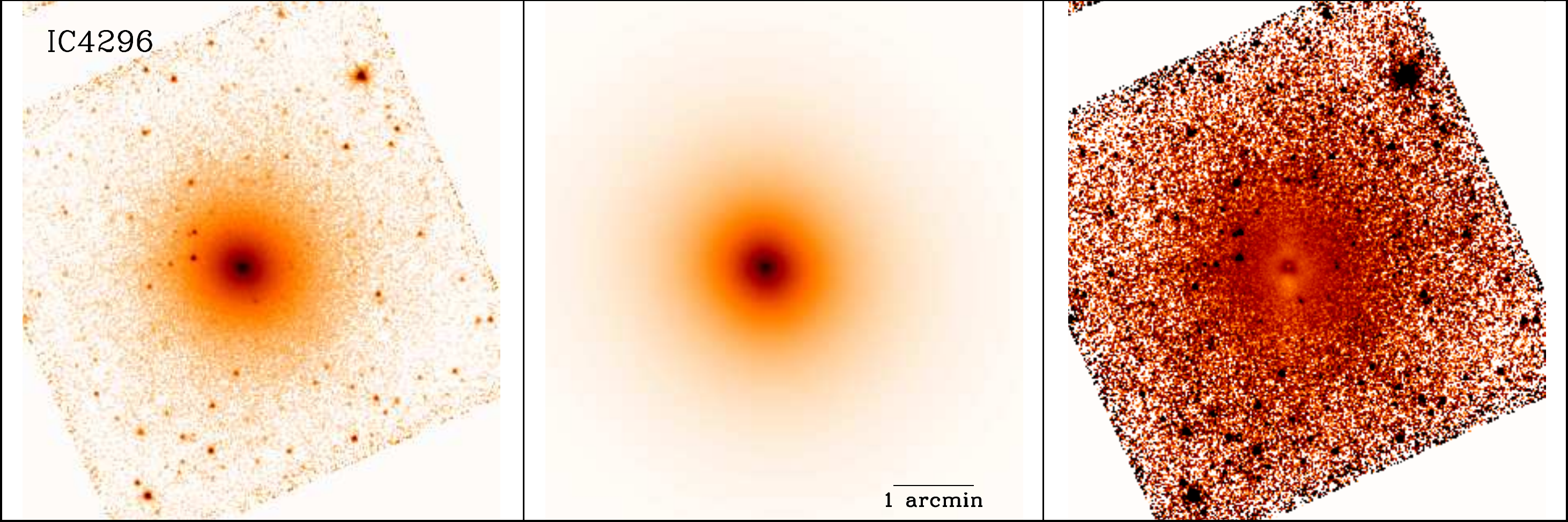}
\includegraphics[height=0.25\linewidth]{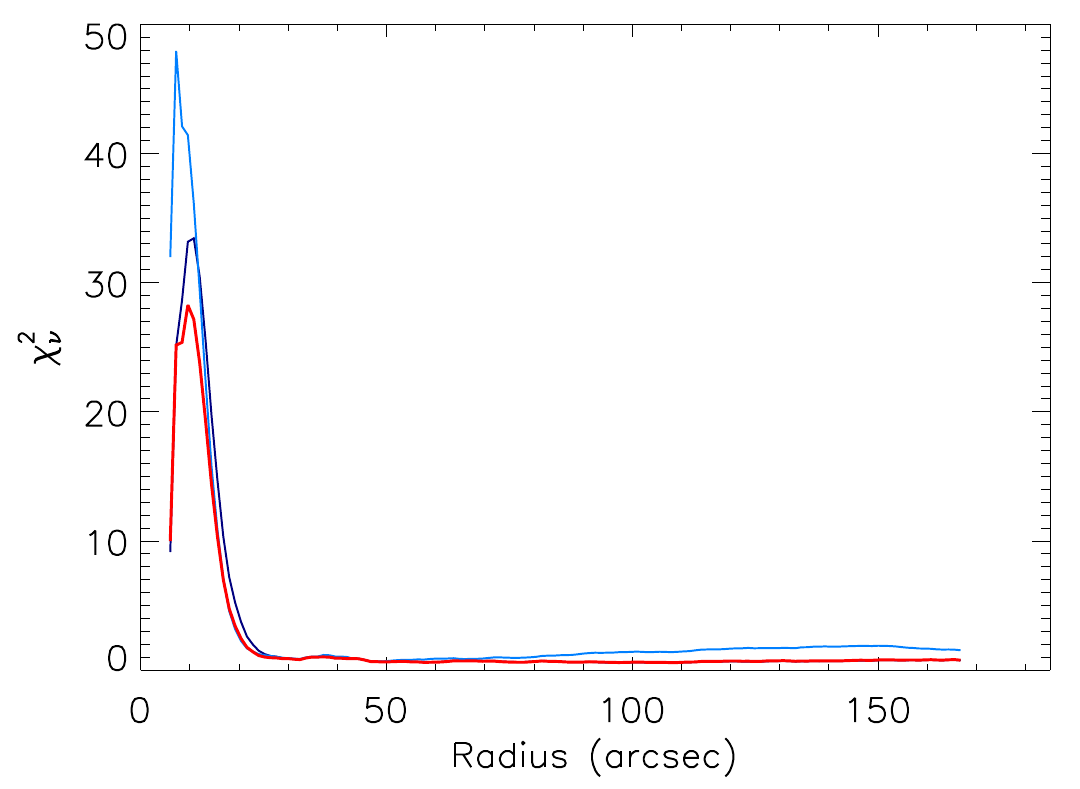}
}
}
\centerline{
\hbox{
\includegraphics[width=0.25\linewidth,angle=90]{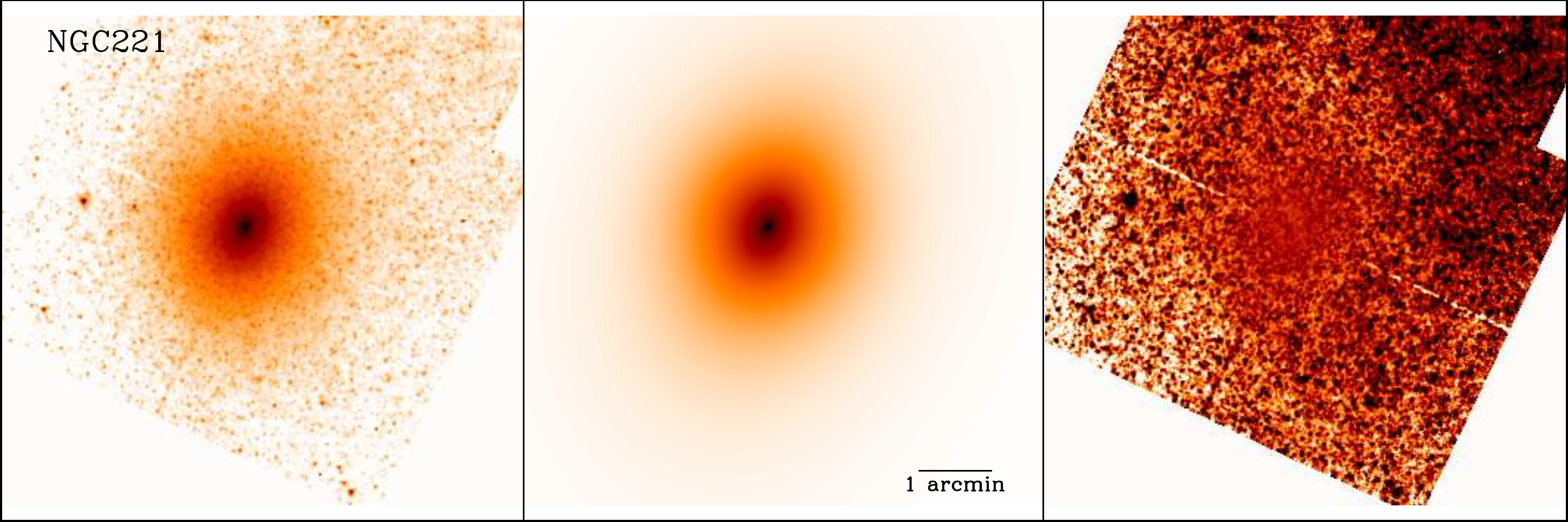}
\includegraphics[height=0.25\linewidth]{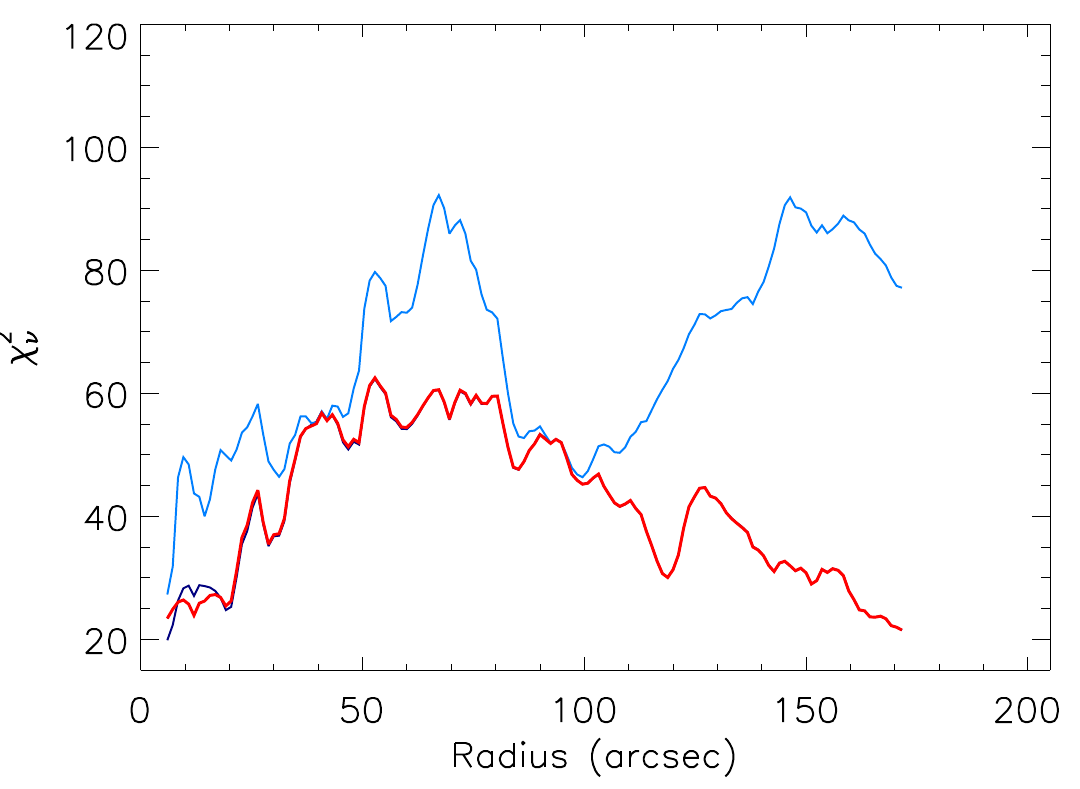}
}
}
\caption{Two-dimensional decomposition.The image, best-fit model and residuals are shown in logarithmic scale. 
The residuals are stretched ($\pm0.25$dex) to highlight 
the finest details.
On the right, the $\chi^2$ is plotted as a function of the radius (in arcsec). Red: best model. 
Dark blue: the fit with closest $n$ and the lowest $\chi^2$ within $R^{opt}_{0.5}$. 
Light blue: same for the closest sky ($s\pm\sigma_s$, see Section~2.3 and Appendix~A for details). 
A color version of this figure is available online.}
\label{fg:ap}
\end{figure*}
\addtocounter{figure}{-1}
\begin{figure*}
\centerline{
\hbox{
\includegraphics[width=0.25\linewidth,angle=90]{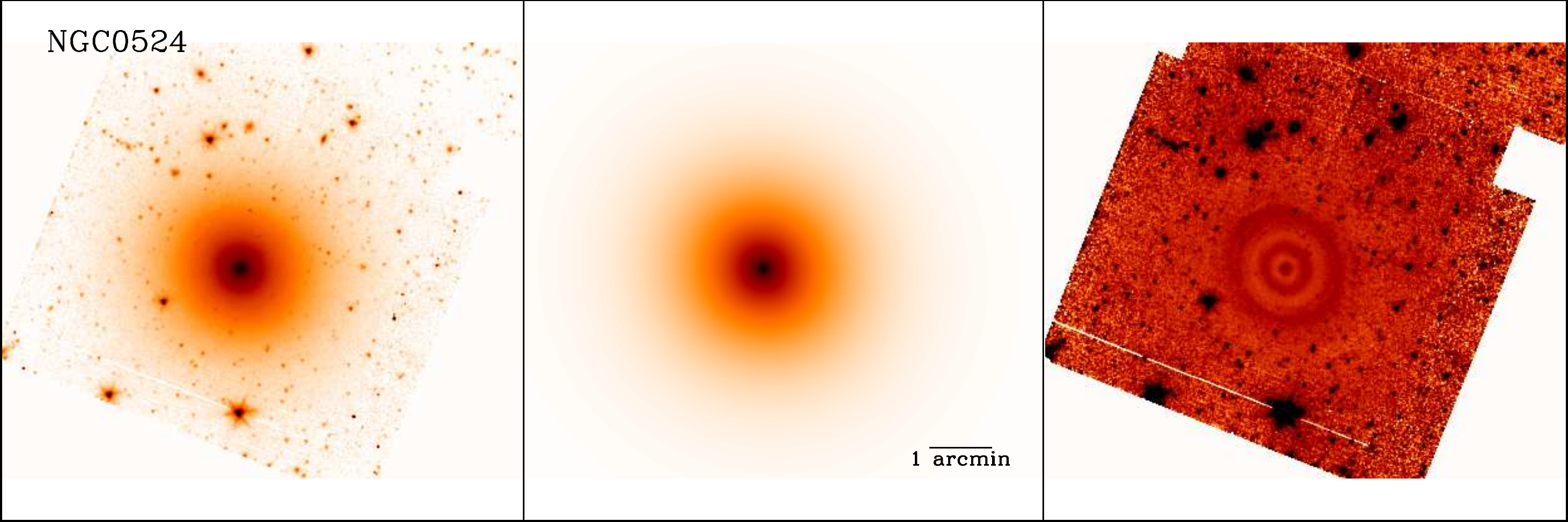}
\includegraphics[height=0.25\linewidth]{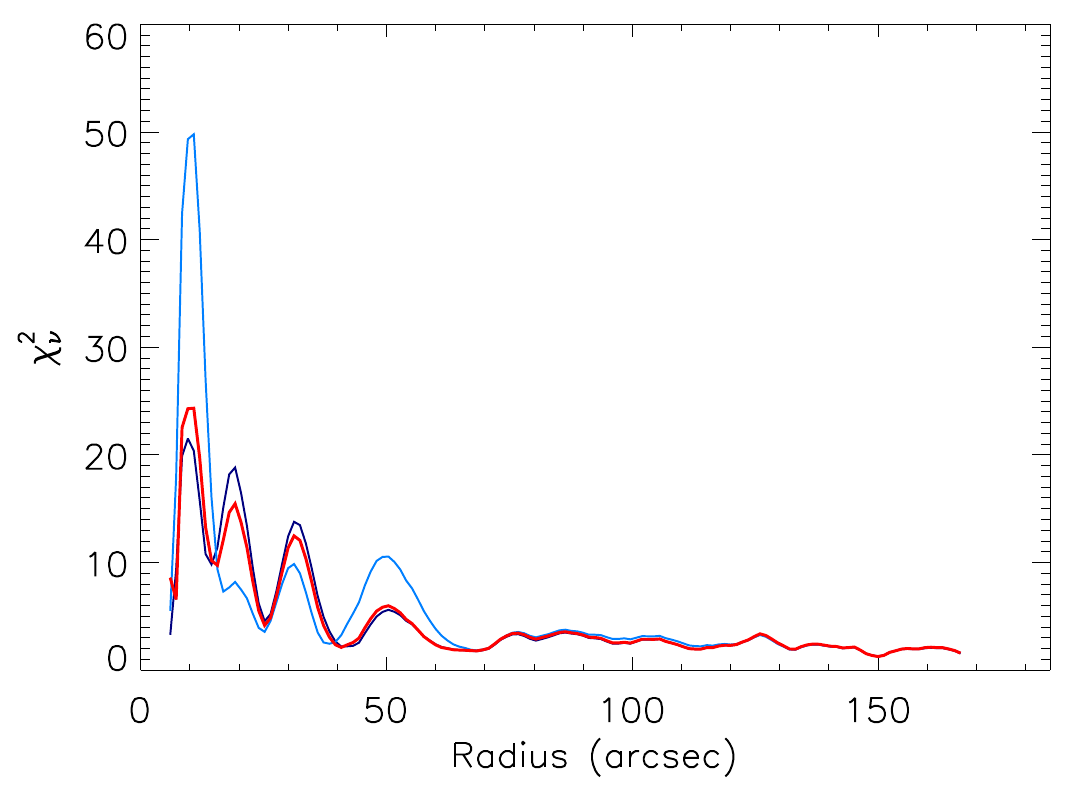}
}
}
\centerline{
\hbox{
\includegraphics[width=0.25\linewidth,angle=90]{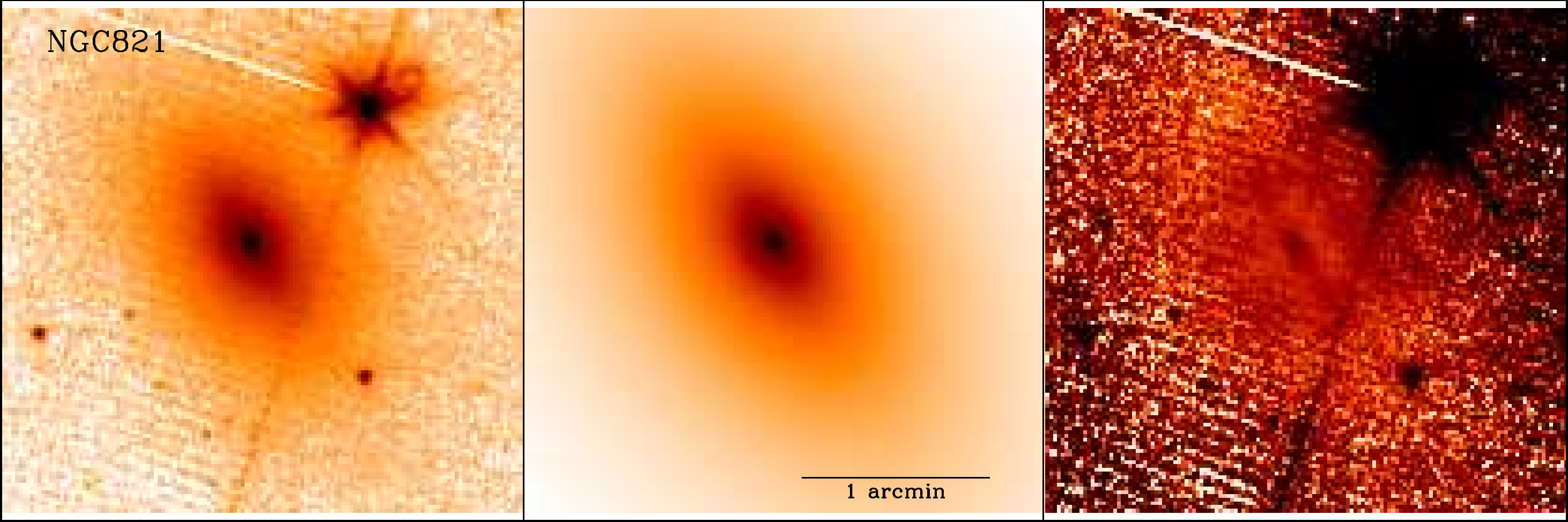}
\includegraphics[height=0.25\linewidth]{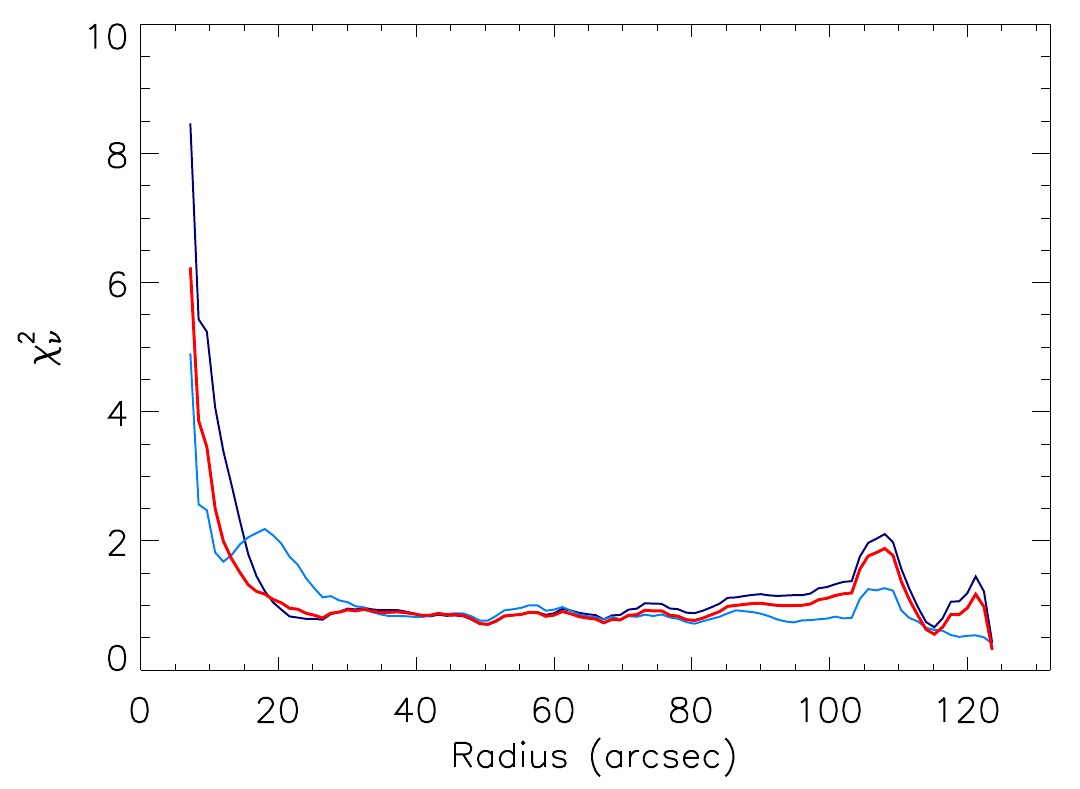}
}
}
\centerline{
\hbox{
\includegraphics[width=0.25\linewidth,angle=90]{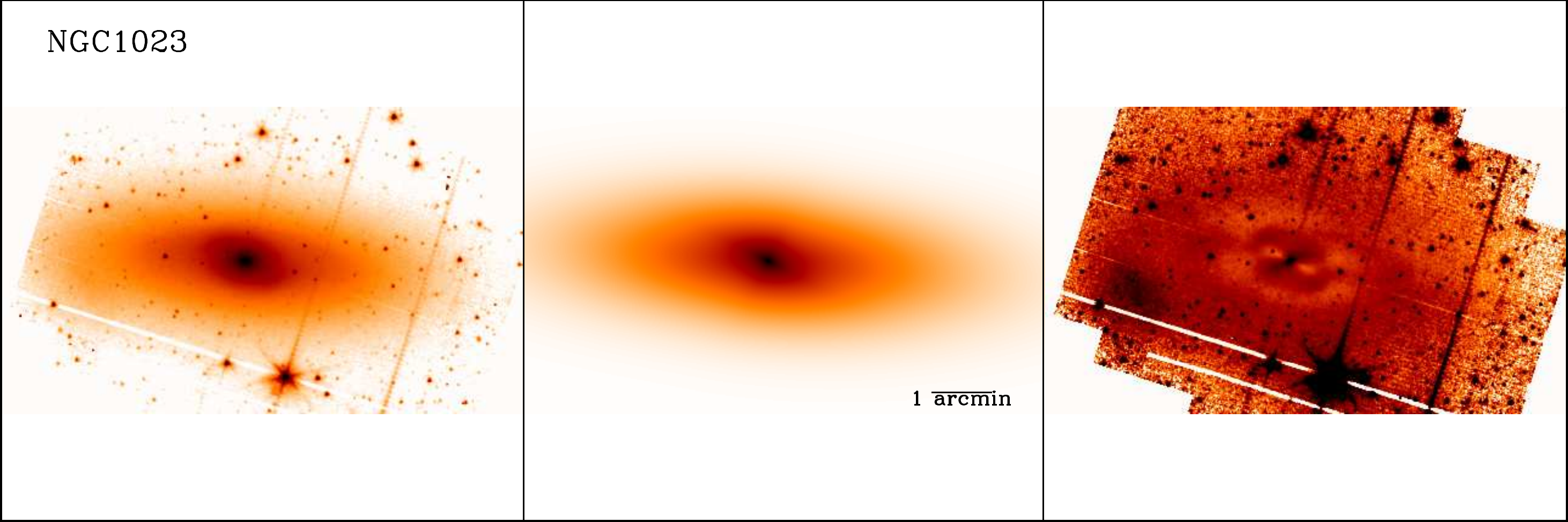}
\includegraphics[height=0.25\linewidth]{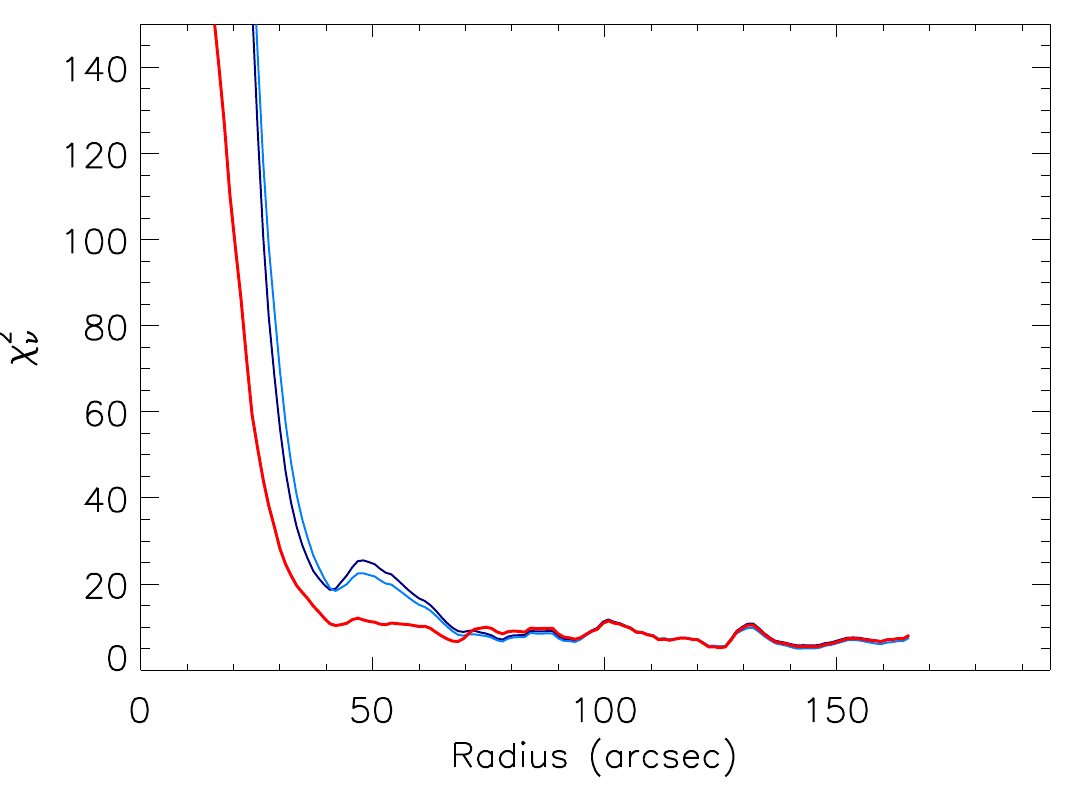}
}
}
\centerline{
\hbox{
\includegraphics[width=0.25\linewidth,angle=90]{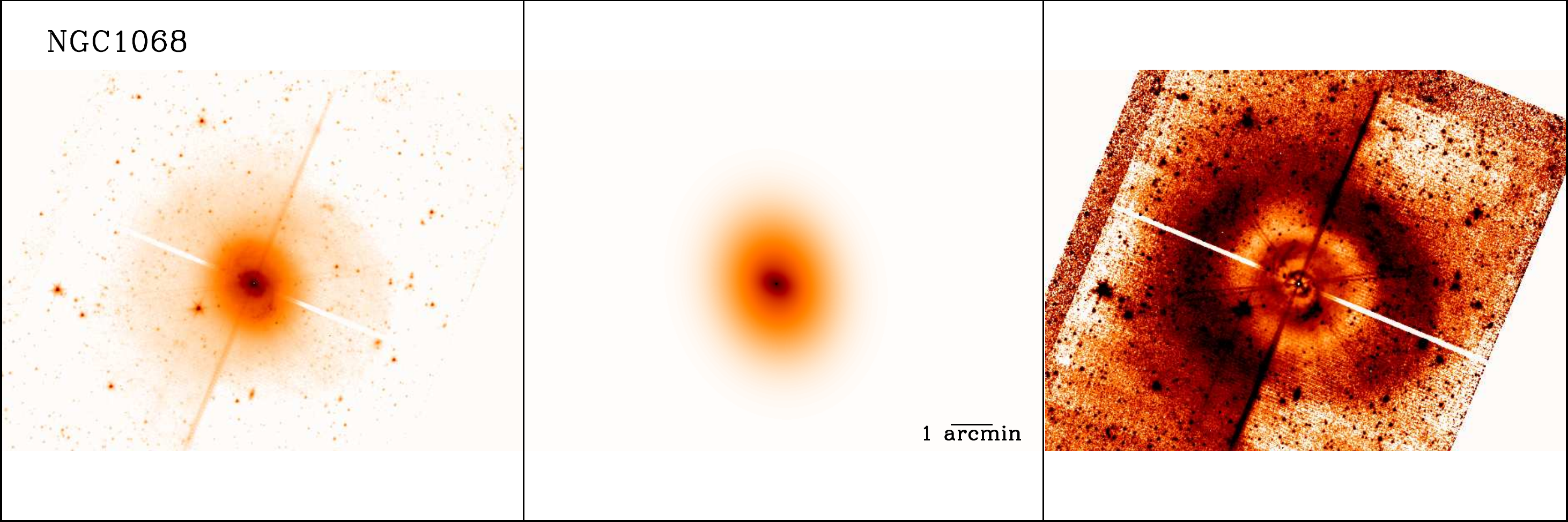}
\includegraphics[height=0.25\linewidth]{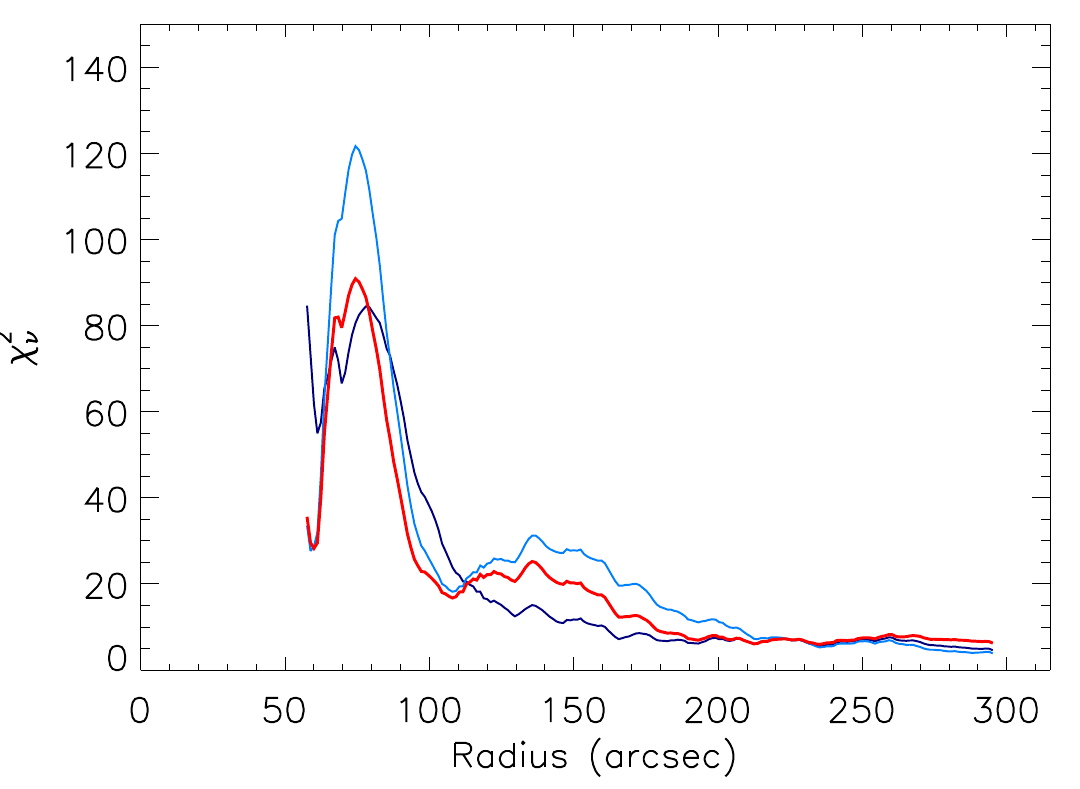}
}
}
\centerline{
\hbox{
\includegraphics[width=0.25\linewidth,angle=90]{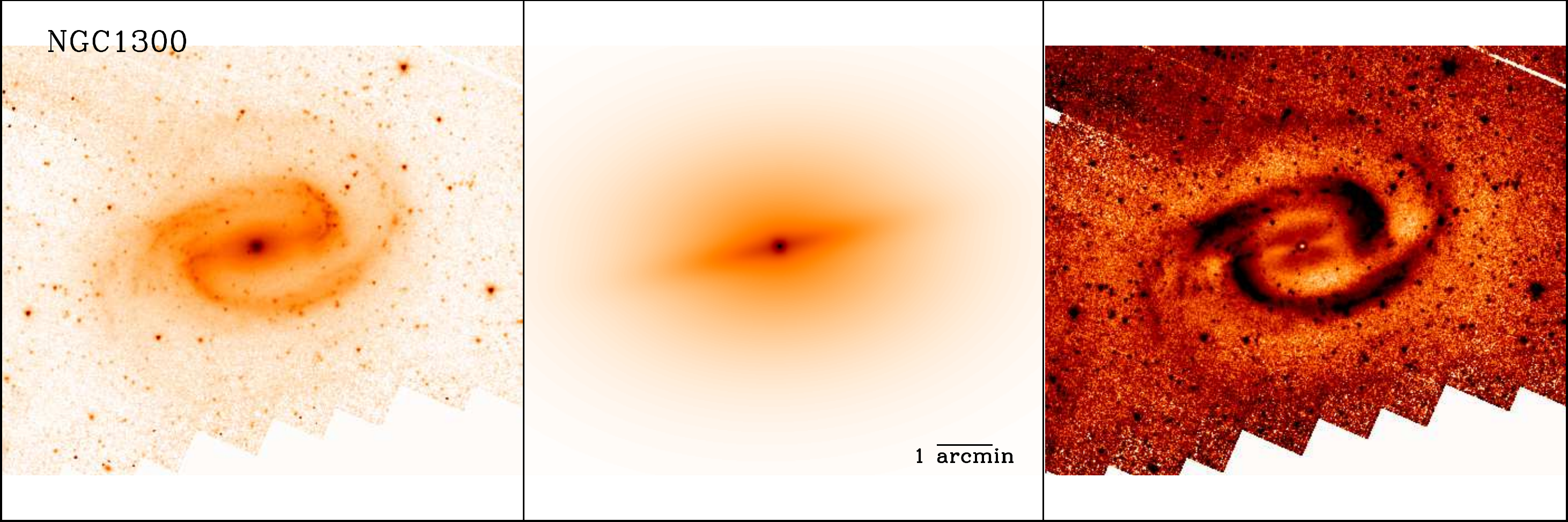}
\includegraphics[height=0.25\linewidth]{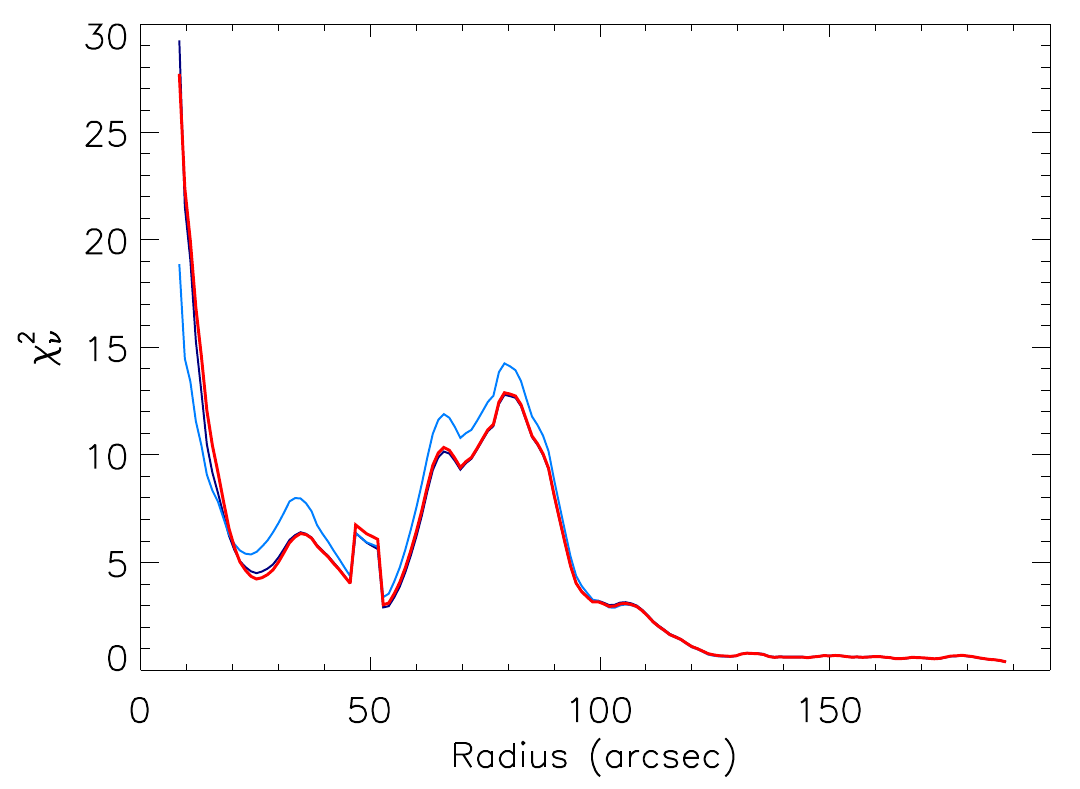}
}
}
\caption{Continued.}
\end{figure*}
\addtocounter{figure}{-1}
\begin{figure*}
\centerline{
\hbox{
\includegraphics[width=0.25\linewidth,angle=90]{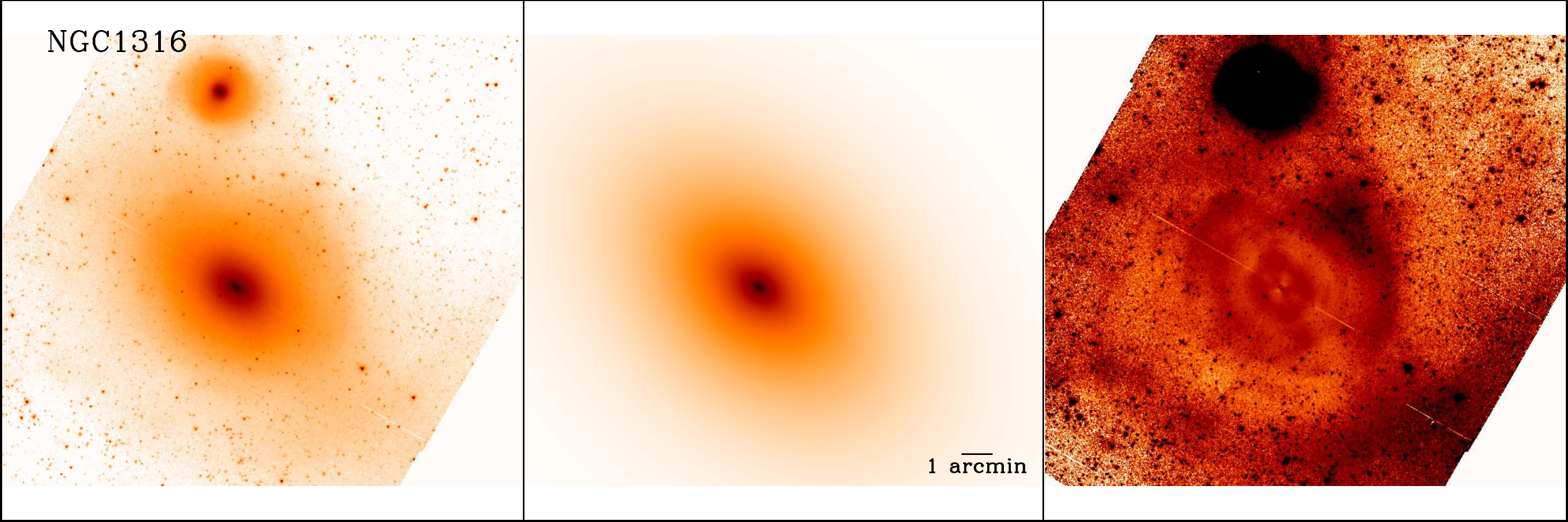}
\includegraphics[height=0.25\linewidth]{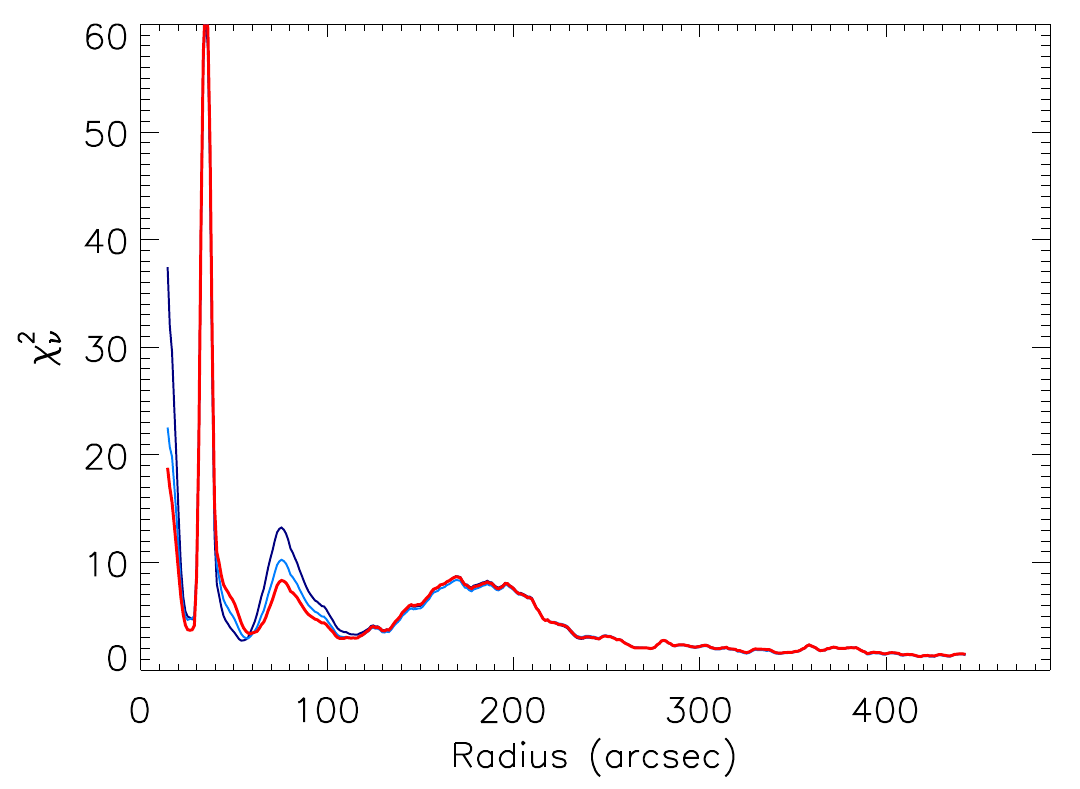}
}
}
\centerline{
\hbox{
\includegraphics[width=0.25\linewidth,angle=90]{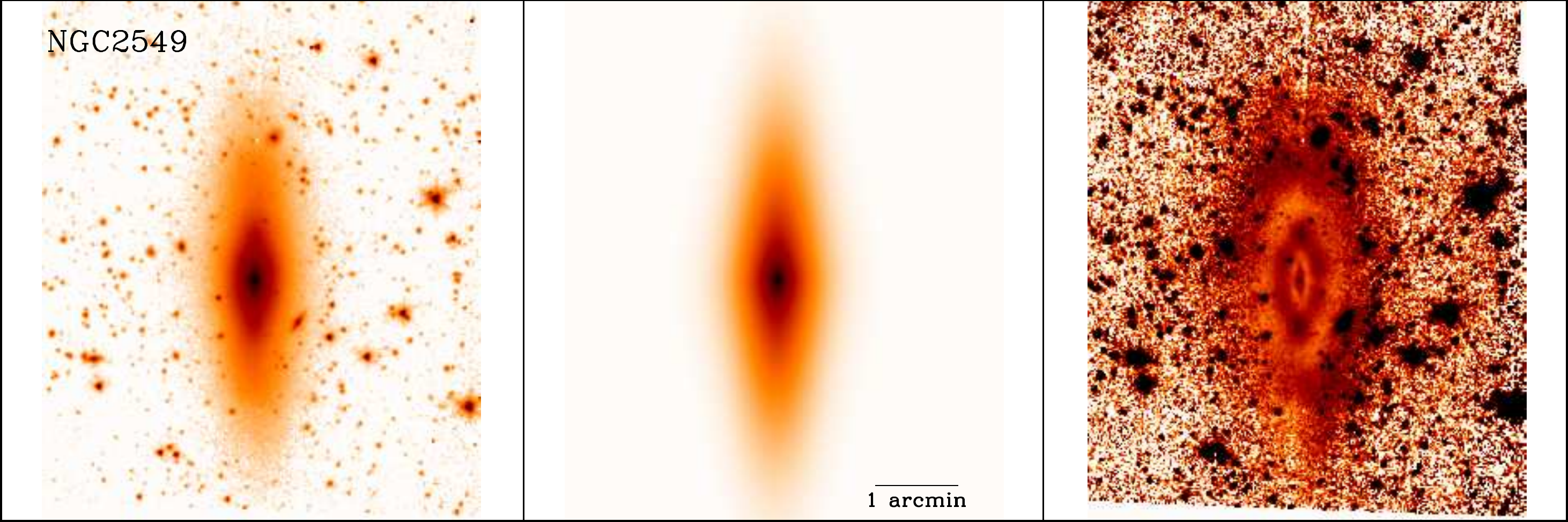}
\includegraphics[height=0.25\linewidth]{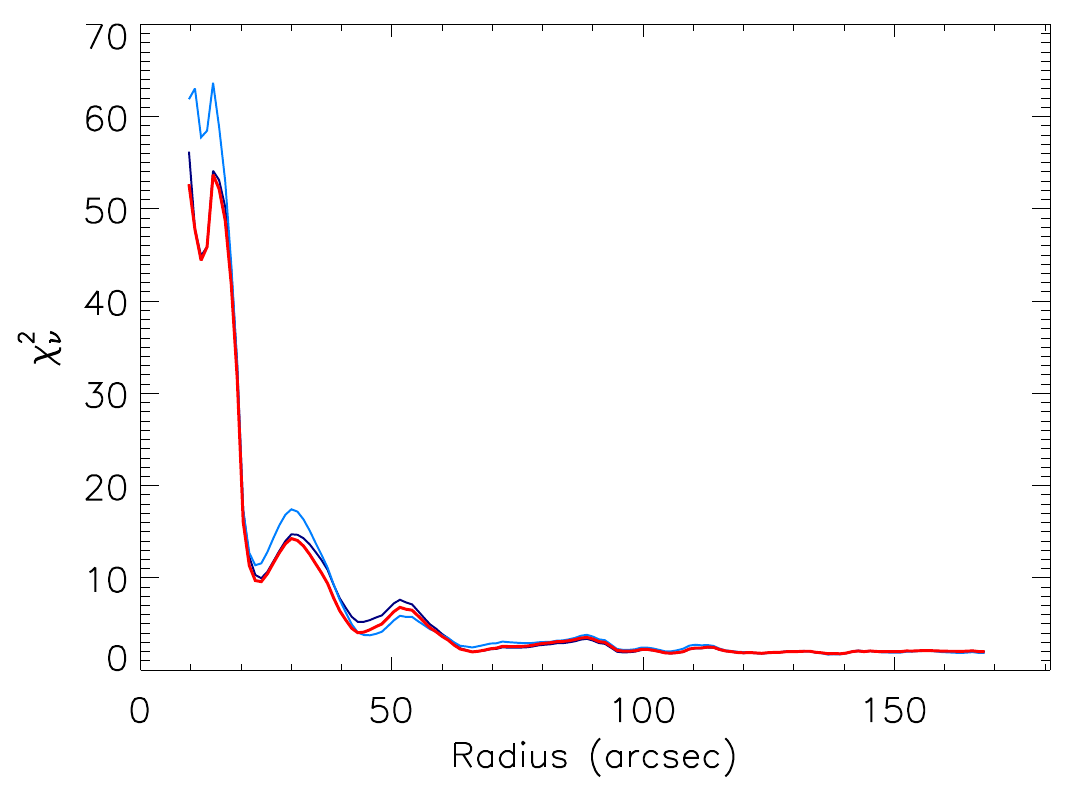}
}
}
\centerline{
\hbox{
\includegraphics[width=0.25\linewidth,angle=90]{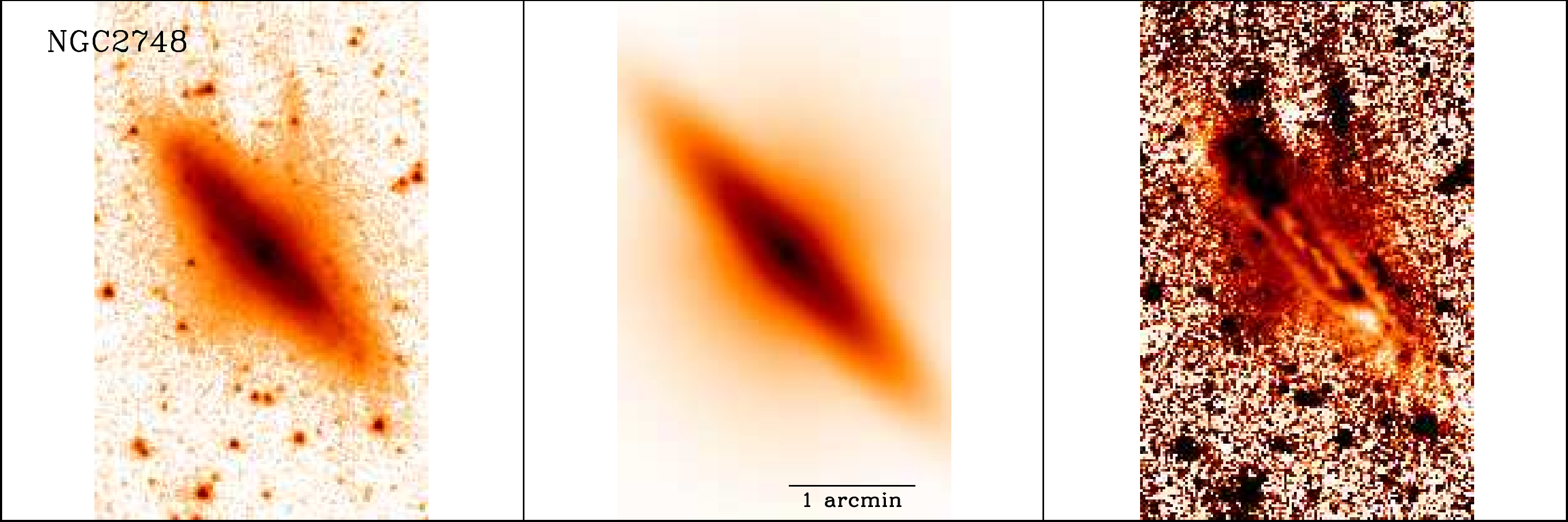}
\includegraphics[height=0.25\linewidth]{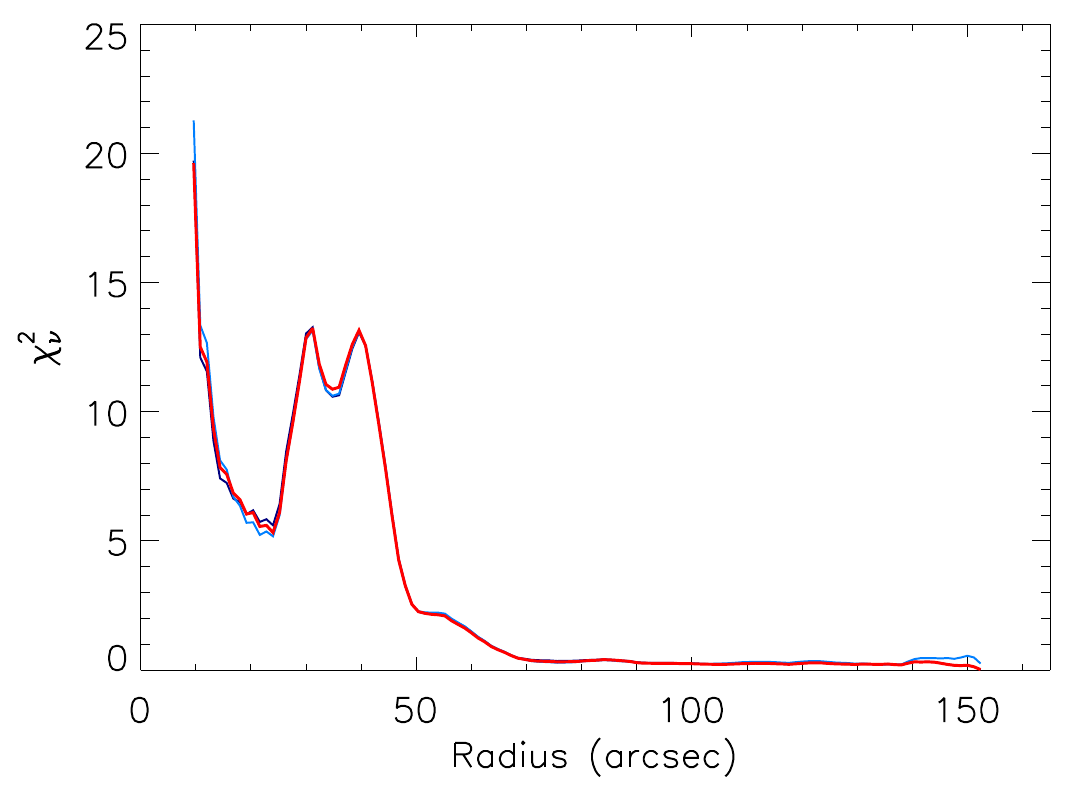}
}
}
\centerline{
\hbox{
\includegraphics[width=0.25\linewidth,angle=90]{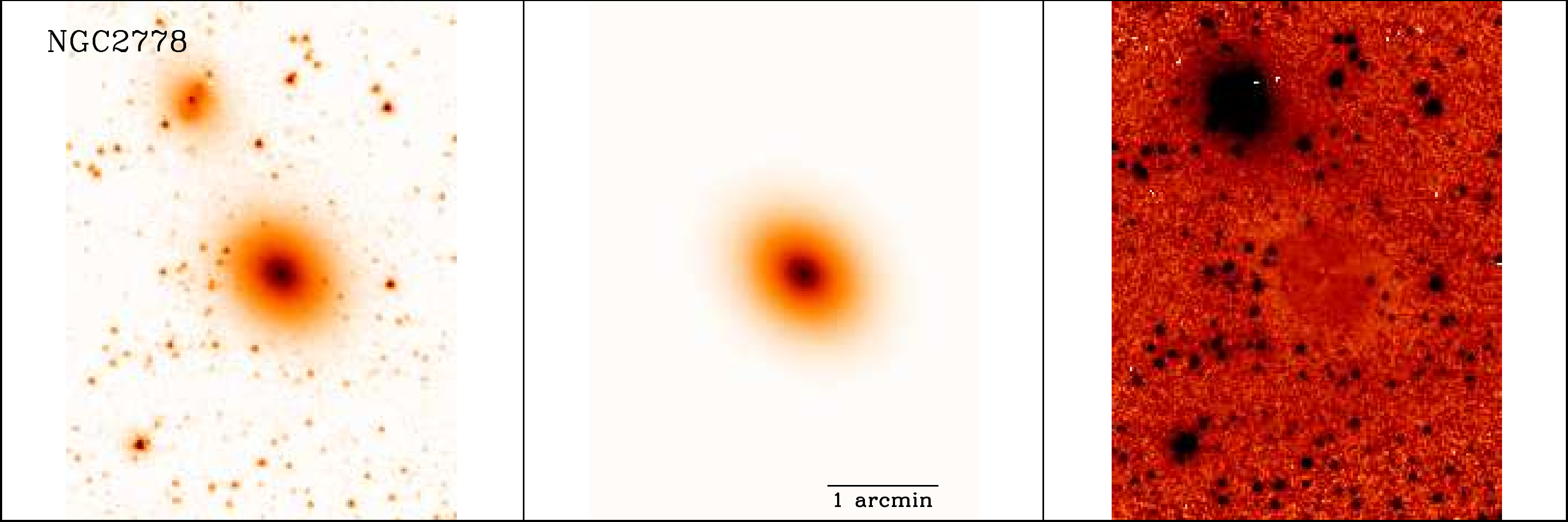}
\includegraphics[height=0.25\linewidth]{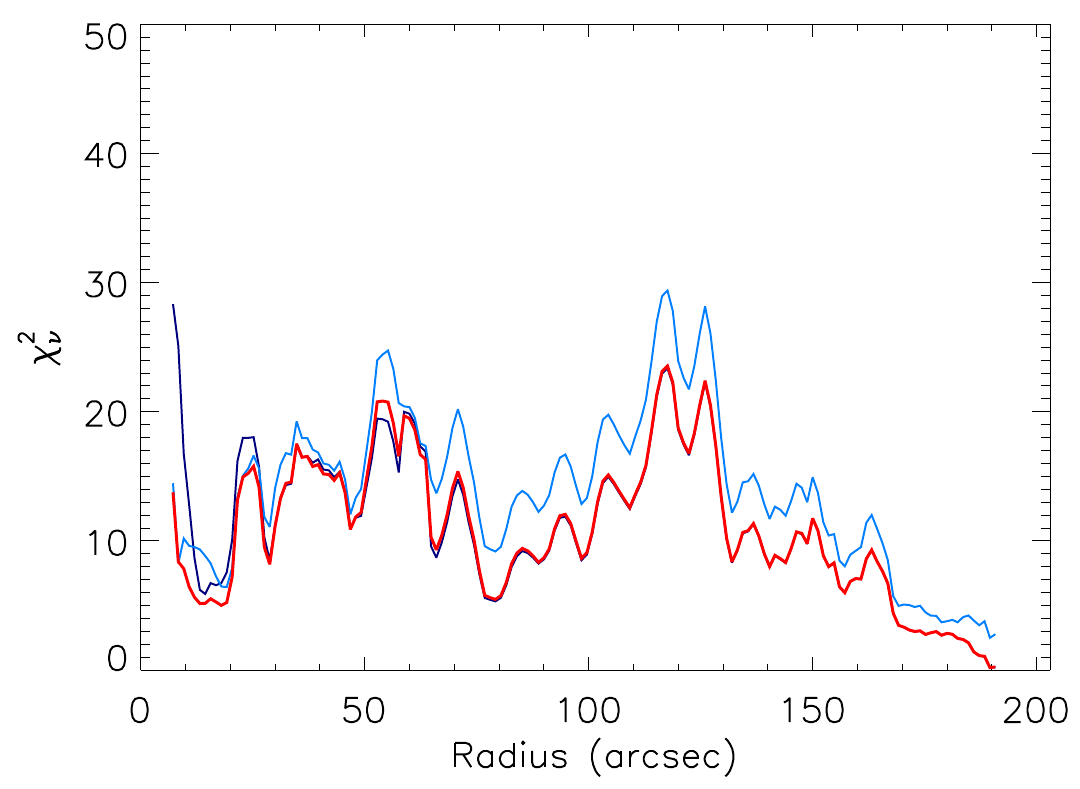}
}
}
\centerline{
\hbox{
\includegraphics[width=0.25\linewidth,angle=90]{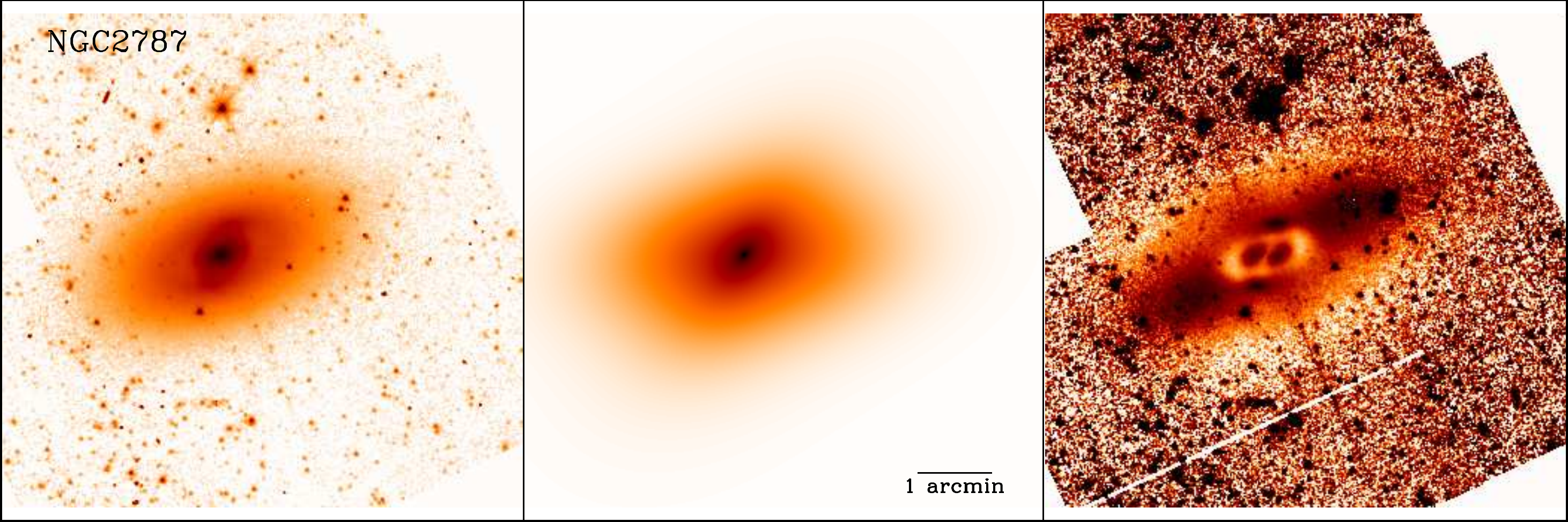}
\includegraphics[height=0.25\linewidth]{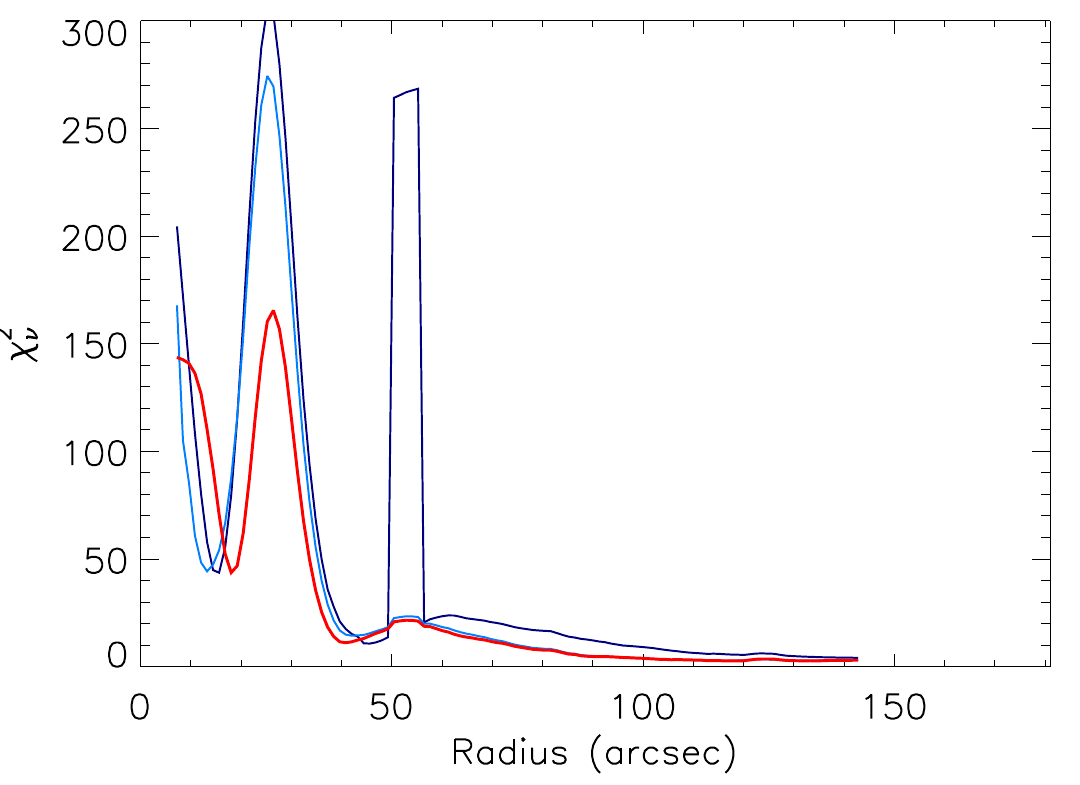}
}
}
\caption{Continued.}
\end{figure*}
\addtocounter{figure}{-1}
\begin{figure*}
\centerline{
\hbox{
\includegraphics[width=0.25\linewidth,angle=90]{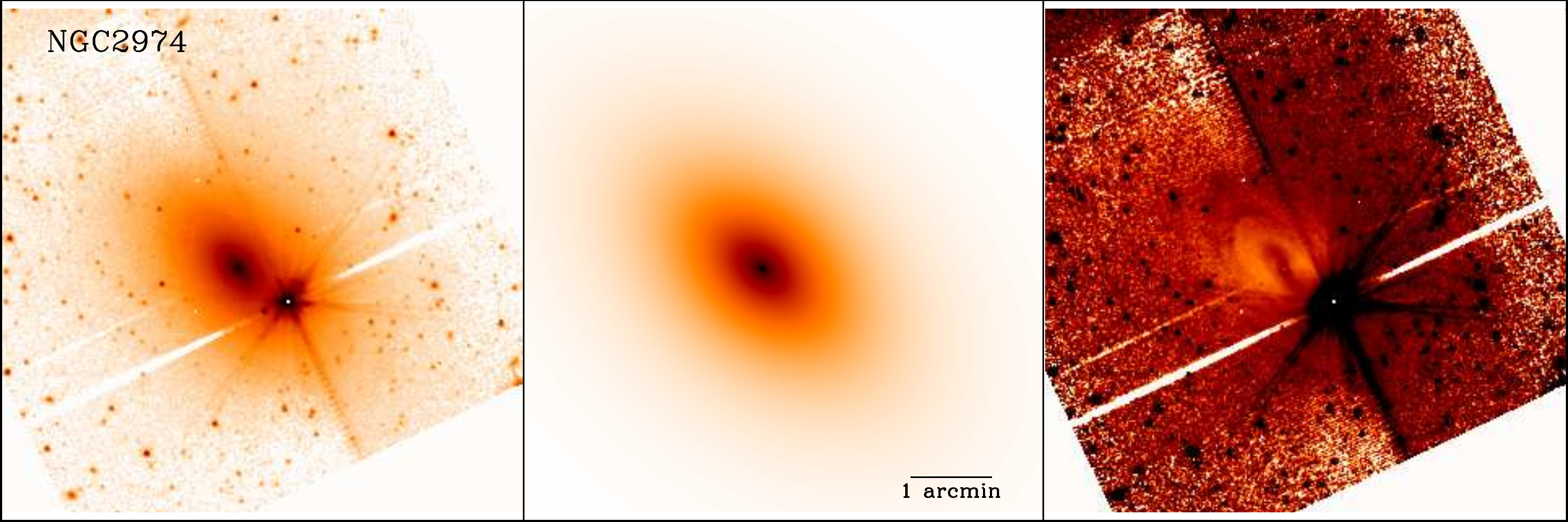}
\includegraphics[height=0.25\linewidth]{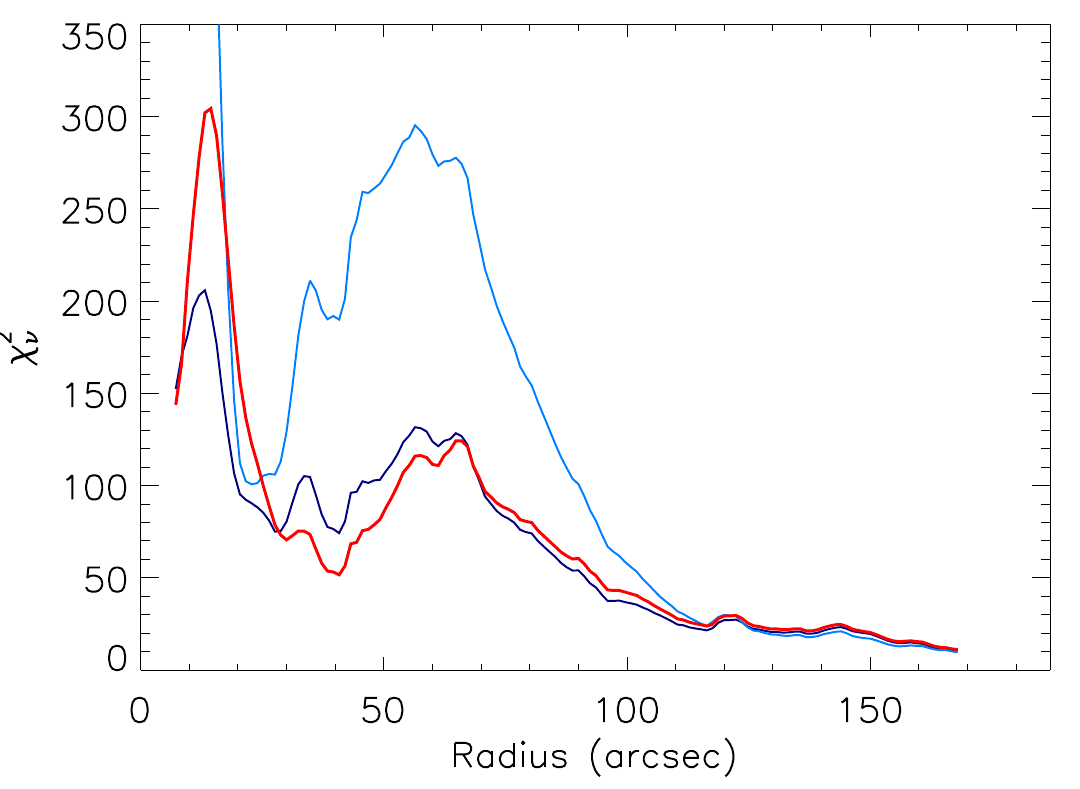}
}
}
\centerline{
\hbox{
\includegraphics[width=0.25\linewidth,angle=90]{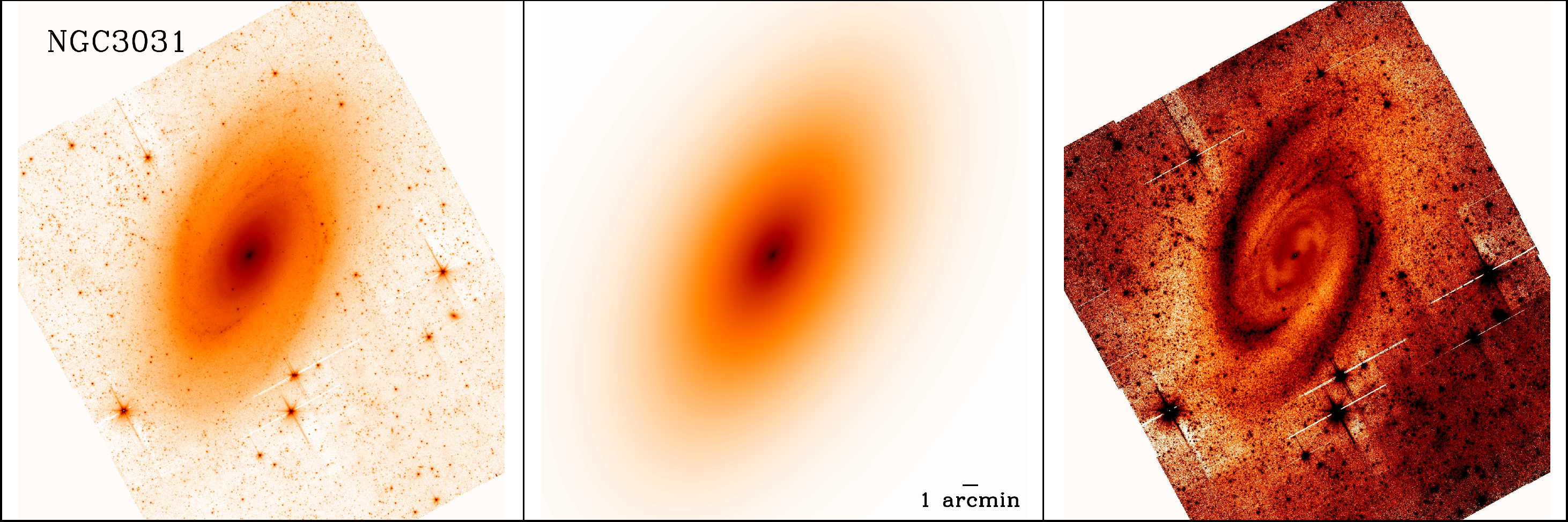}
\includegraphics[height=0.25\linewidth]{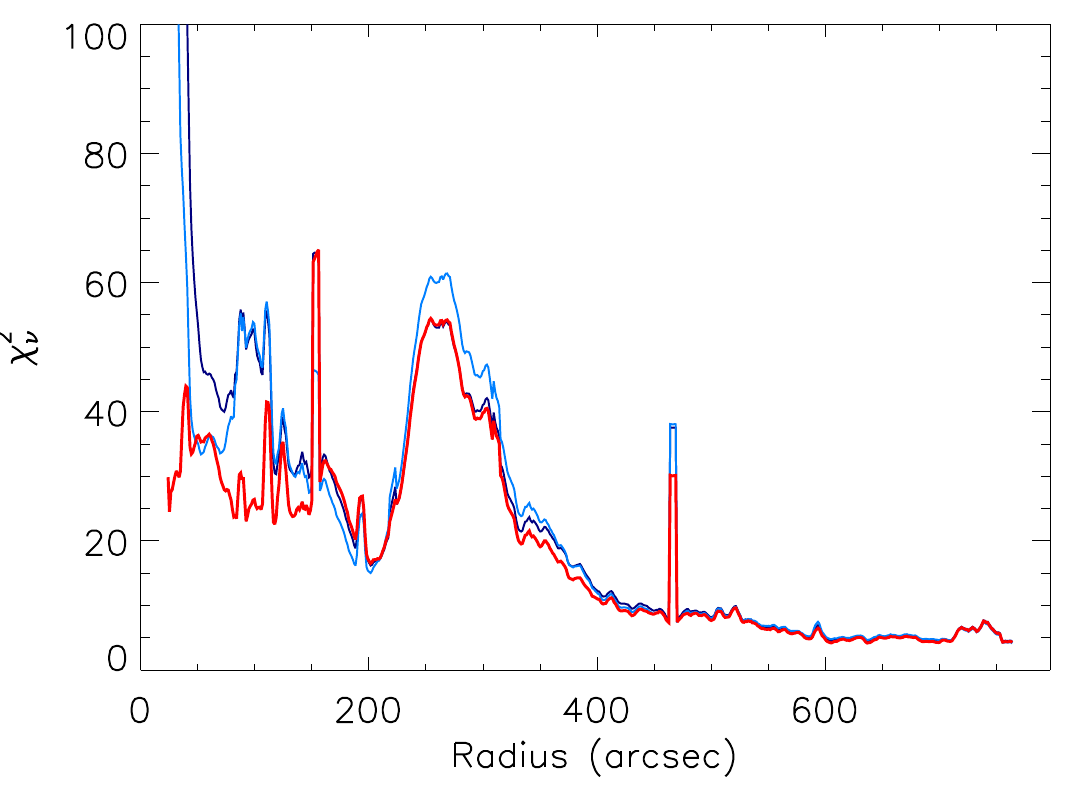}
}
}
\centerline{
\hbox{
\includegraphics[width=0.25\linewidth,angle=90]{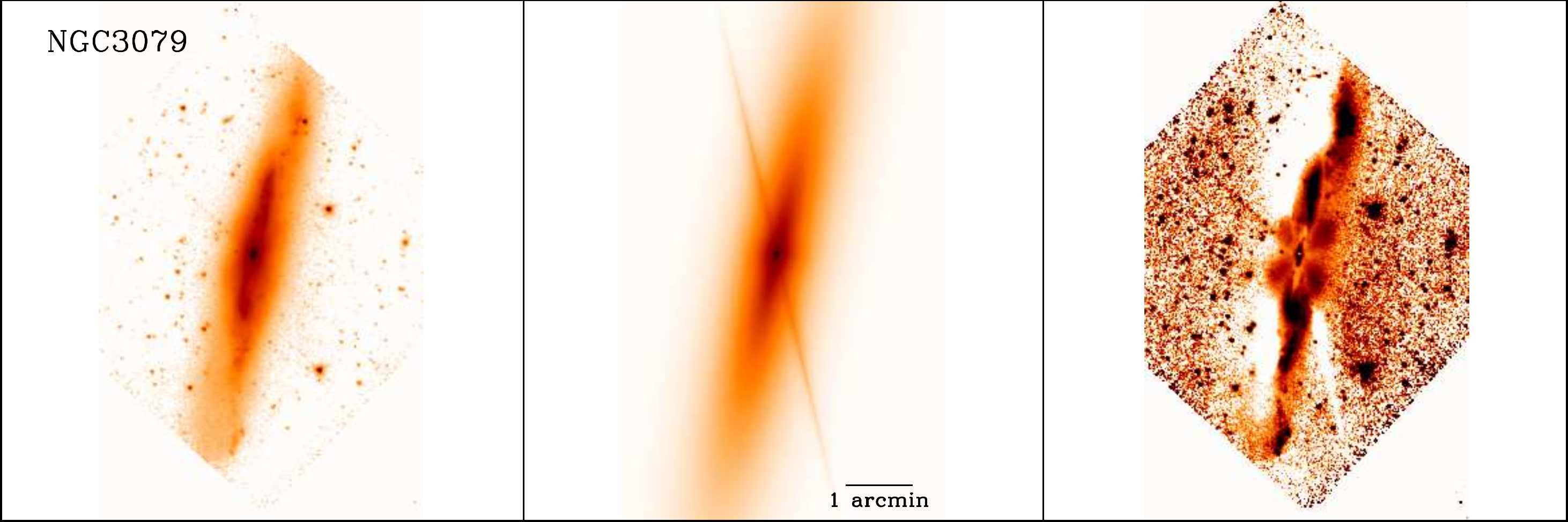}
\includegraphics[height=0.25\linewidth]{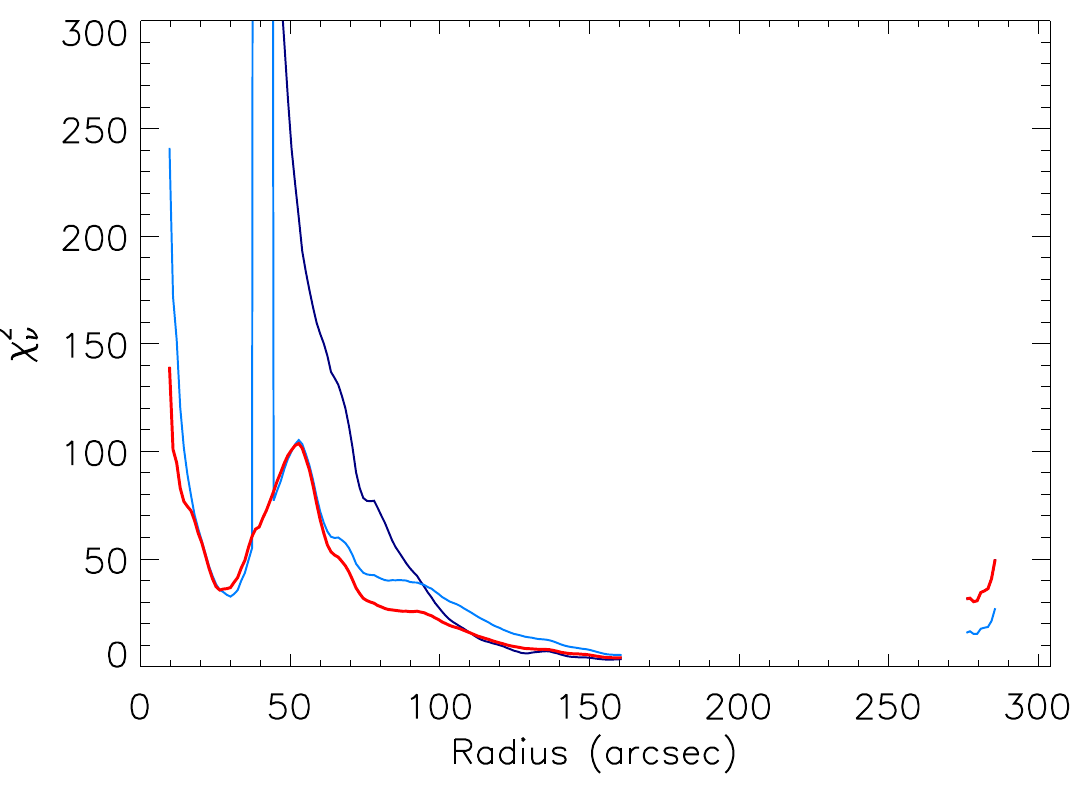}
}
}
\centerline{
\hbox{
\includegraphics[width=0.25\linewidth,angle=90]{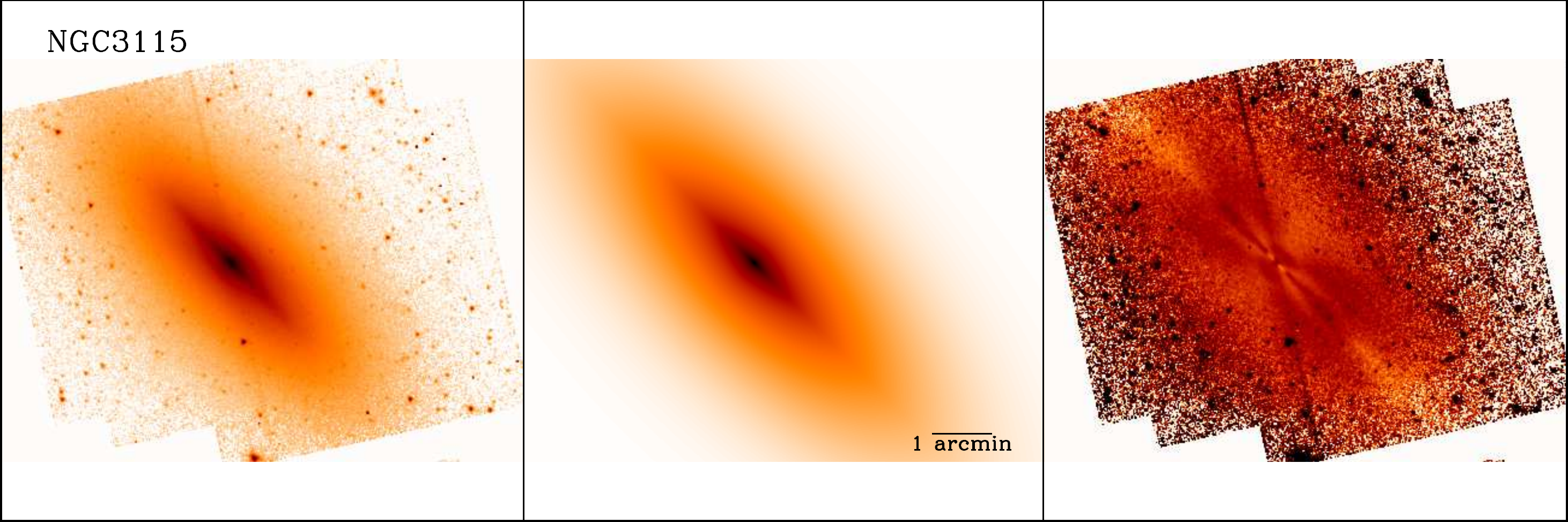}
\includegraphics[height=0.25\linewidth]{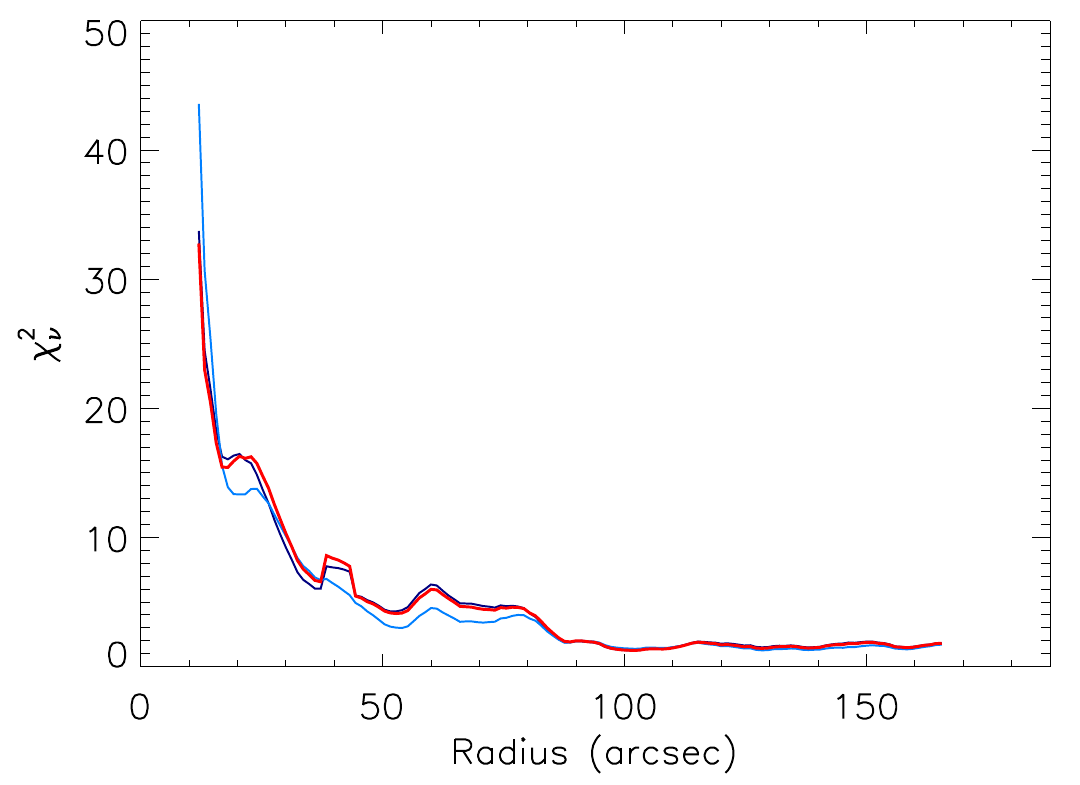}
}
}
\centerline{
\hbox{
\includegraphics[width=0.25\linewidth,angle=90]{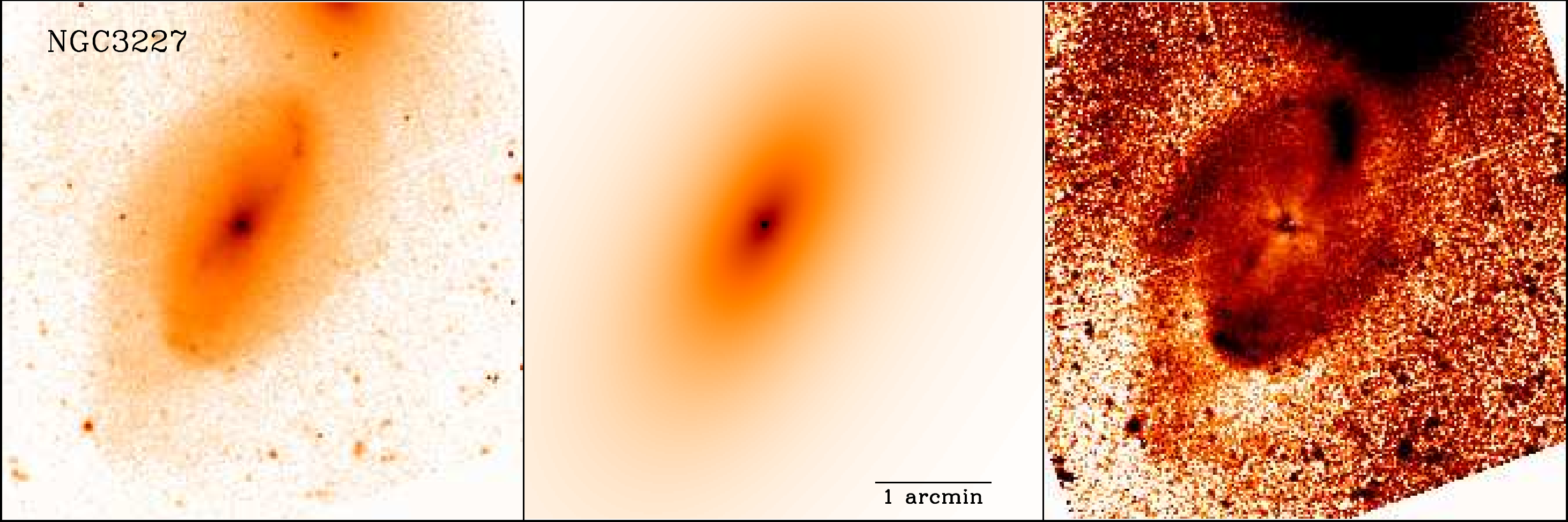}
\includegraphics[height=0.25\linewidth]{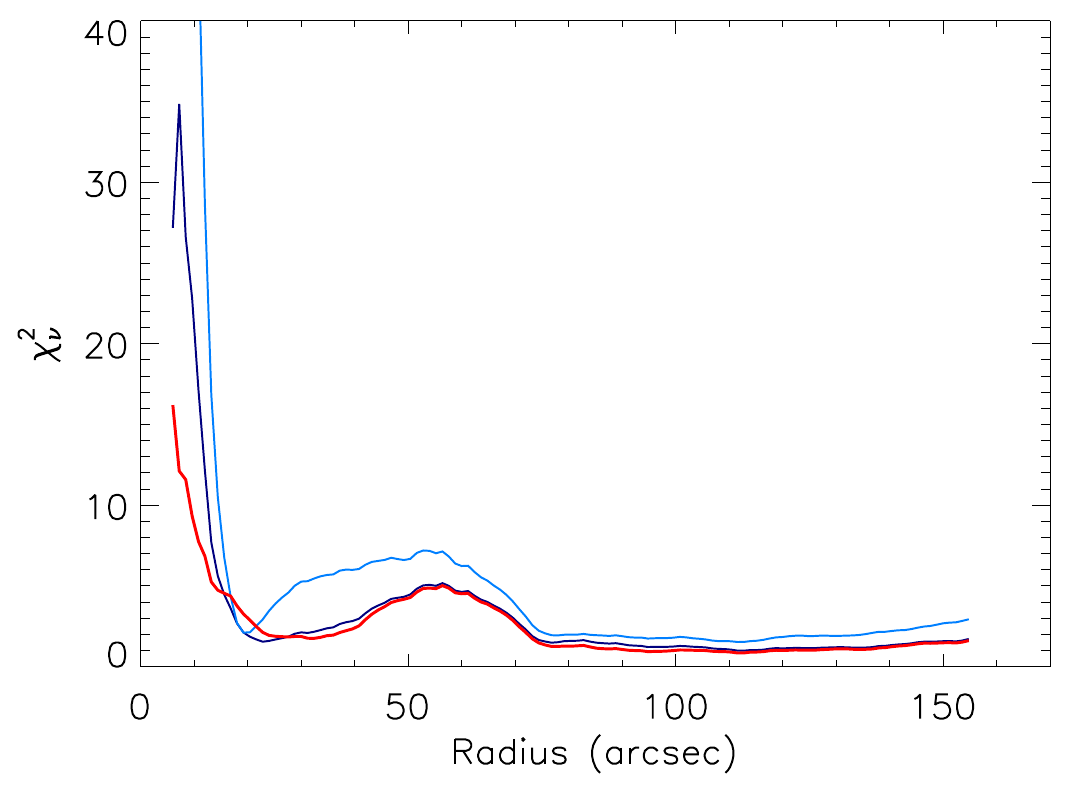}
}
}
\caption{Continued.}
\end{figure*}
\addtocounter{figure}{-1}
\begin{figure*}
\centerline{
\hbox{
\includegraphics[width=0.25\linewidth,angle=90]{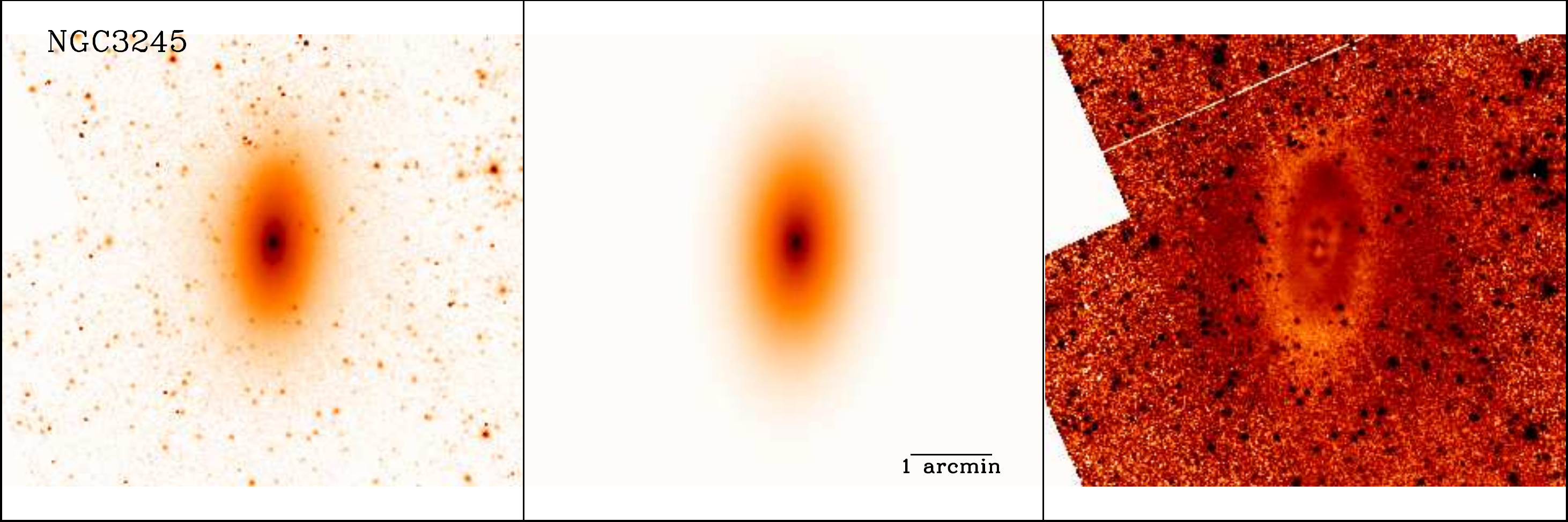}
\includegraphics[height=0.25\linewidth]{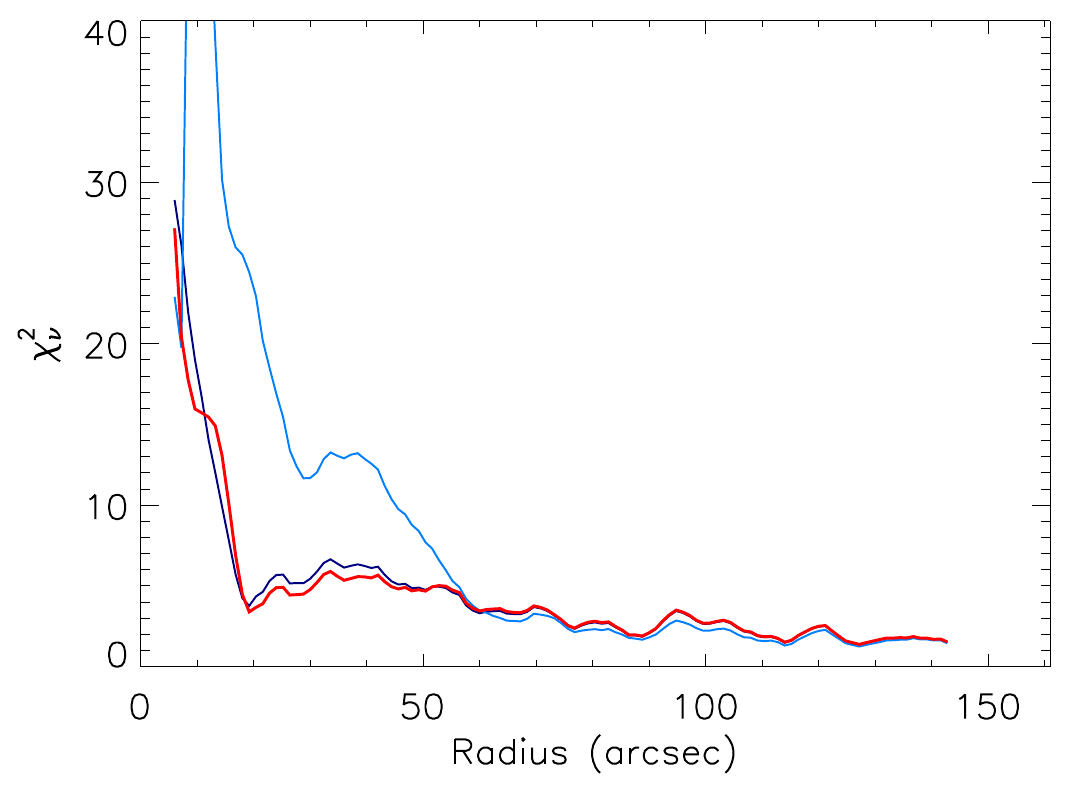}
}
}
\centerline{
\hbox{
\includegraphics[width=0.25\linewidth,angle=90]{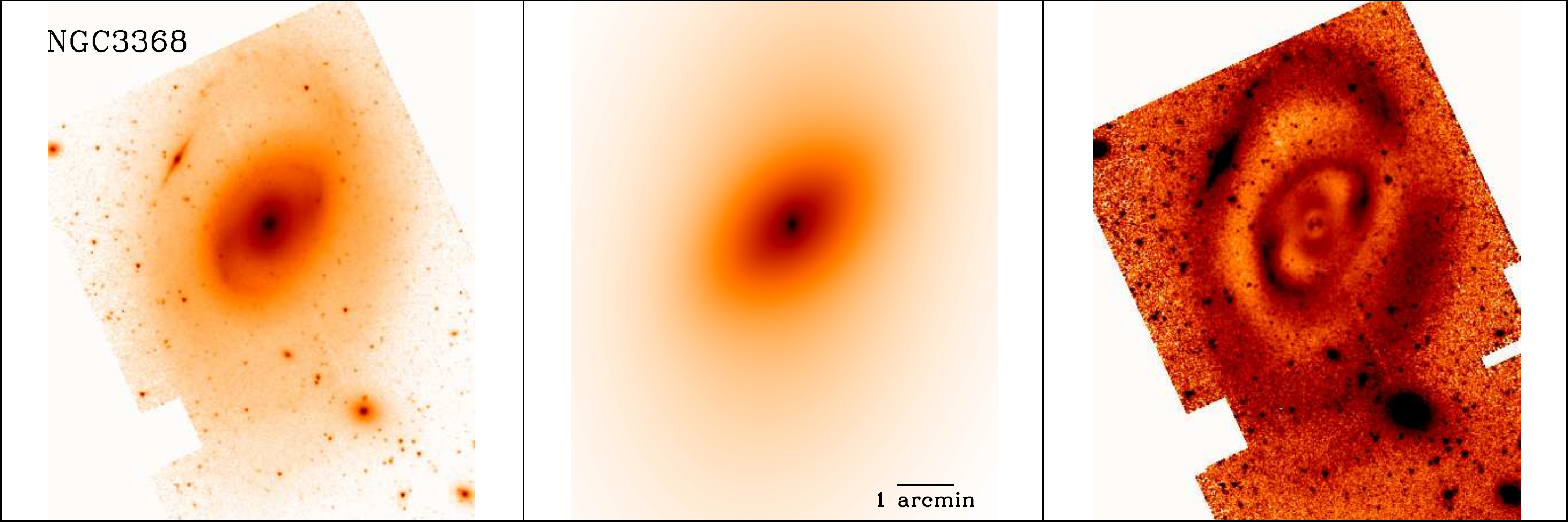}
\includegraphics[height=0.25\linewidth]{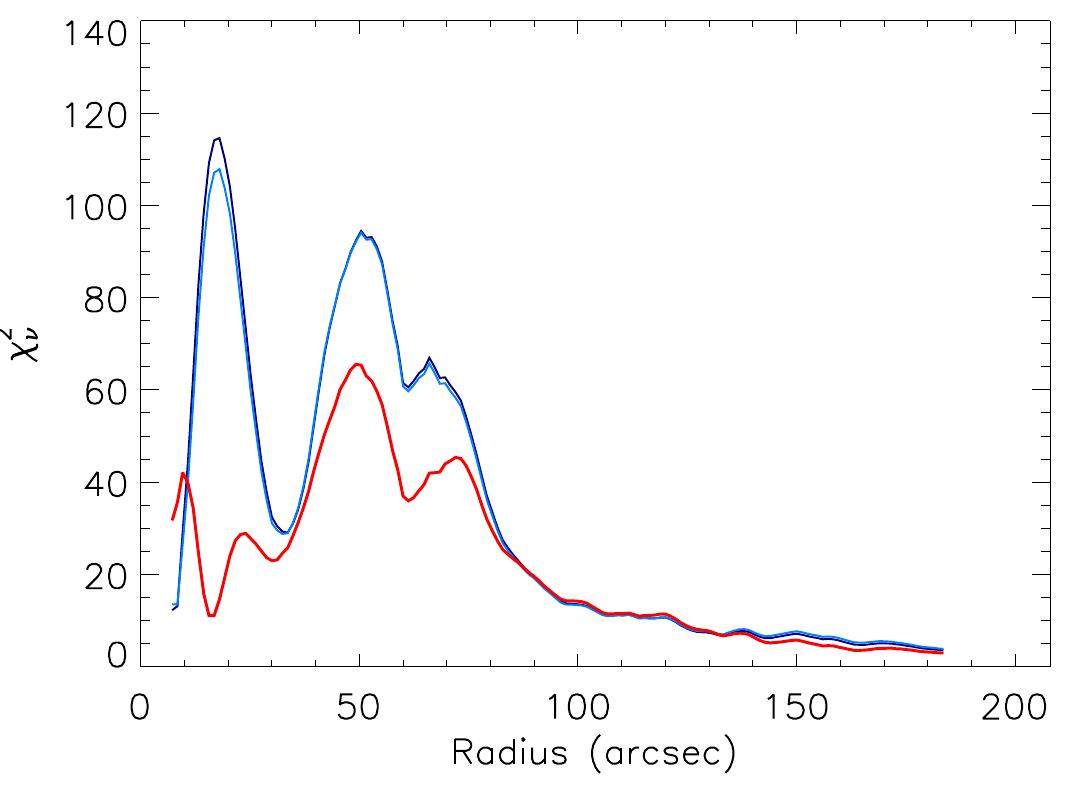}
}
}
\centerline{
\hbox{
\includegraphics[width=0.25\linewidth,angle=90]{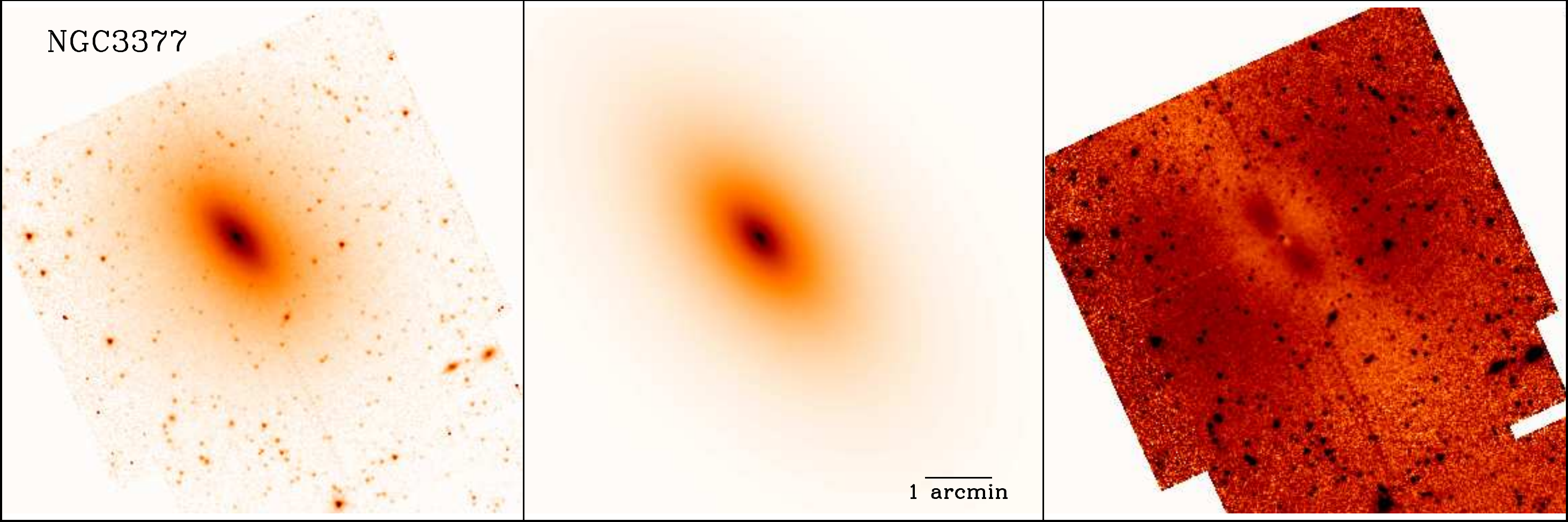}
\includegraphics[height=0.25\linewidth]{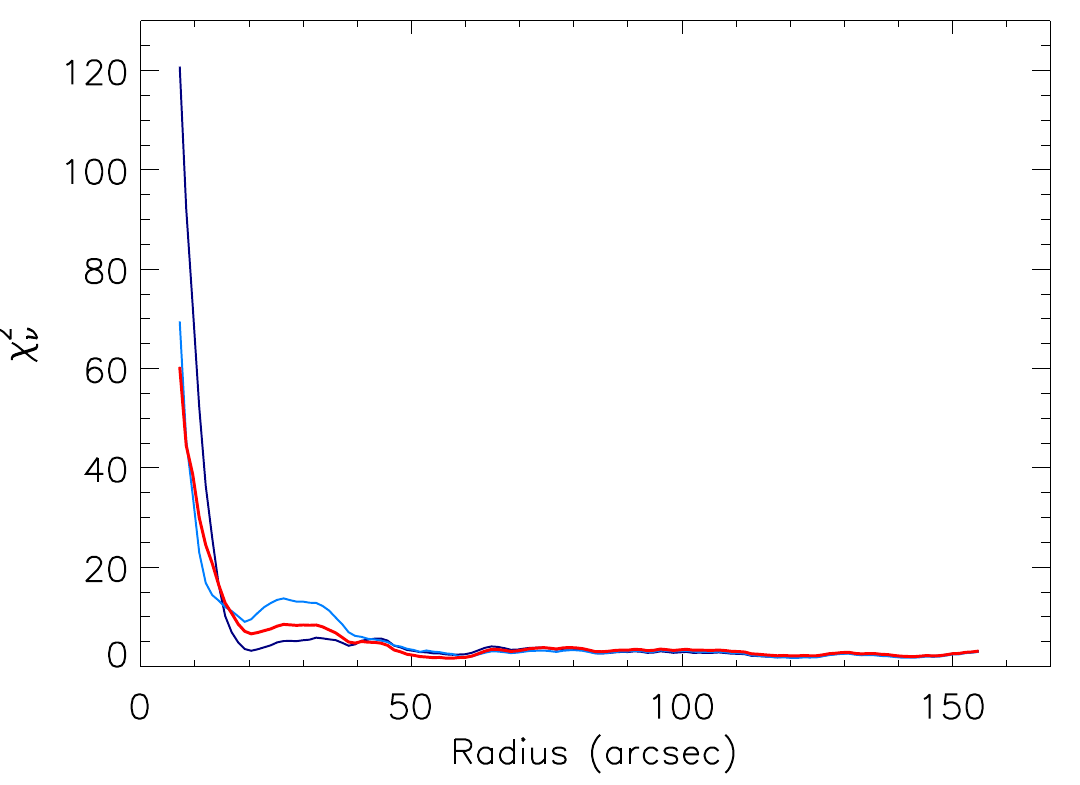}
}
}
\centerline{
\hbox{
\includegraphics[width=0.25\linewidth,angle=90]{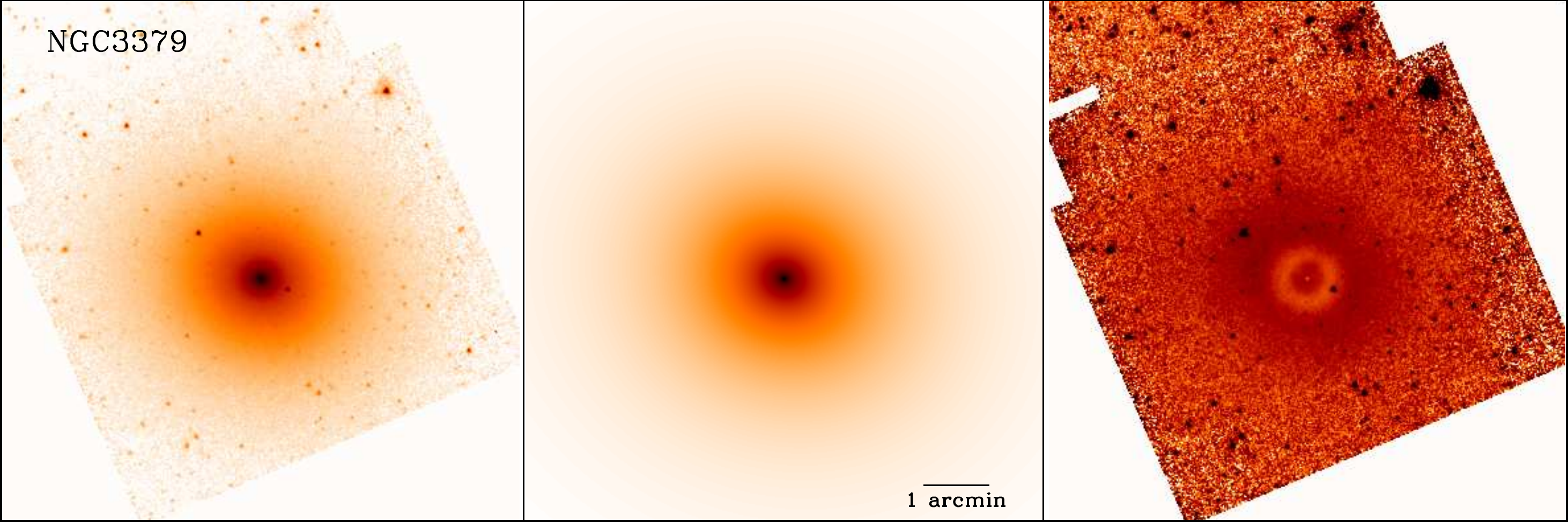}
\includegraphics[height=0.25\linewidth]{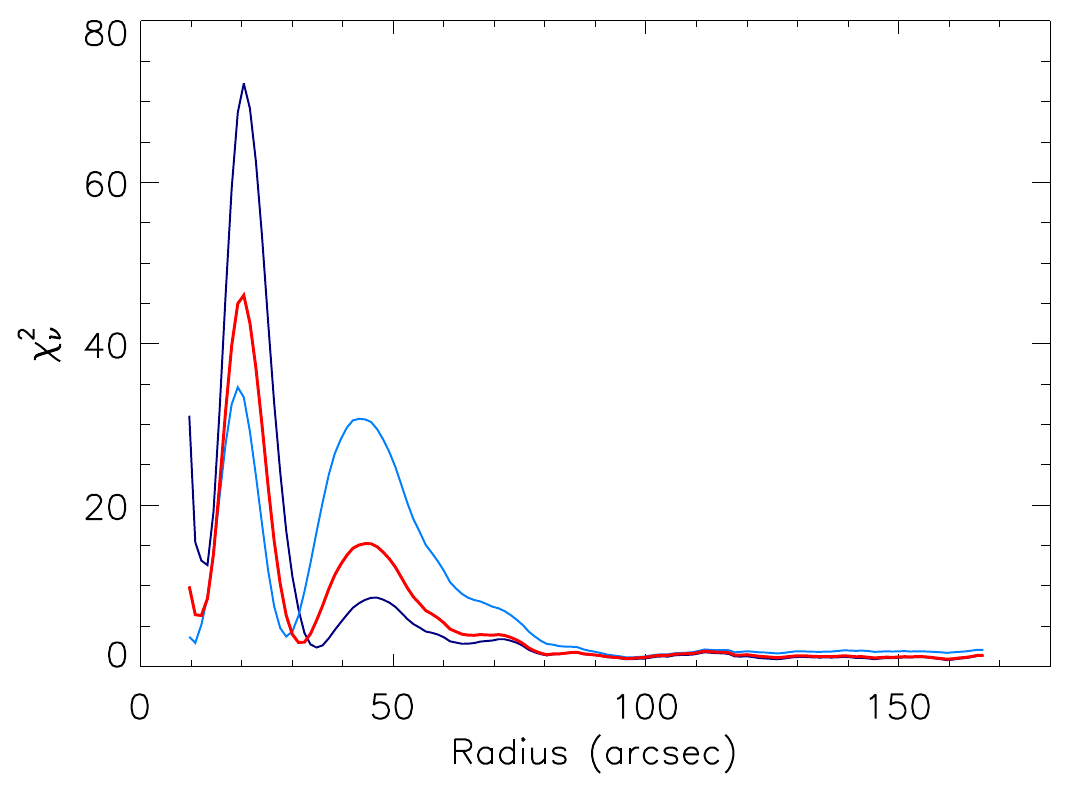}
}
}
\centerline{
\hbox{
\includegraphics[width=0.25\linewidth,angle=90]{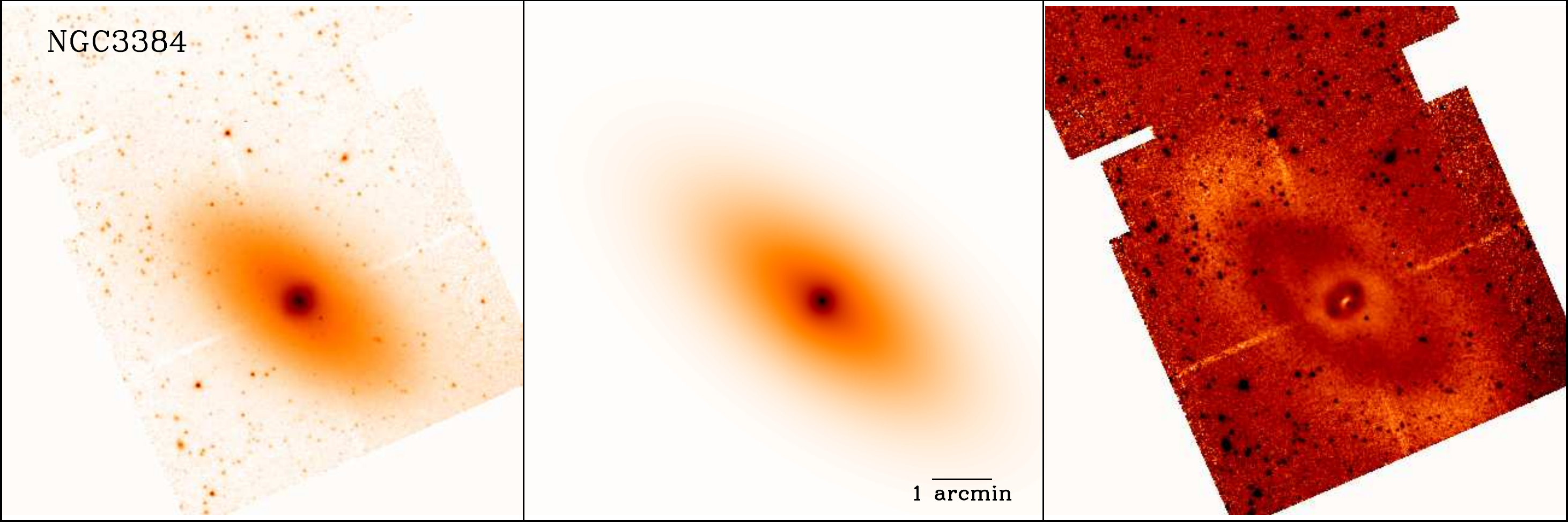}
\includegraphics[height=0.25\linewidth]{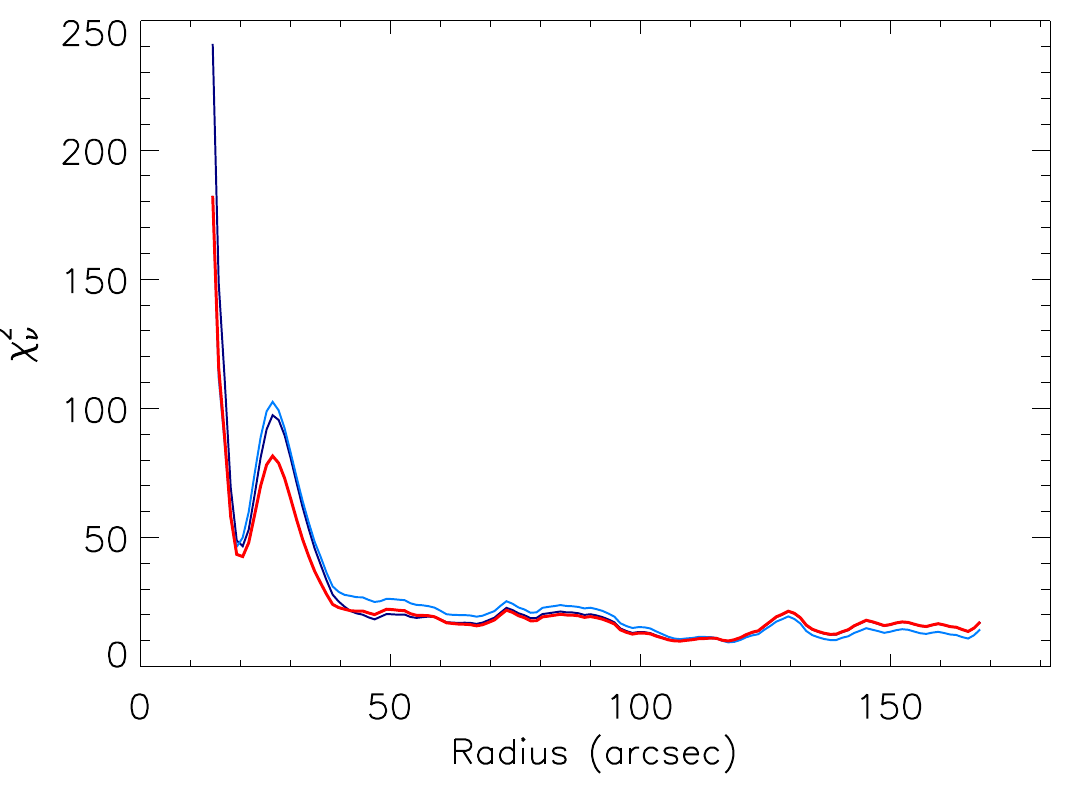}
}
}
\caption{Continued.}
\end{figure*}
\addtocounter{figure}{-1}
\begin{figure*}
\centerline{
\hbox{
\includegraphics[width=0.25\linewidth,angle=90]{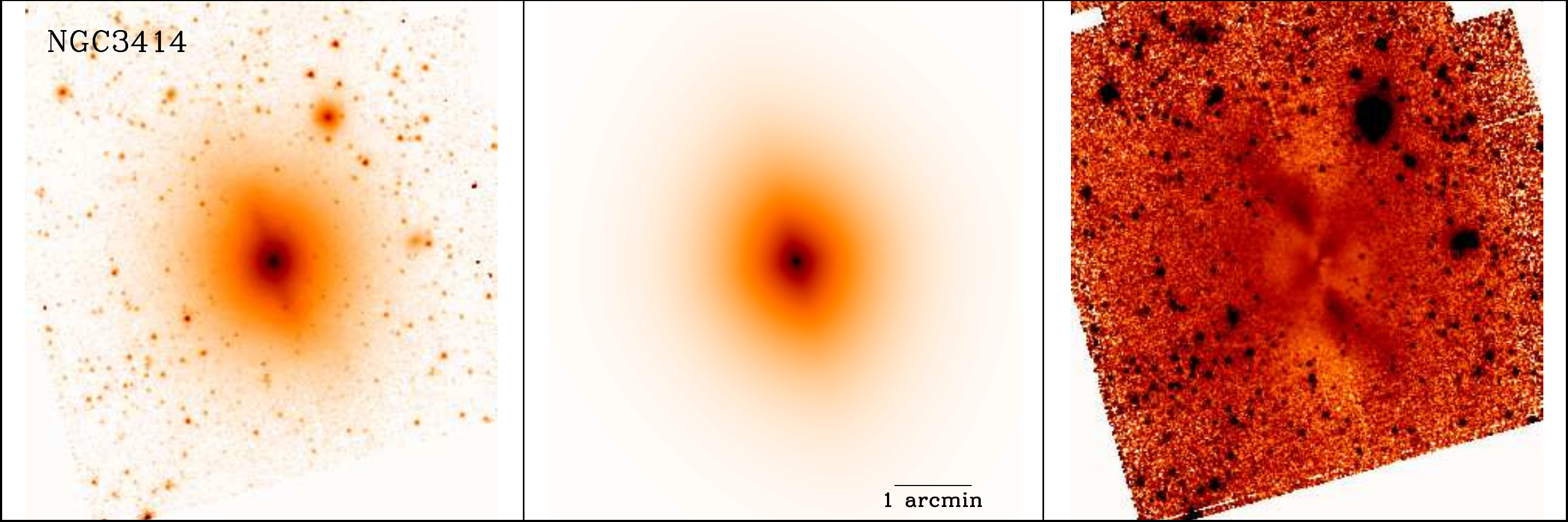}
\includegraphics[height=0.25\linewidth]{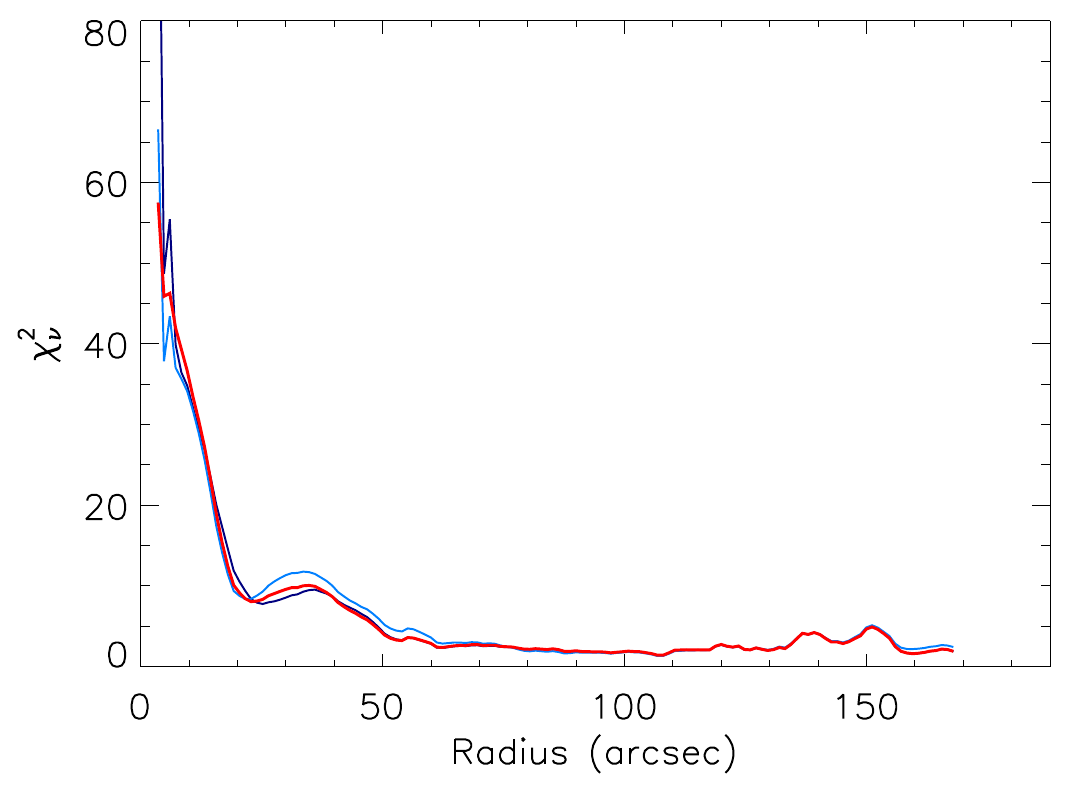}
}
}
\centerline{
\hbox{
\includegraphics[width=0.25\linewidth,angle=90]{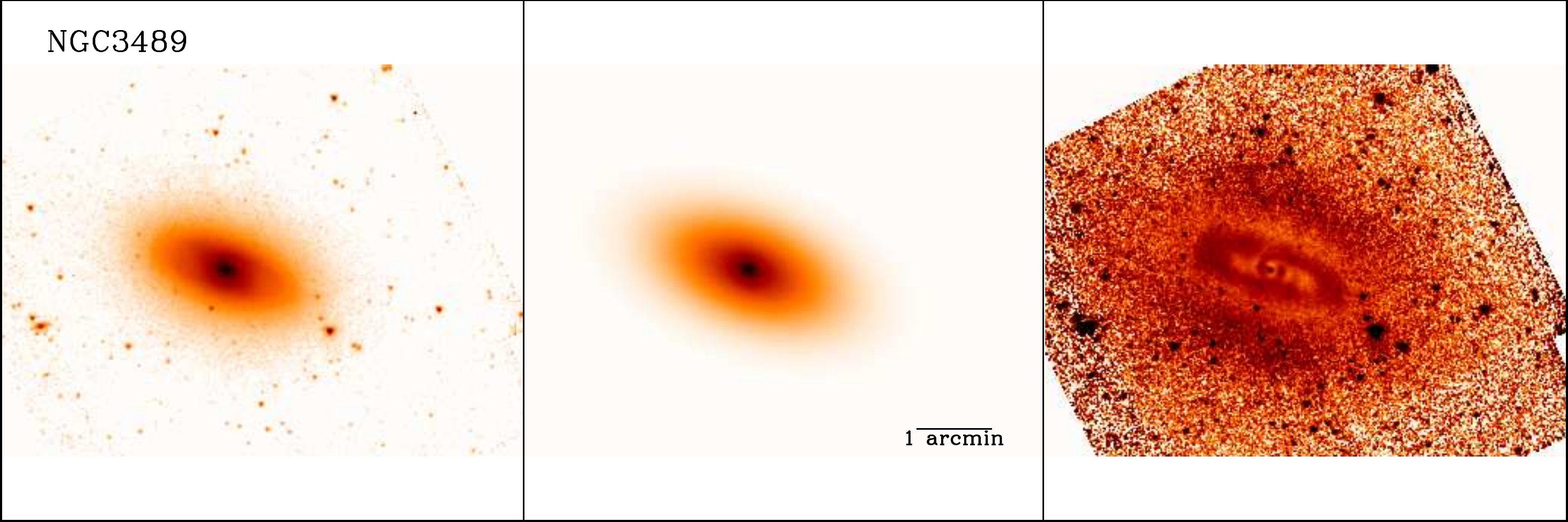}
\includegraphics[height=0.25\linewidth]{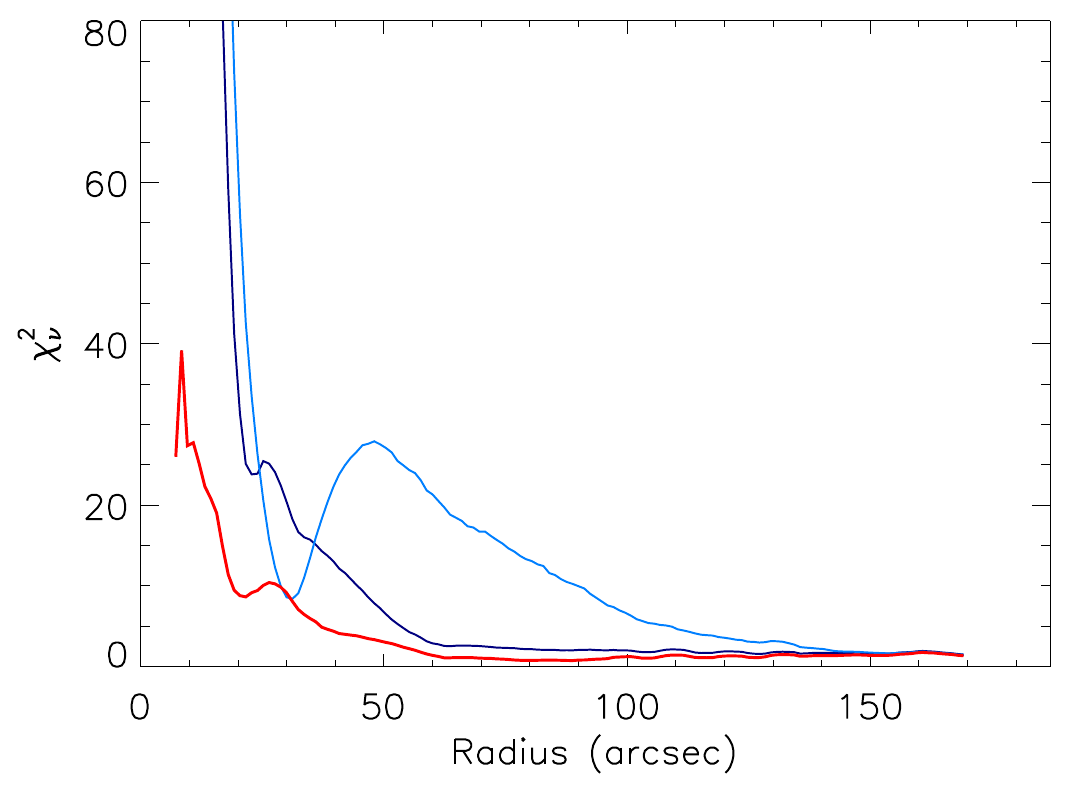}
}
}
\centerline{
\hbox{
\includegraphics[width=0.25\linewidth,angle=90]{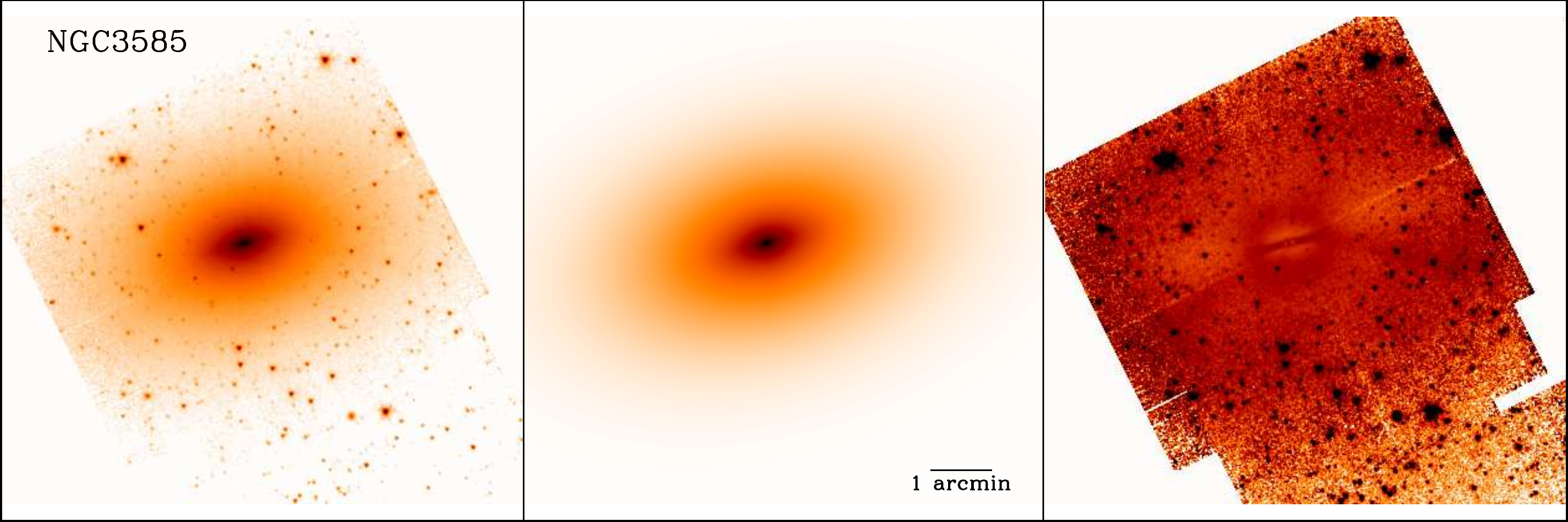}
\includegraphics[height=0.25\linewidth]{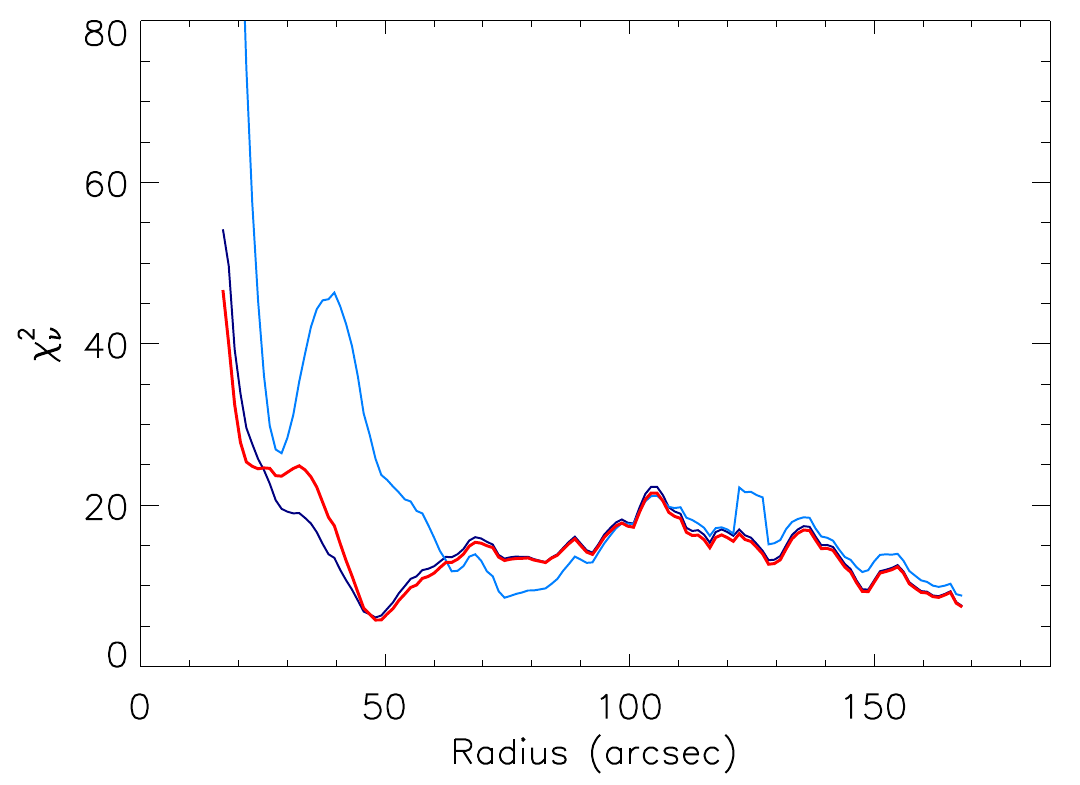}
}
}
\centerline{
\hbox{
\includegraphics[width=0.25\linewidth,angle=90]{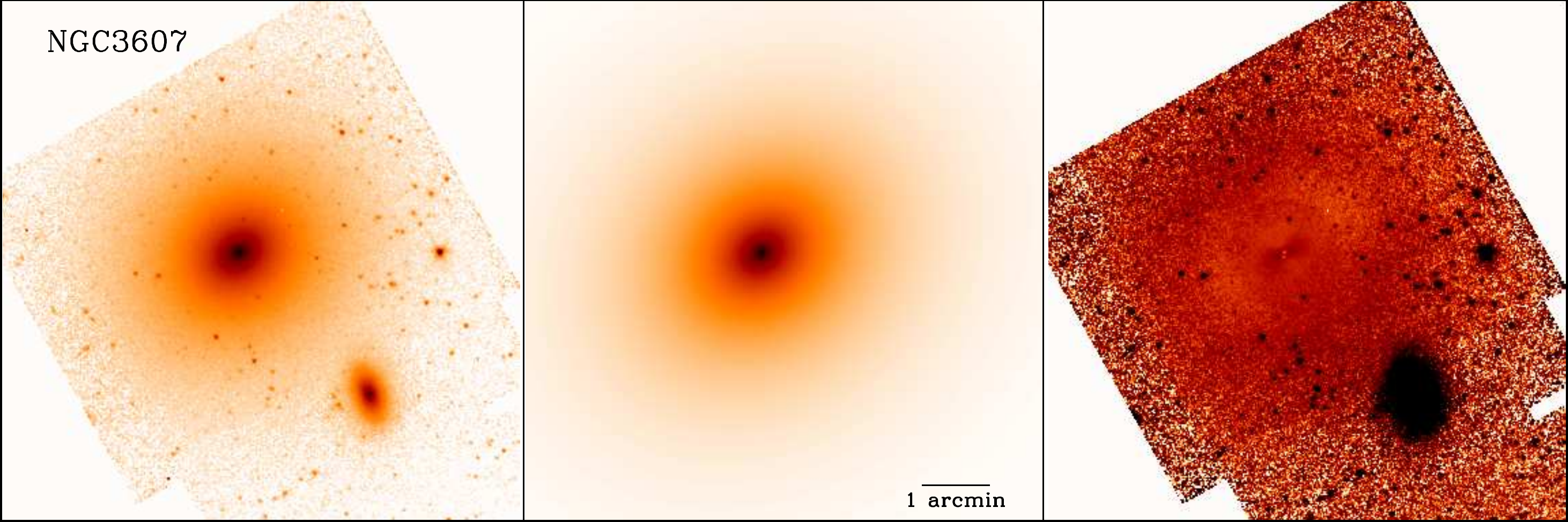}
\includegraphics[height=0.25\linewidth]{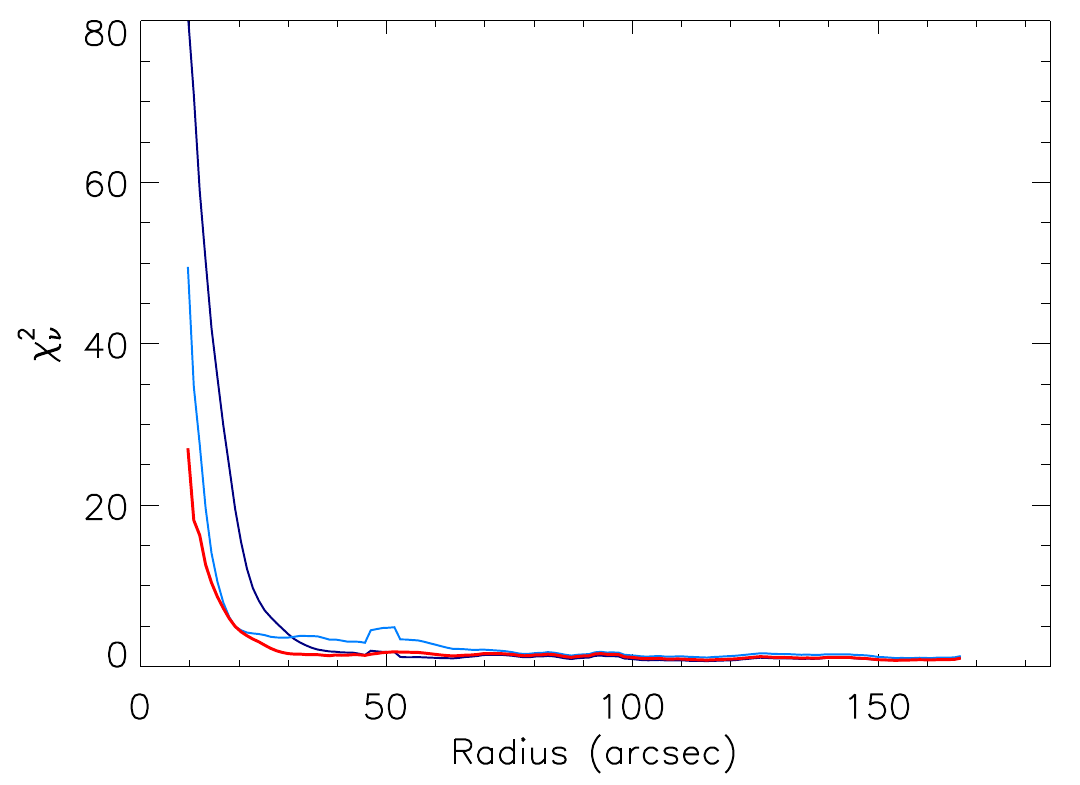}
}
}
\centerline{
\hbox{
\includegraphics[width=0.25\linewidth,angle=90]{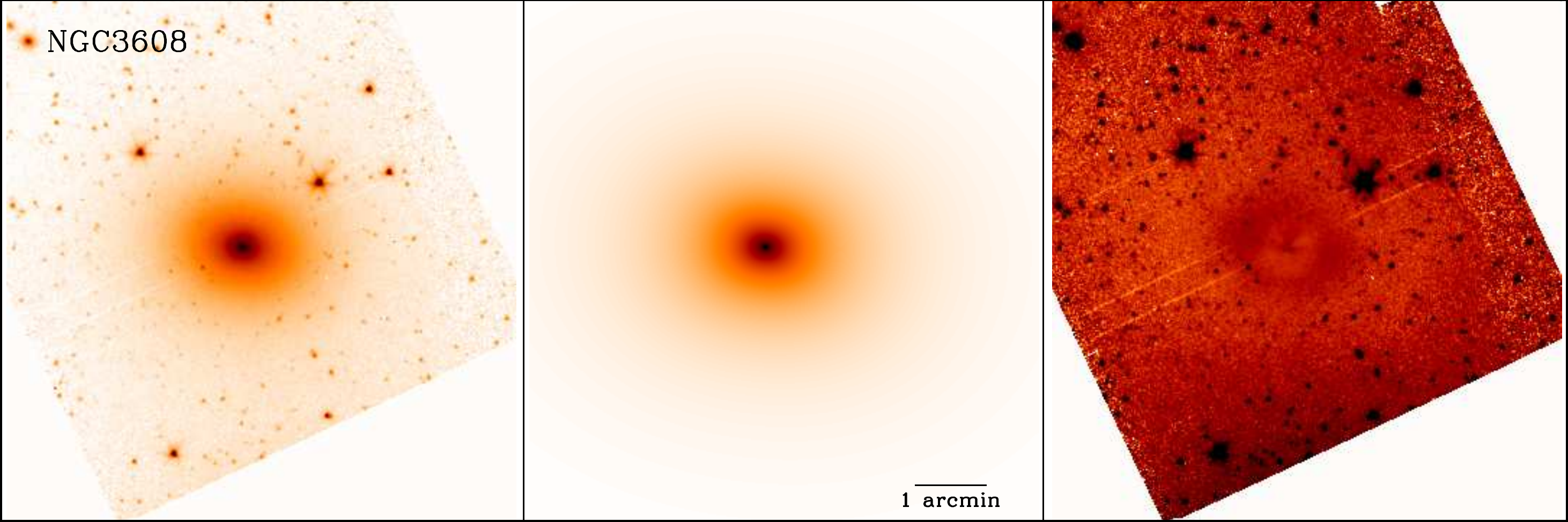}
\includegraphics[height=0.25\linewidth]{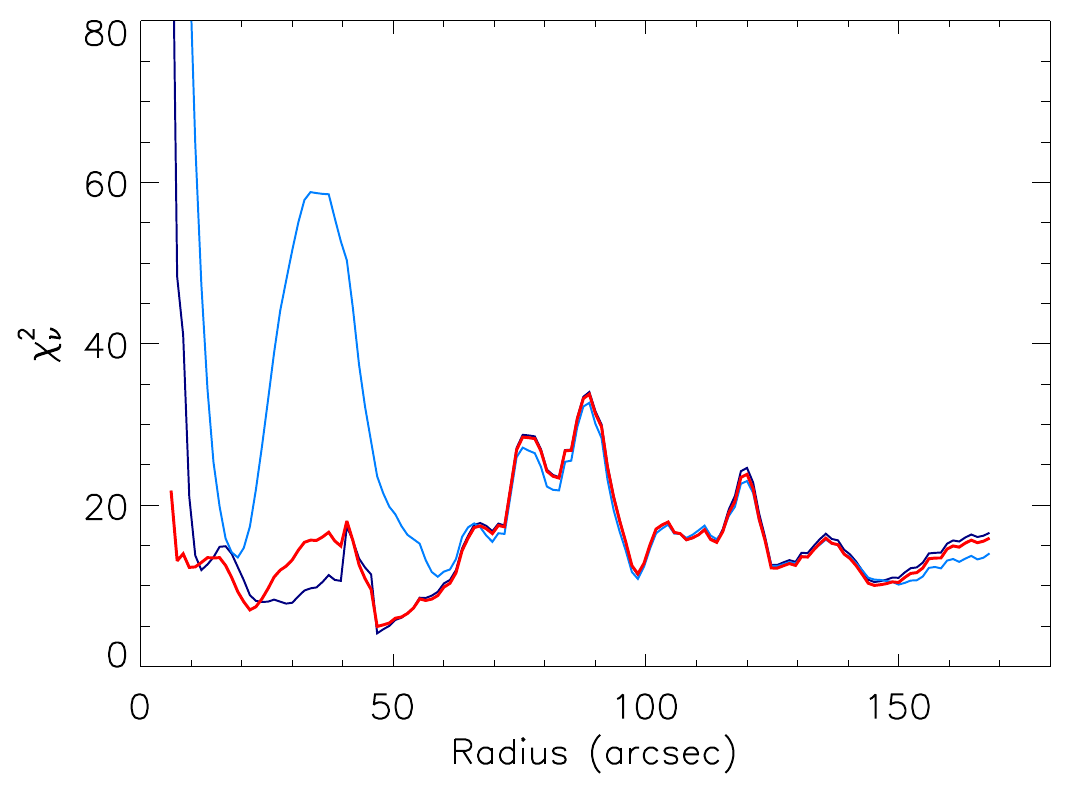}
}
}
\caption{Continued.}
\end{figure*}
\addtocounter{figure}{-1}
\begin{figure*}
\centerline{
\hbox{
\includegraphics[width=0.25\linewidth,angle=90]{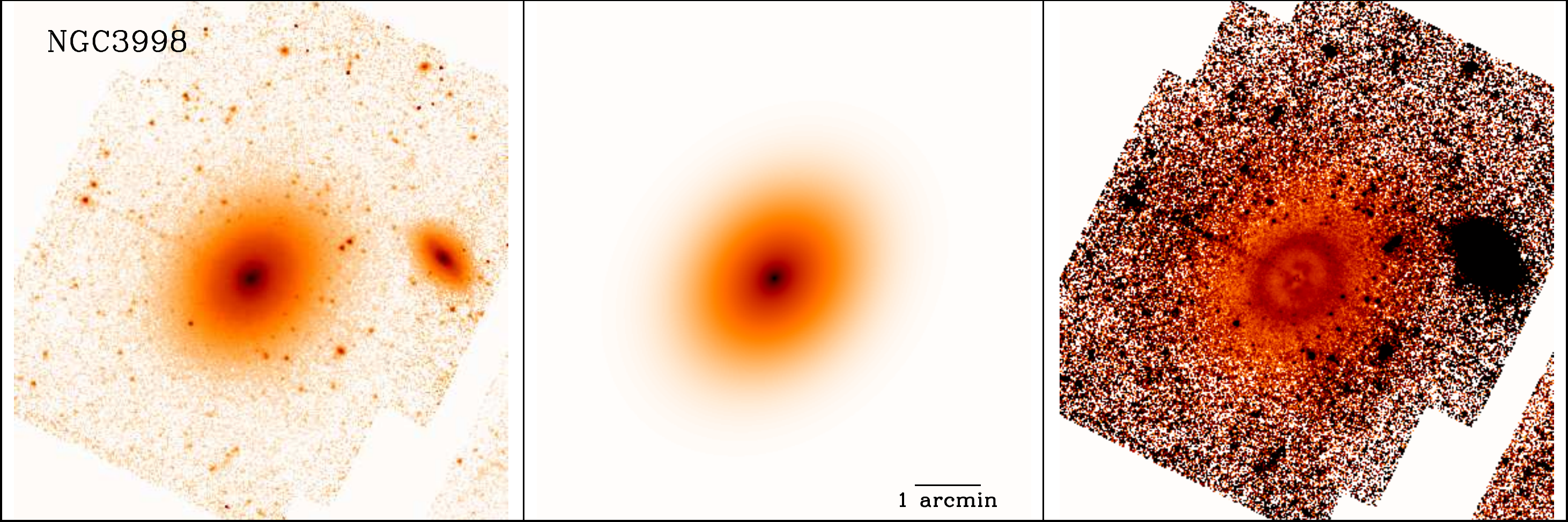}
\includegraphics[height=0.25\linewidth]{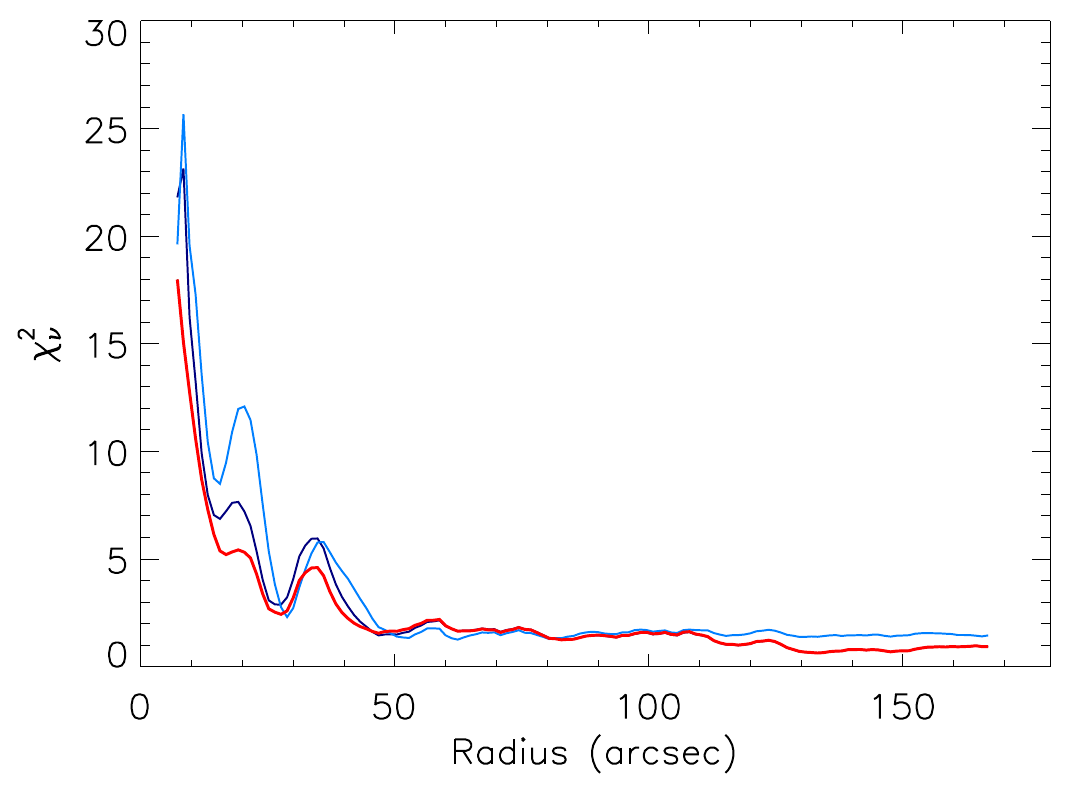}
}
}
\centerline{
\hbox{
\includegraphics[width=0.25\linewidth,angle=90]{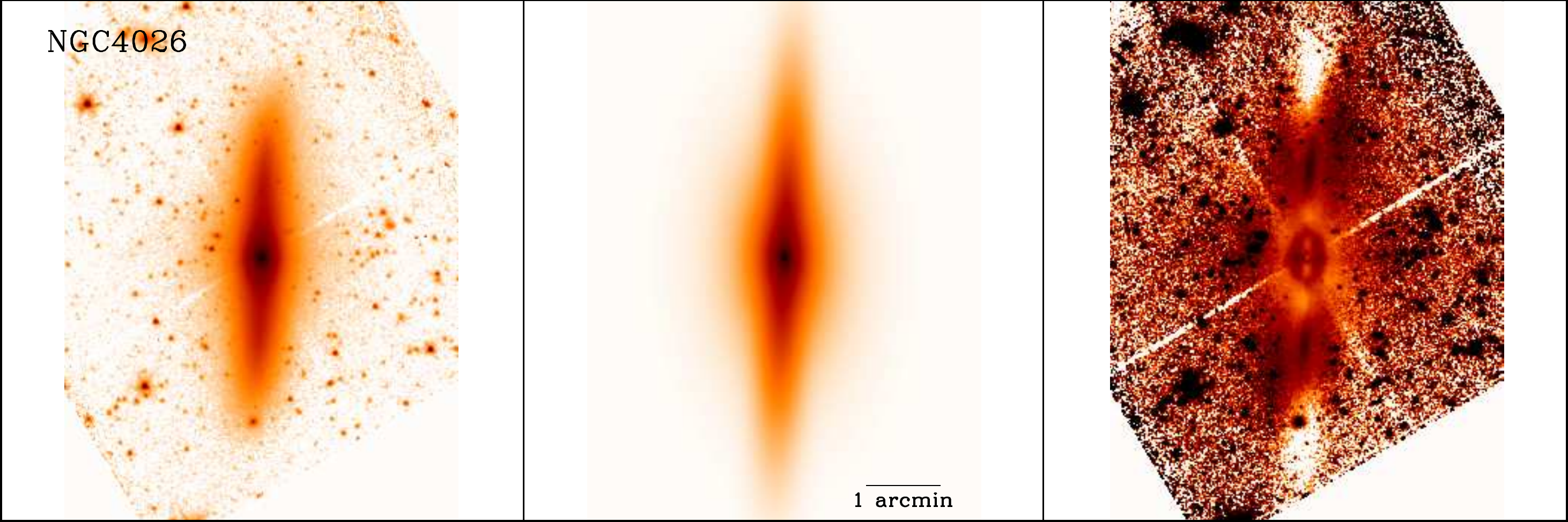}
\includegraphics[height=0.25\linewidth]{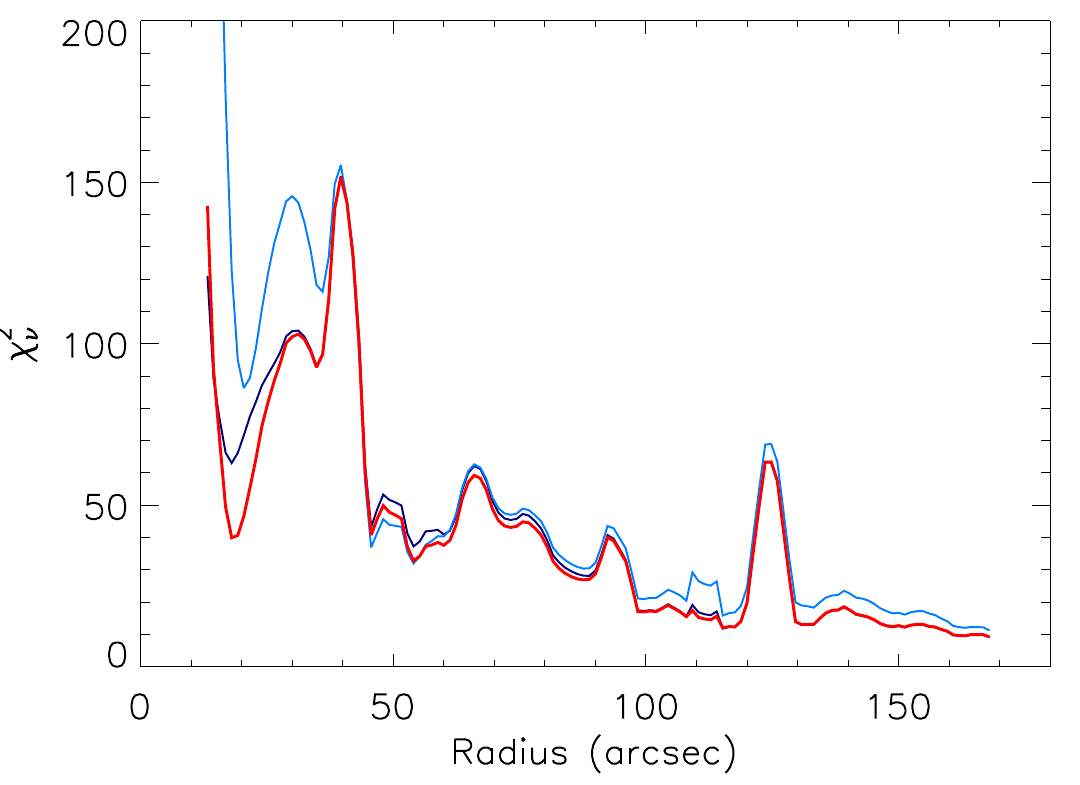}
}
}
\centerline{
\hbox{
\includegraphics[width=0.25\linewidth,angle=90]{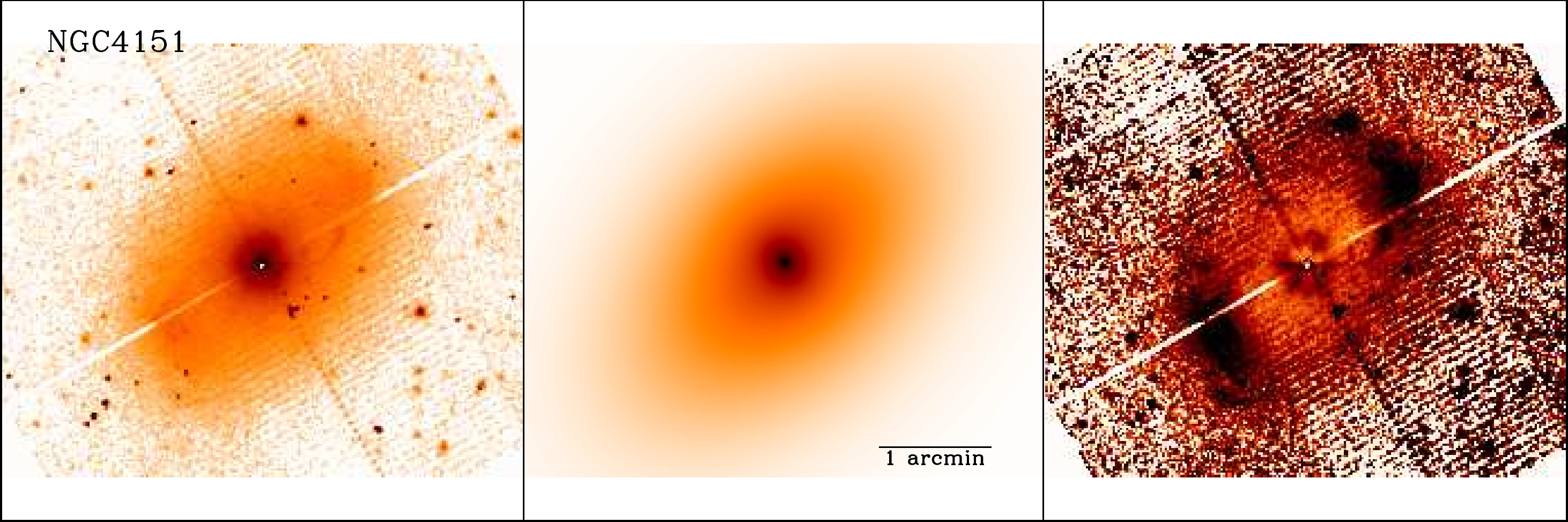}
\includegraphics[height=0.25\linewidth]{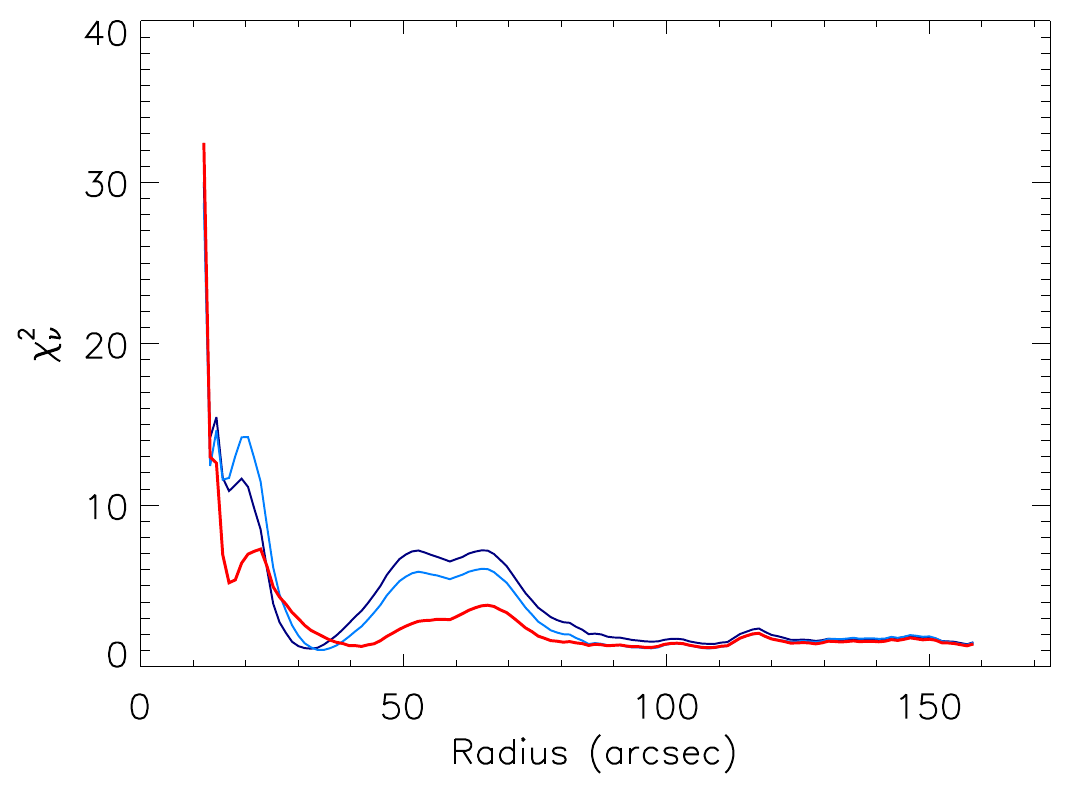}
}
}
\centerline{
\hbox{
\includegraphics[width=0.25\linewidth,angle=90]{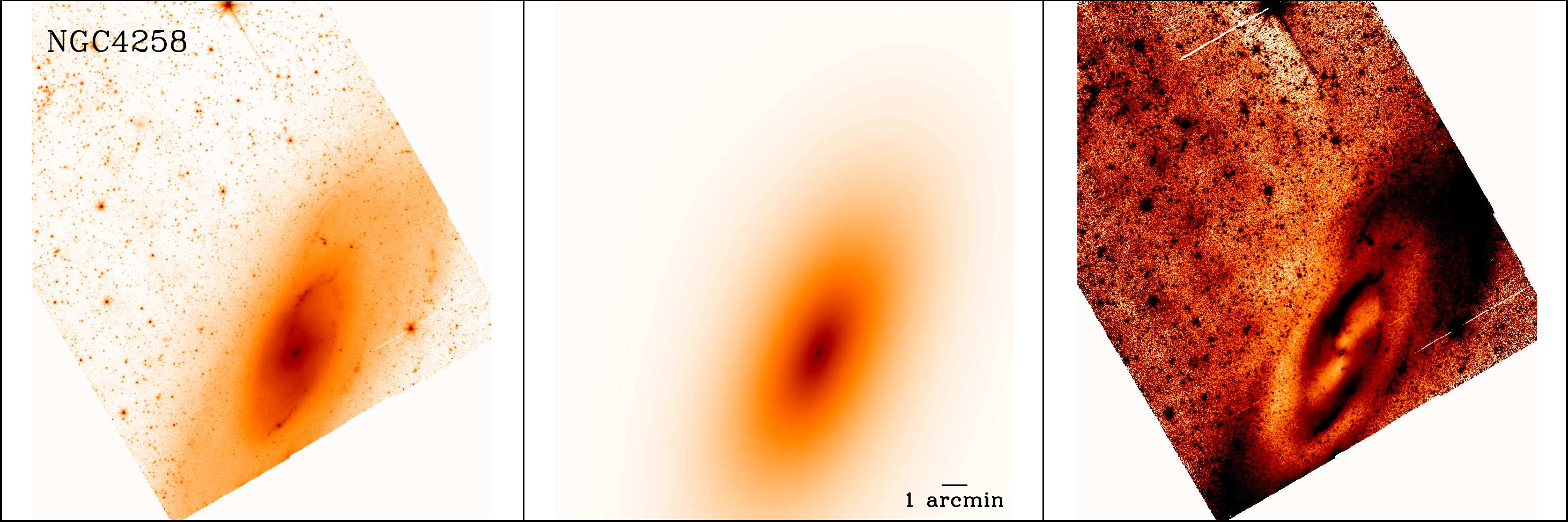}
\includegraphics[height=0.25\linewidth]{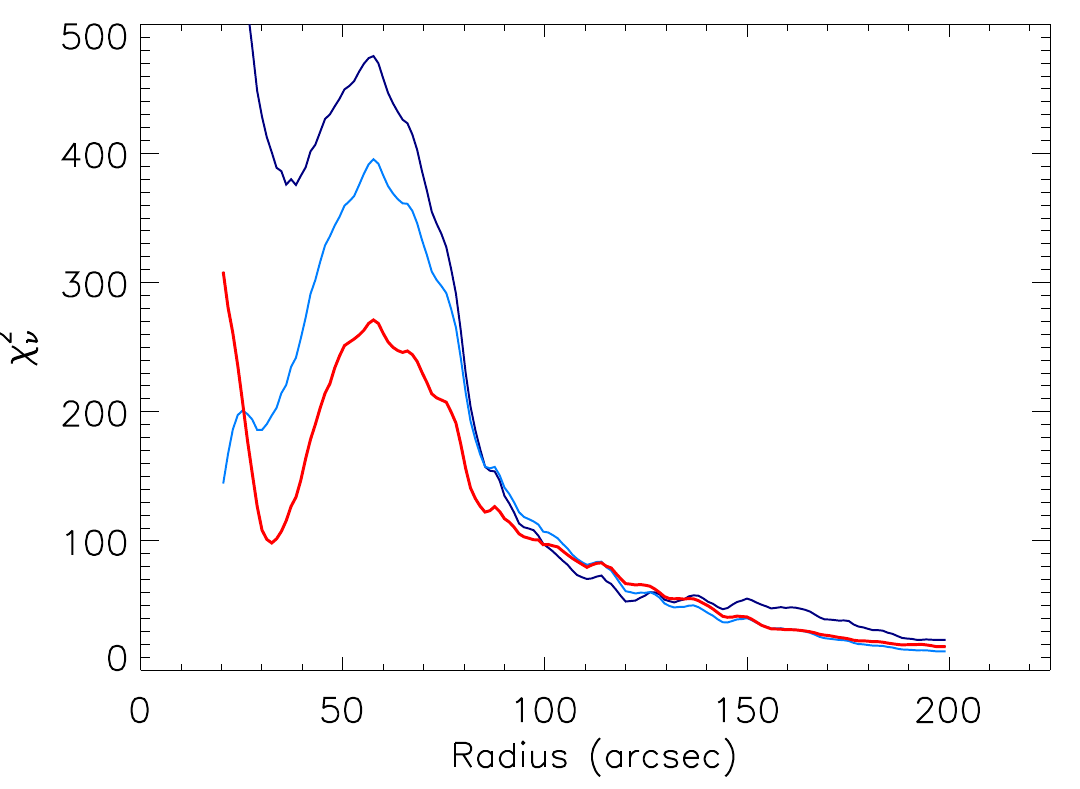}
}
}
\centerline{
\hbox{
\includegraphics[width=0.25\linewidth,angle=90]{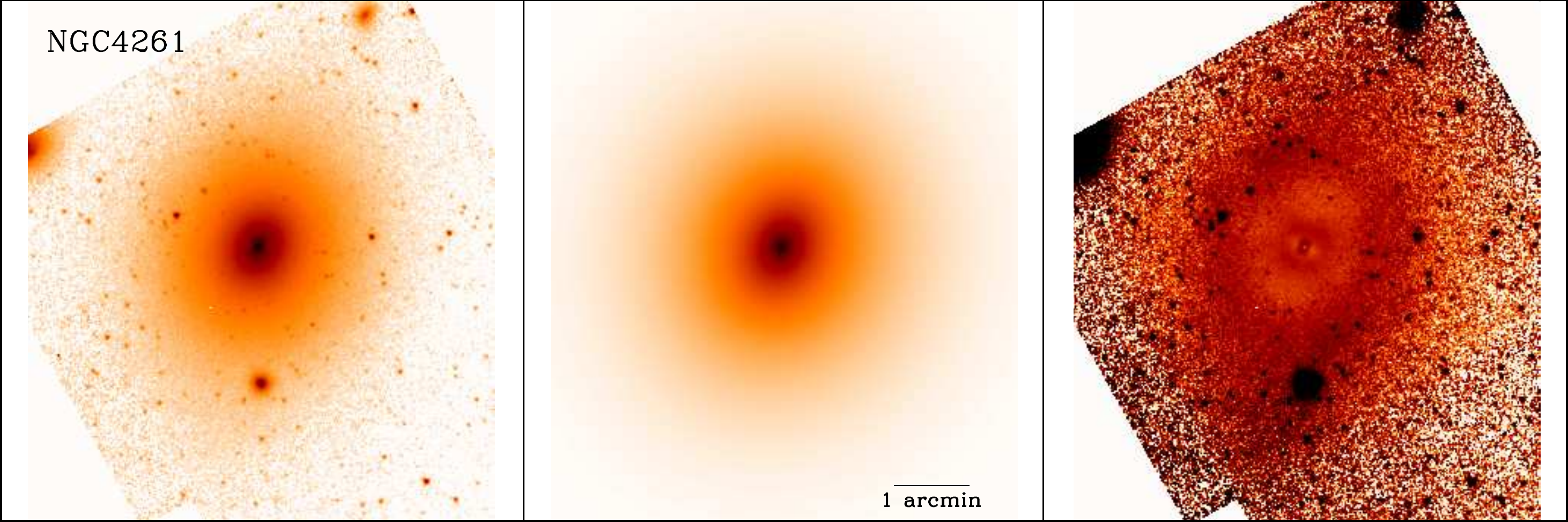}
\includegraphics[height=0.25\linewidth]{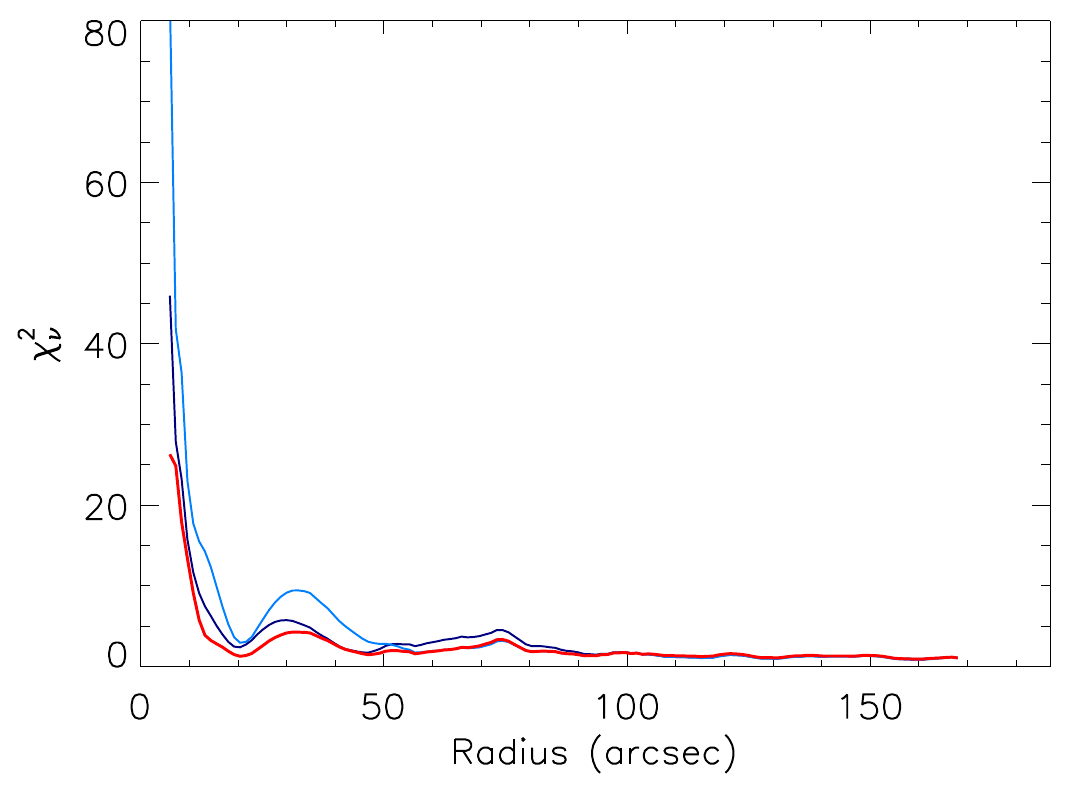}
}
}
\caption{Continued.}
\end{figure*}
\addtocounter{figure}{-1}
\begin{figure*}
\centerline{
\hbox{
\includegraphics[width=0.25\linewidth,angle=90]{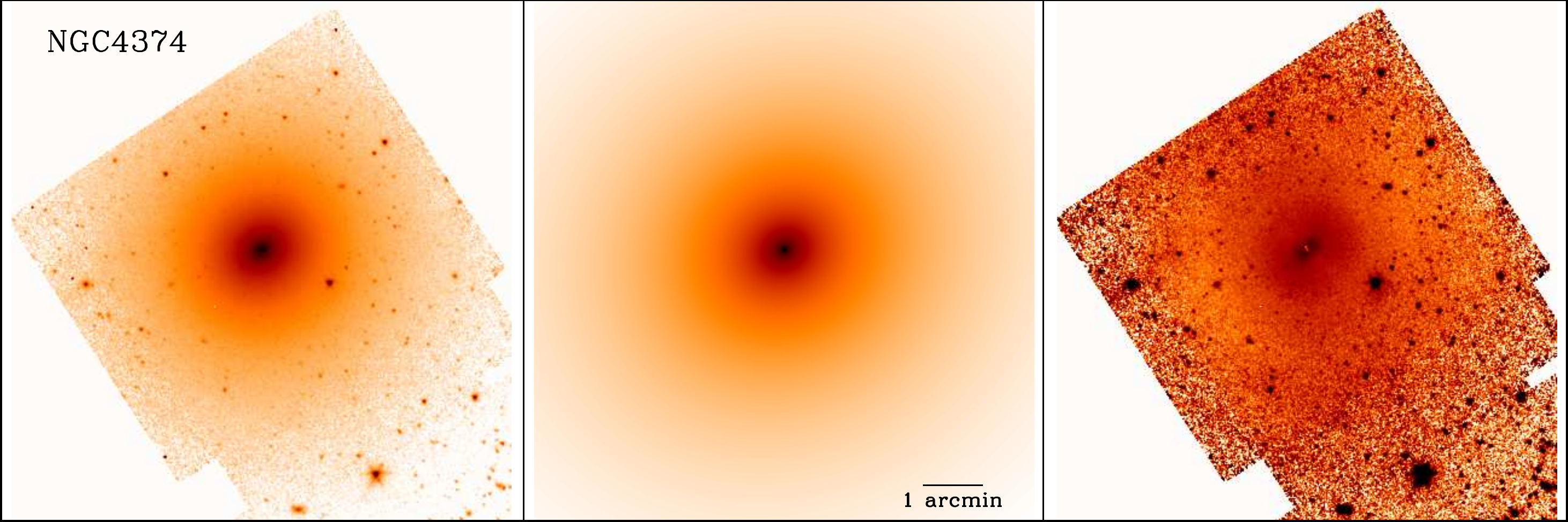}
\includegraphics[height=0.25\linewidth]{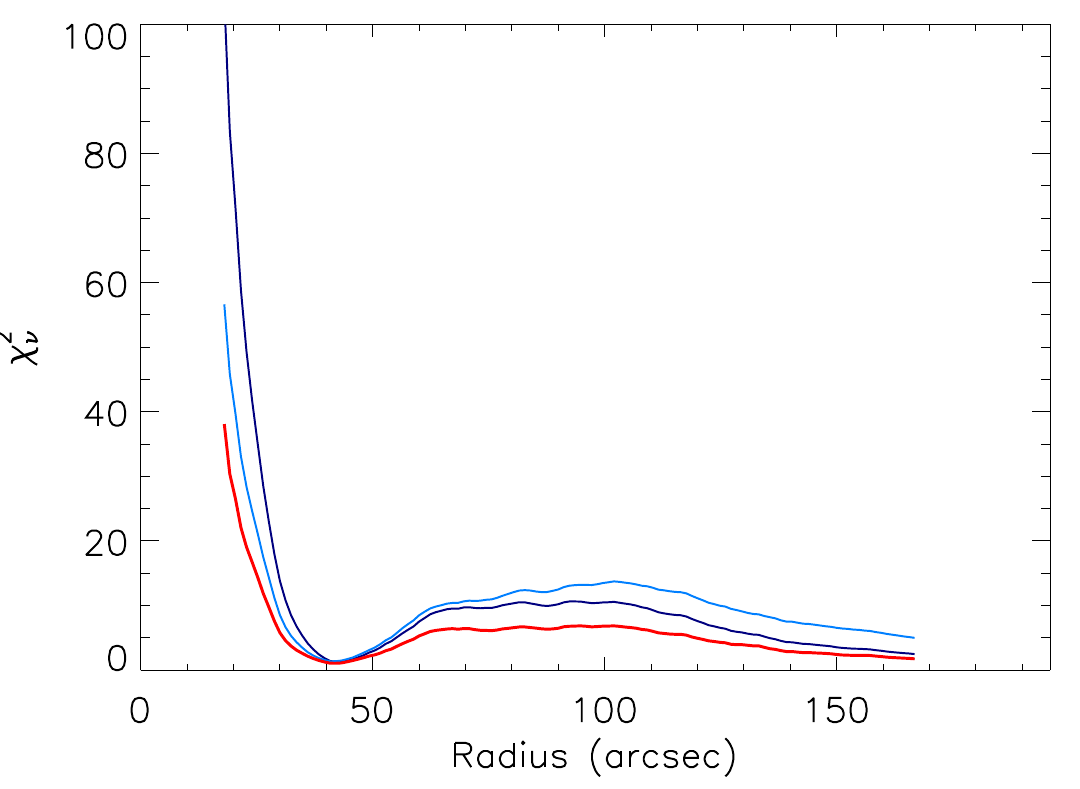}
}
}
\centerline{
\hbox{
\includegraphics[width=0.25\linewidth,angle=90]{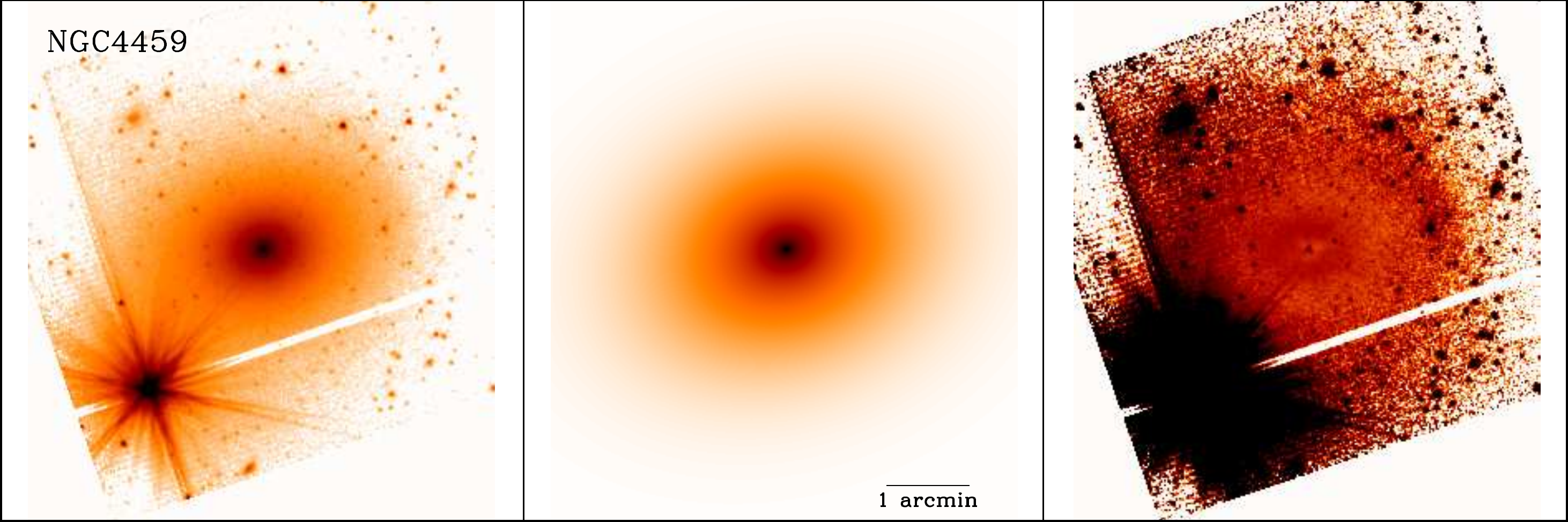}
\includegraphics[height=0.25\linewidth]{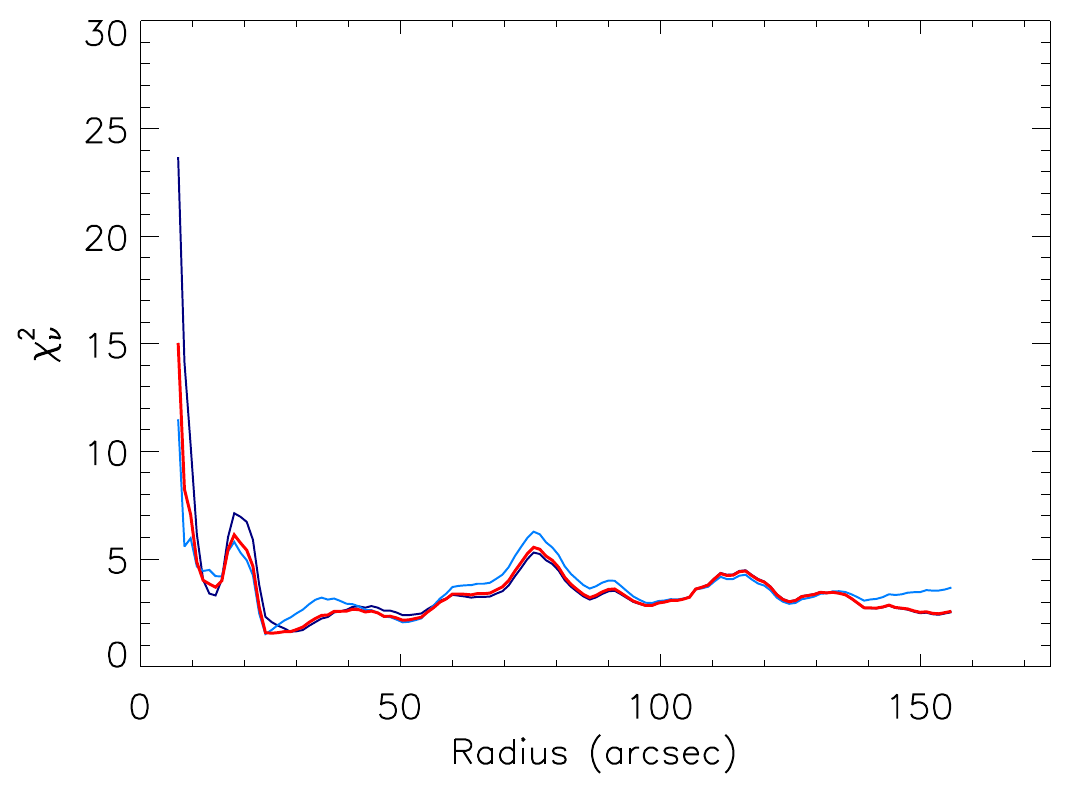}
}
}
\centerline{
\hbox{
\includegraphics[width=0.25\linewidth,angle=90]{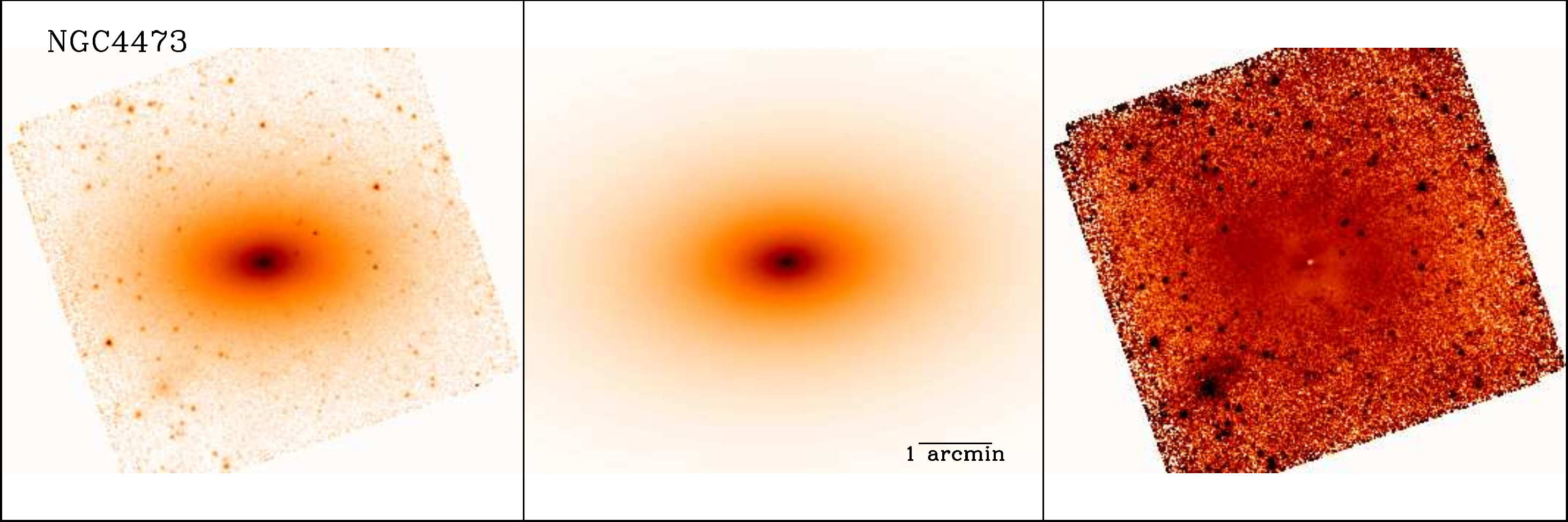}
\includegraphics[height=0.25\linewidth]{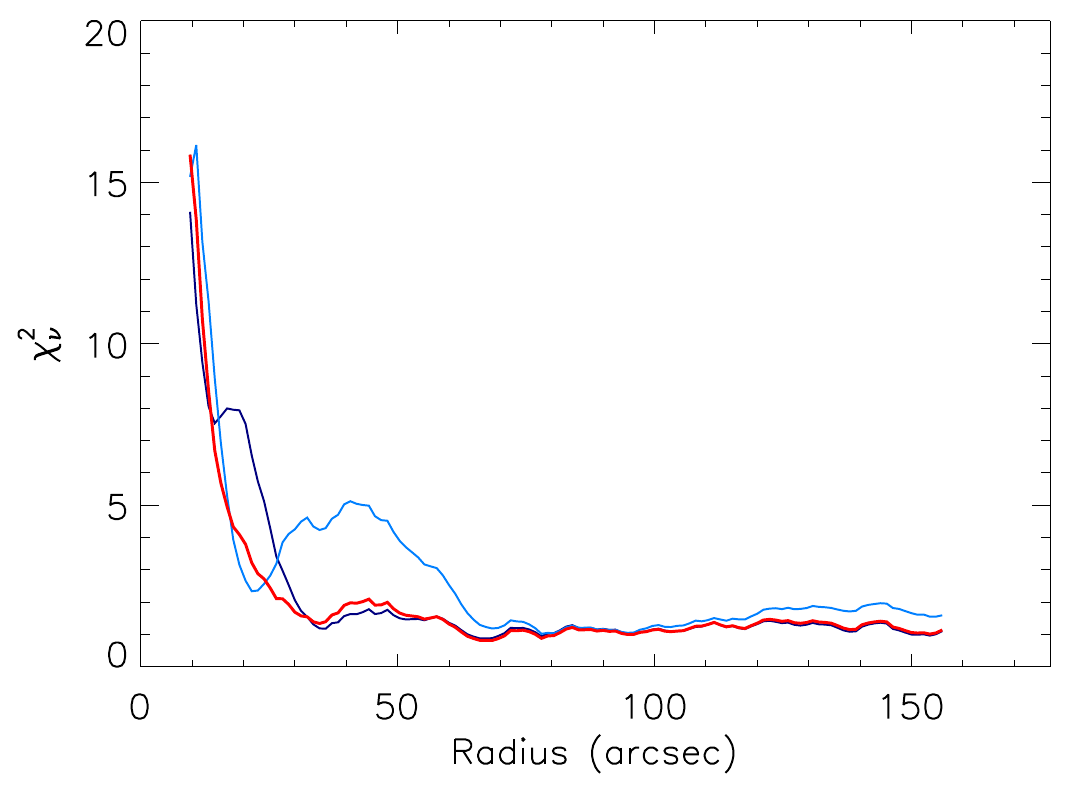}
}
}
\centerline{
\hbox{
\includegraphics[width=0.25\linewidth,angle=90]{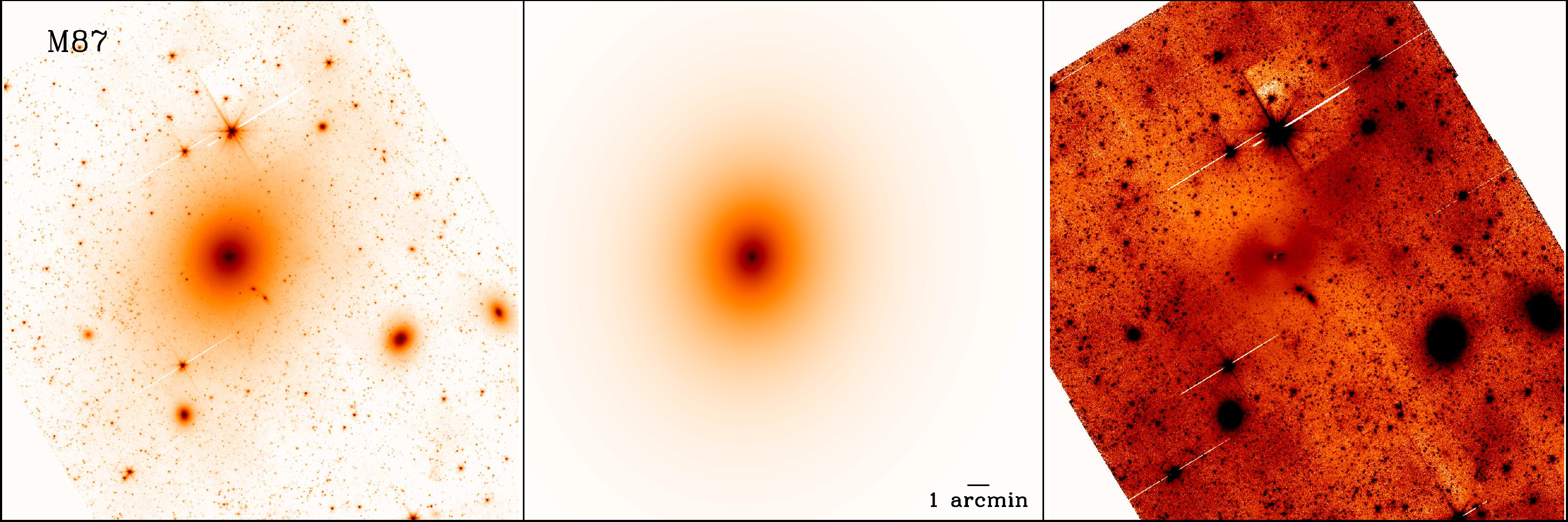}
\includegraphics[height=0.25\linewidth]{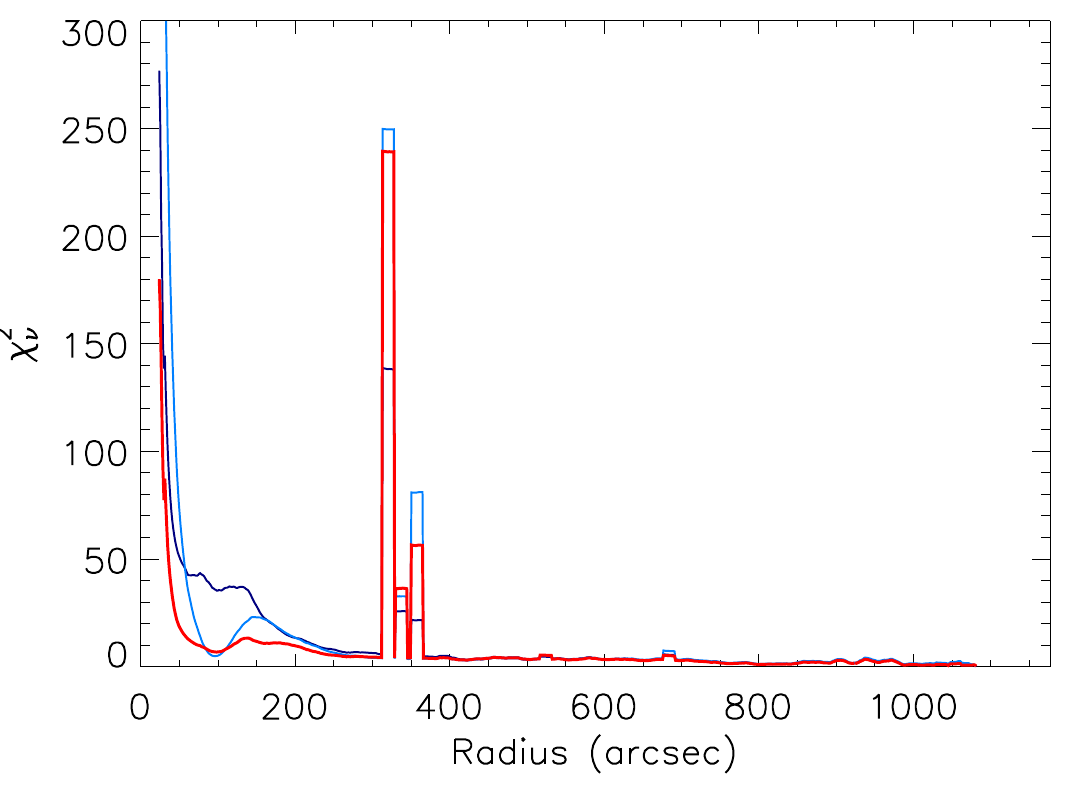}
}
}
\centerline{
\hbox{
\includegraphics[width=0.25\linewidth,angle=90]{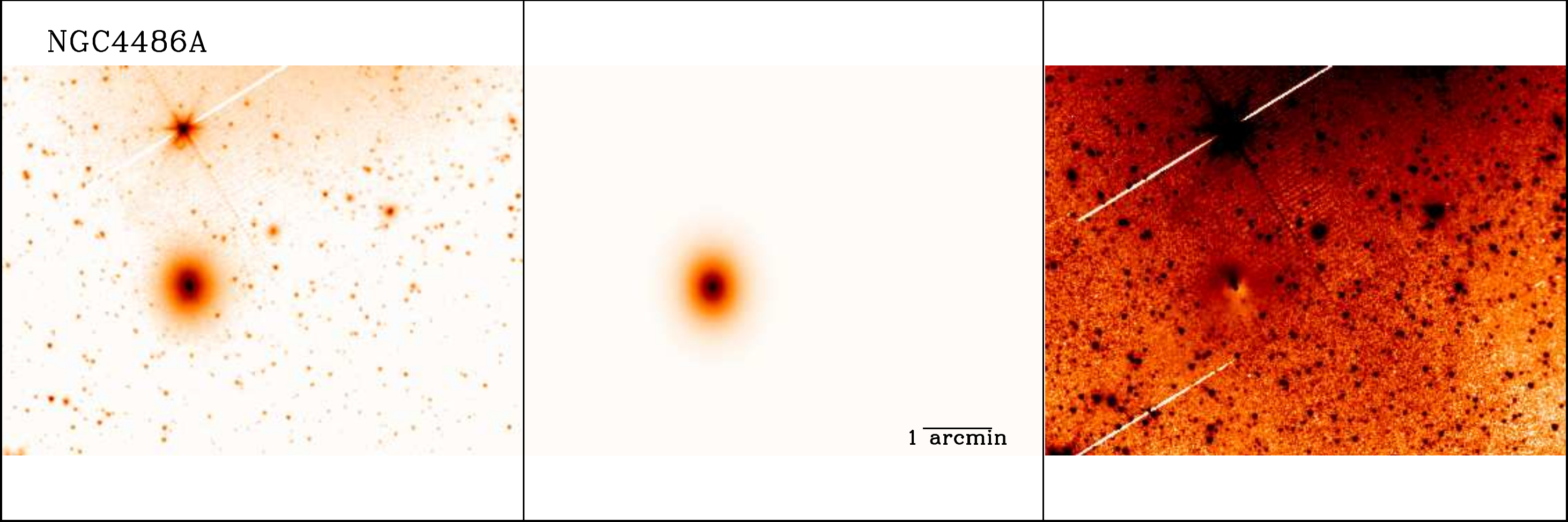}
\includegraphics[height=0.25\linewidth]{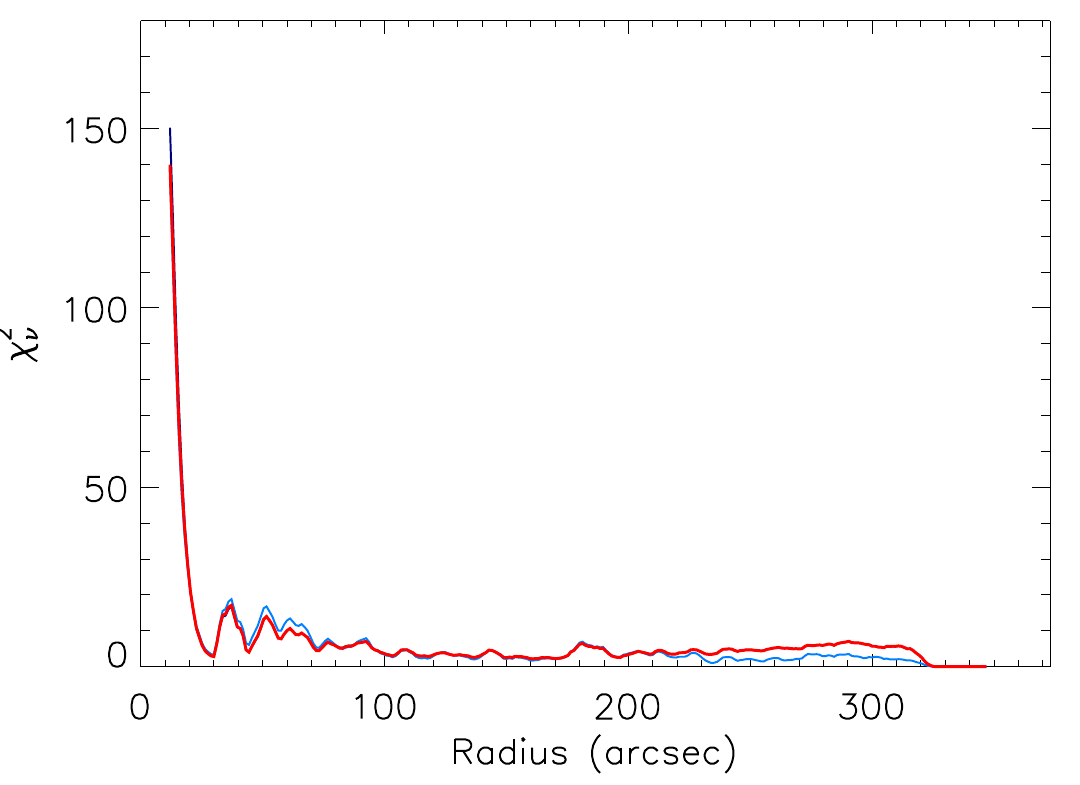}
}
}
\caption{Continued.}
\end{figure*}
\begin{figure*}
\centerline{
\hbox{
\includegraphics[width=0.25\linewidth,angle=90]{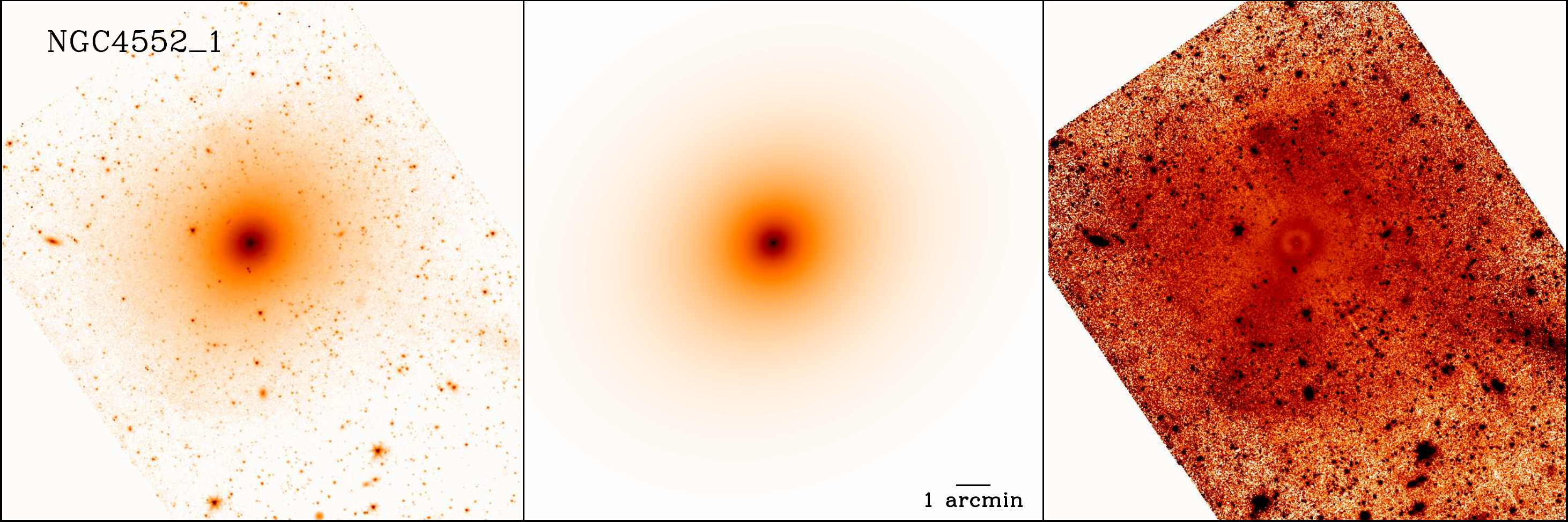}
\includegraphics[height=0.25\linewidth]{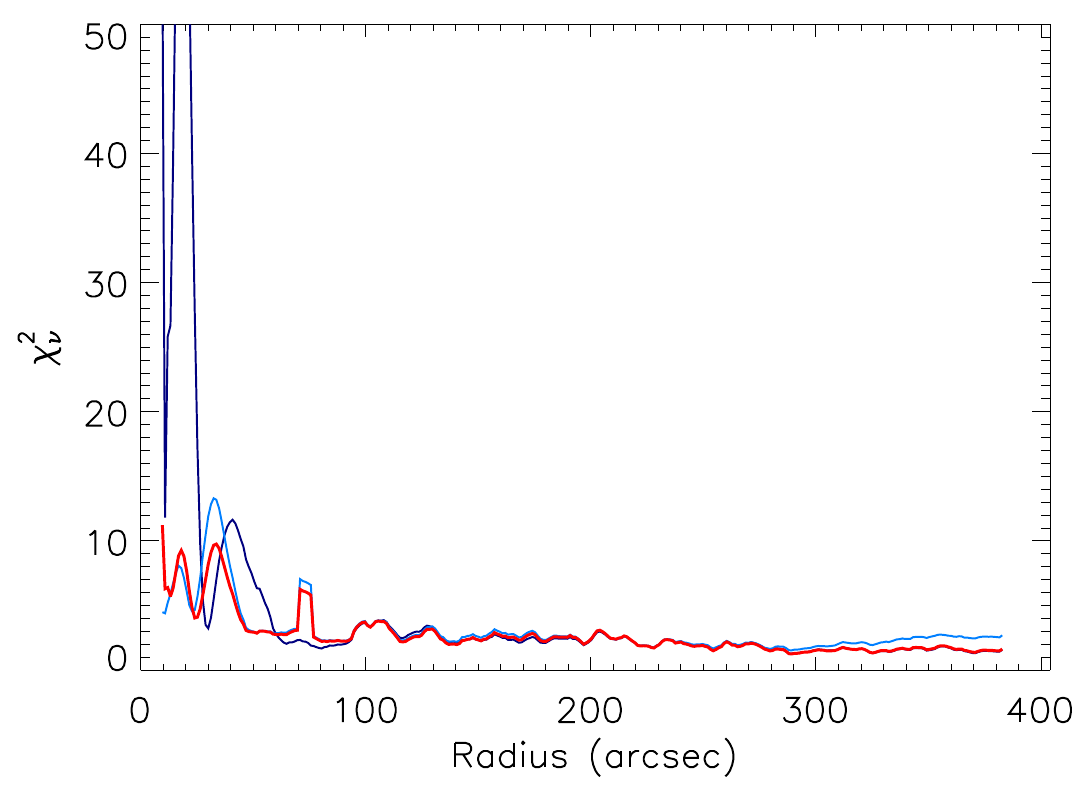}
}
}
\centerline{
\hbox{
\includegraphics[width=0.25\linewidth,angle=90]{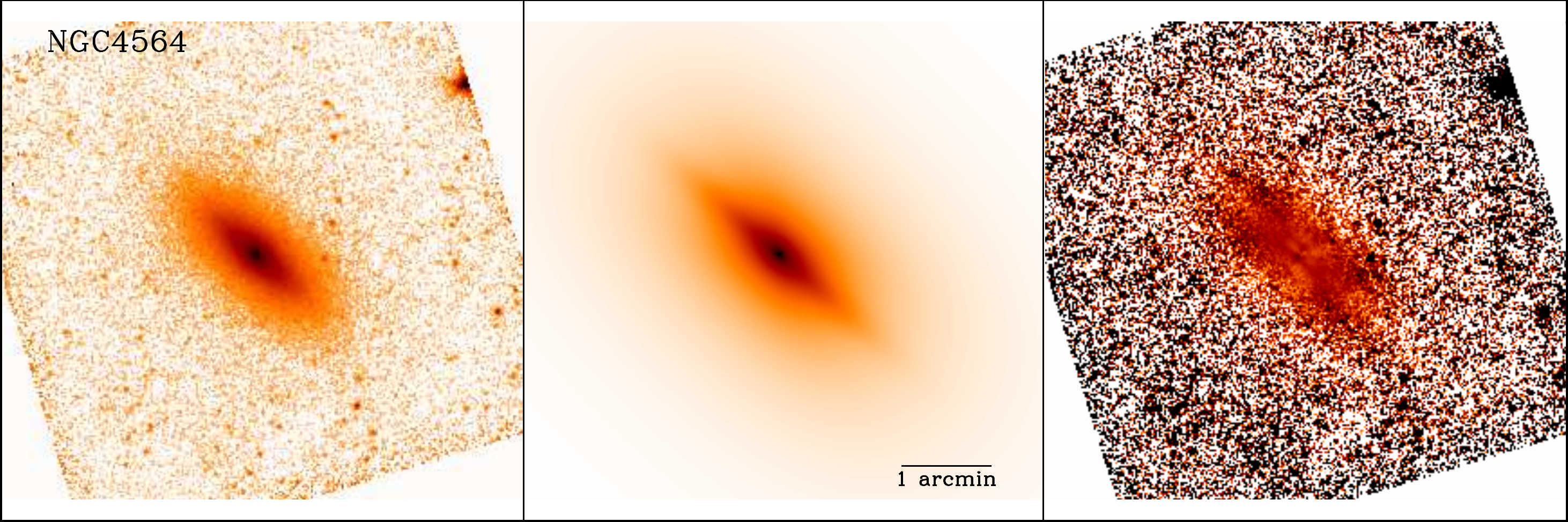}
\includegraphics[height=0.25\linewidth]{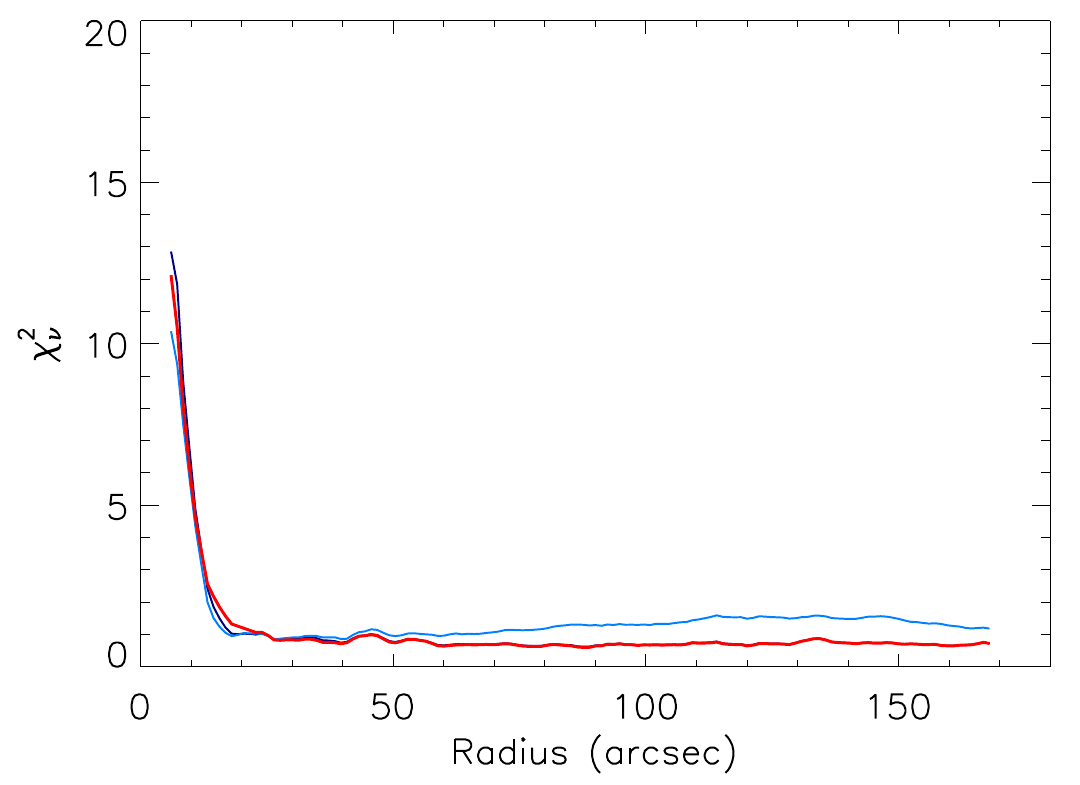}
}
}
\centerline{
\hbox{
\includegraphics[width=0.25\linewidth,angle=90]{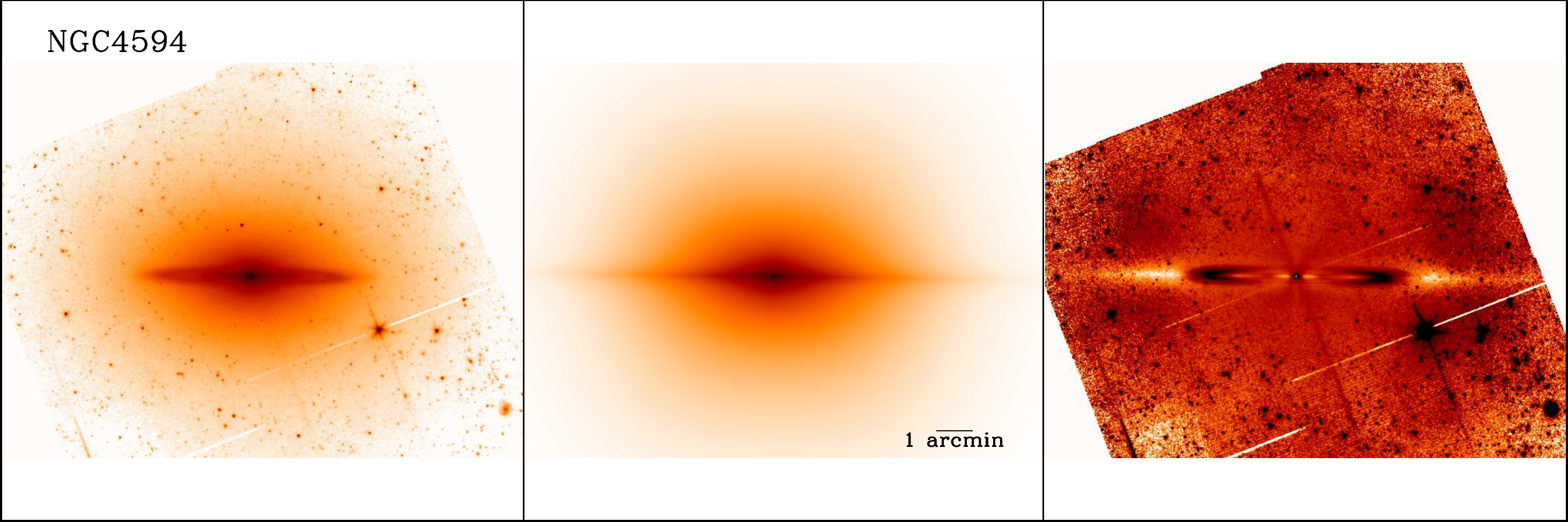}
\includegraphics[height=0.25\linewidth]{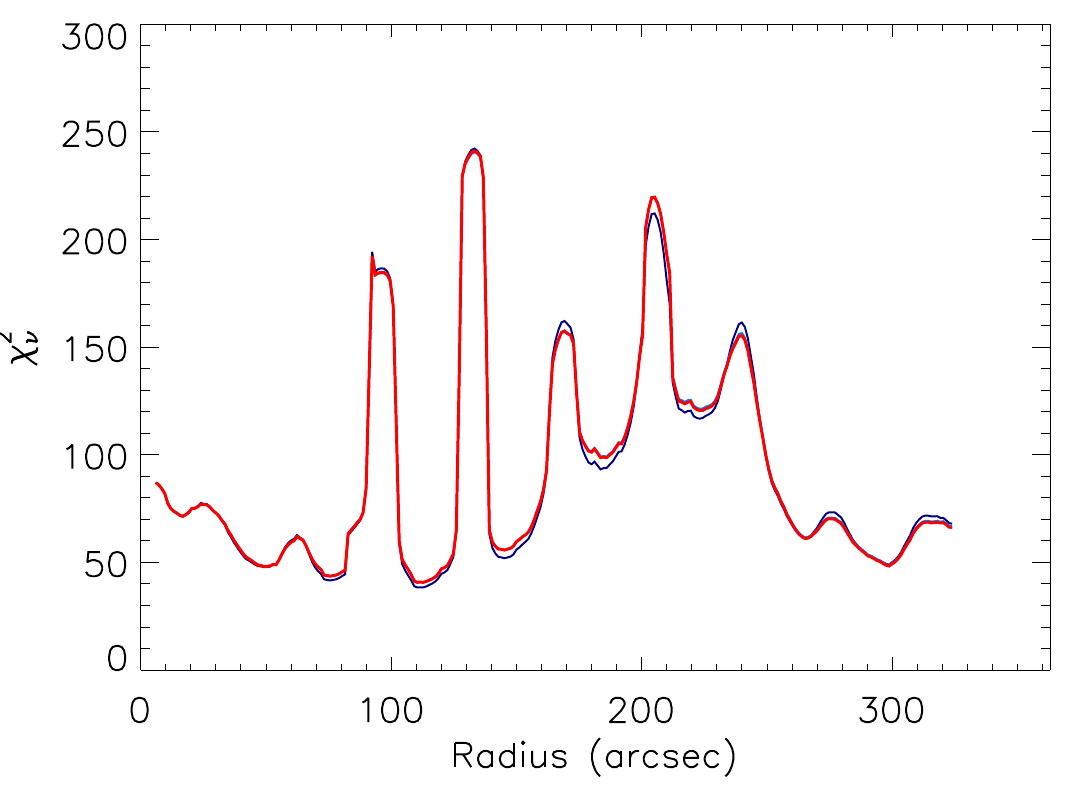}
}
}
\centerline{
\hbox{
\includegraphics[width=0.25\linewidth,angle=90]{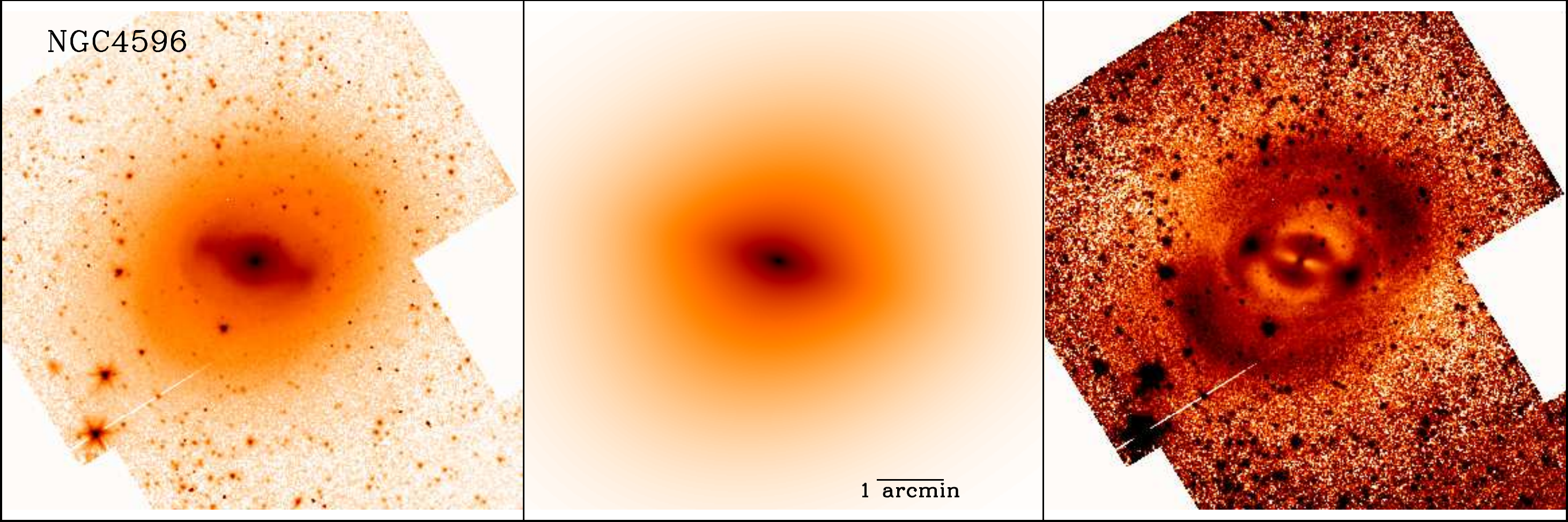}
\includegraphics[height=0.25\linewidth]{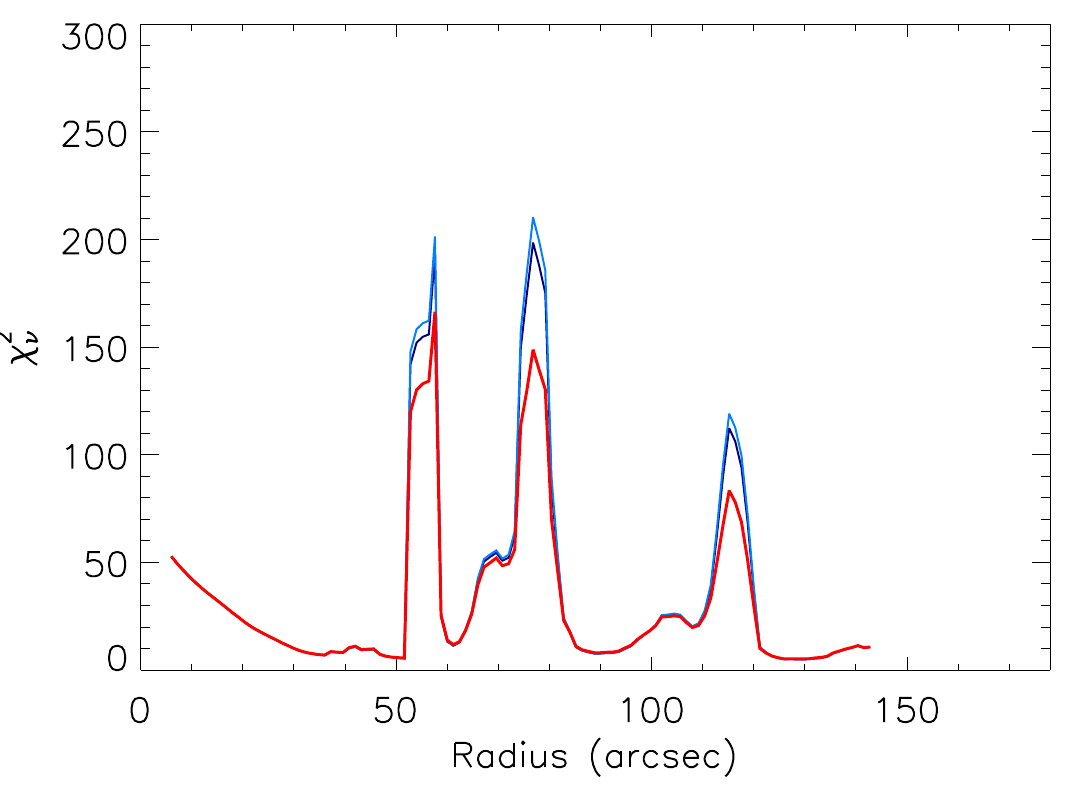}
}
}
\centerline{
\hbox{
\includegraphics[width=0.25\linewidth,angle=90]{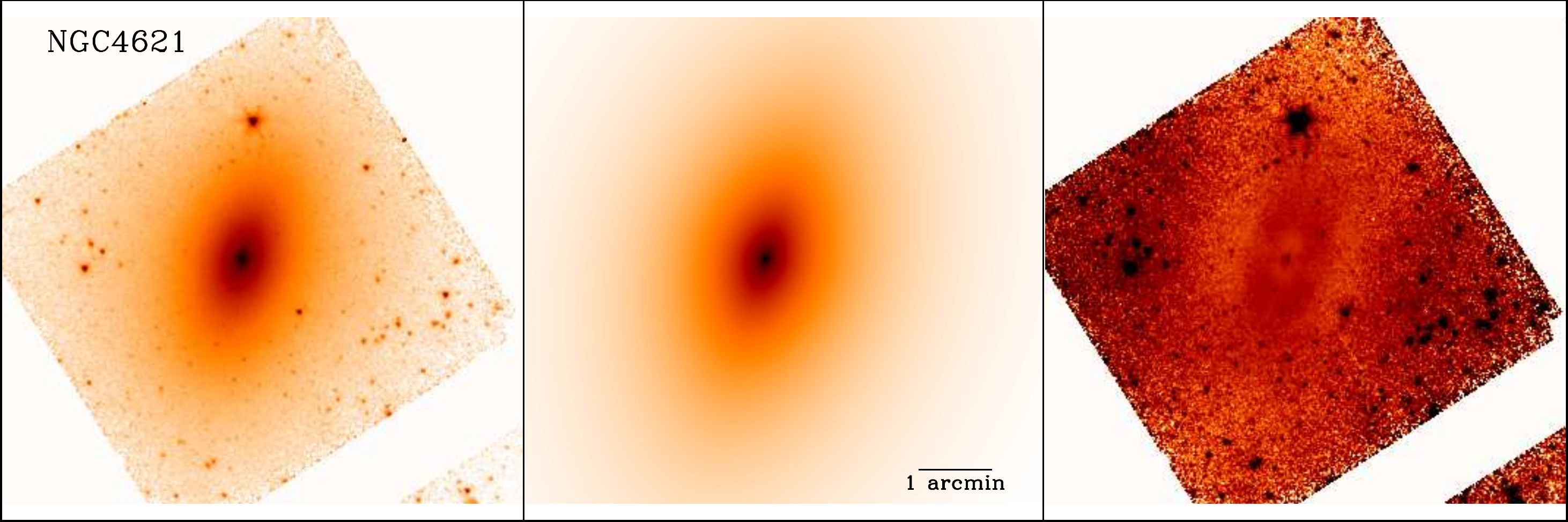}
\includegraphics[height=0.25\linewidth]{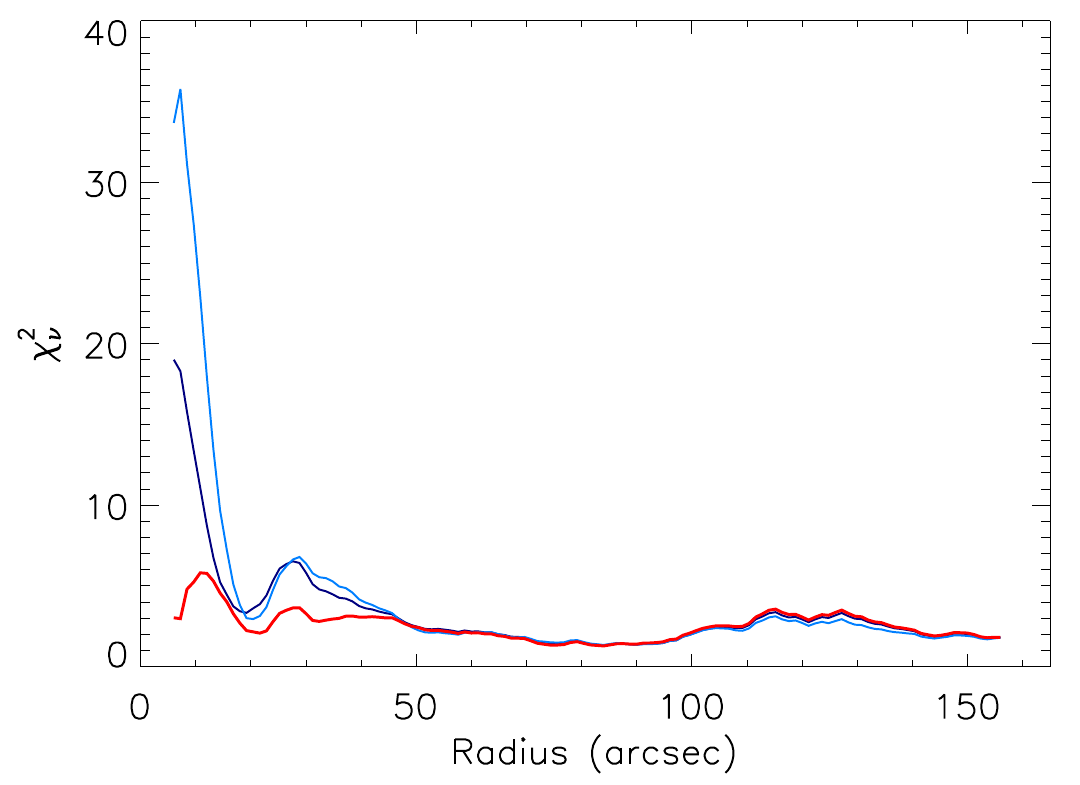}
}
}
\caption{Continued.}
\end{figure*}
\begin{figure*}
\centerline{
\hbox{
\includegraphics[width=0.25\linewidth,angle=90]{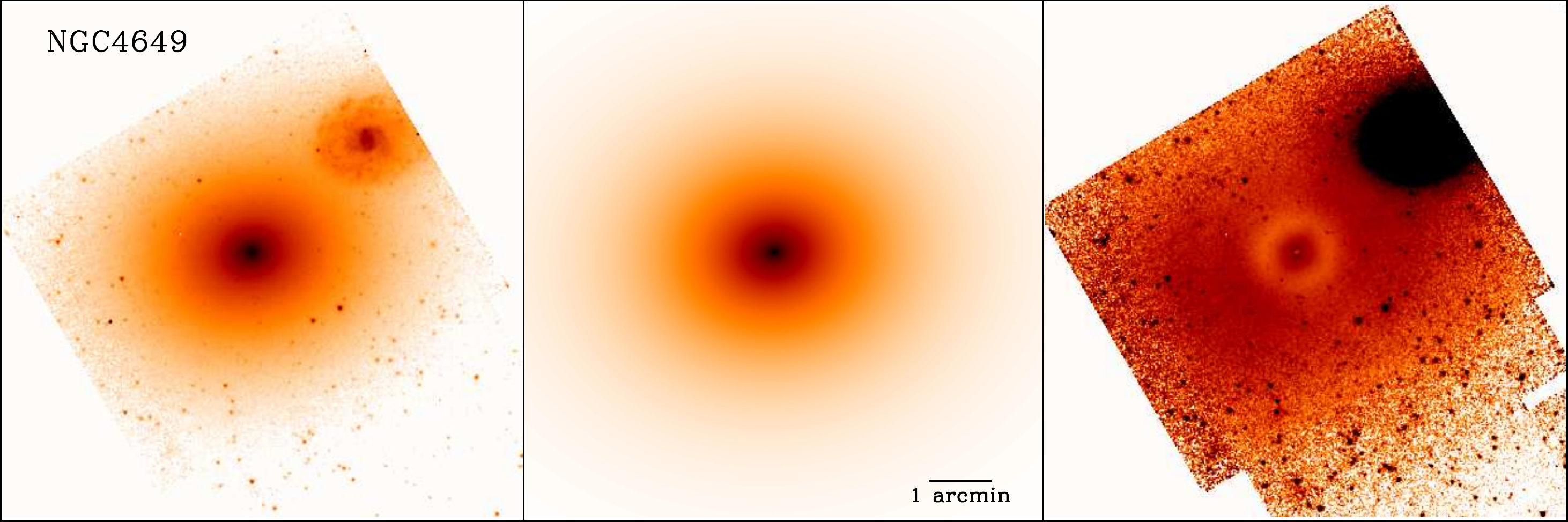}
\includegraphics[height=0.25\linewidth]{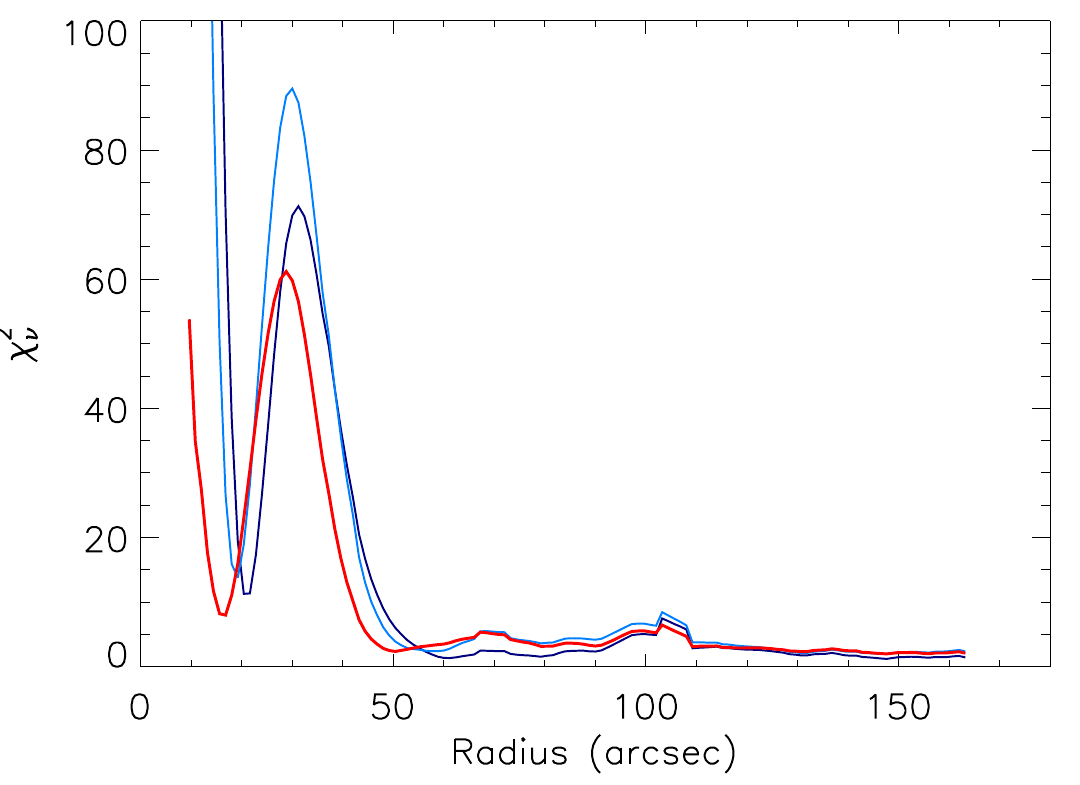}
}
}
\centerline{
\hbox{
\includegraphics[width=0.25\linewidth,angle=90]{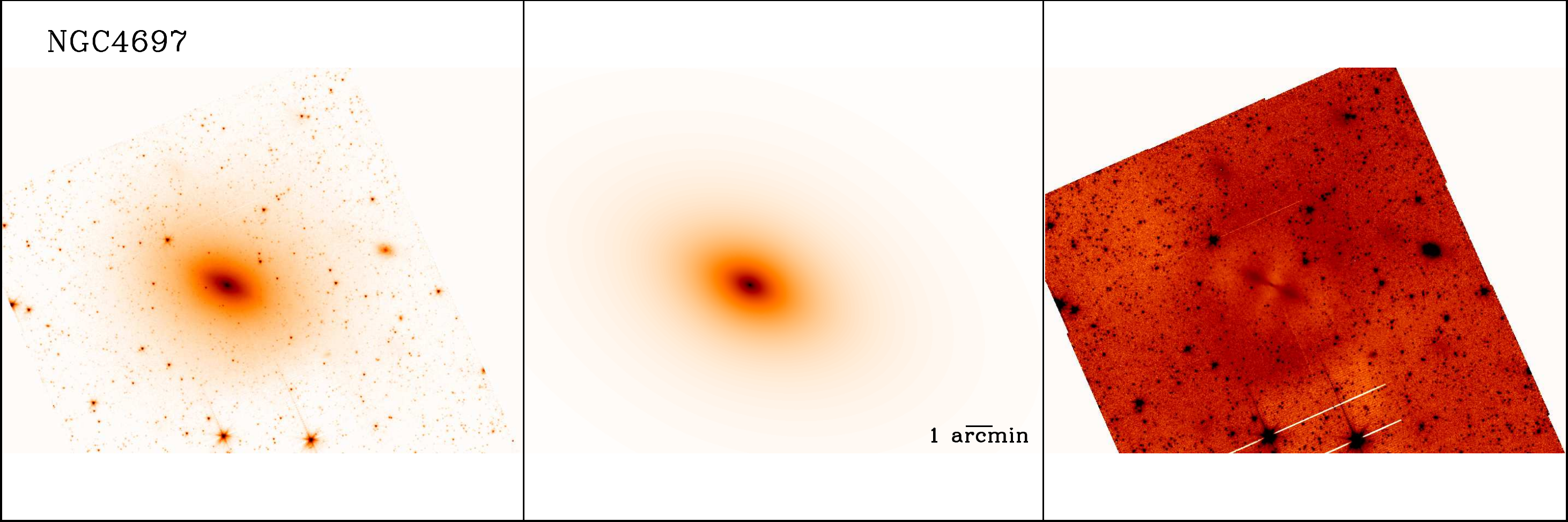}
\includegraphics[height=0.25\linewidth]{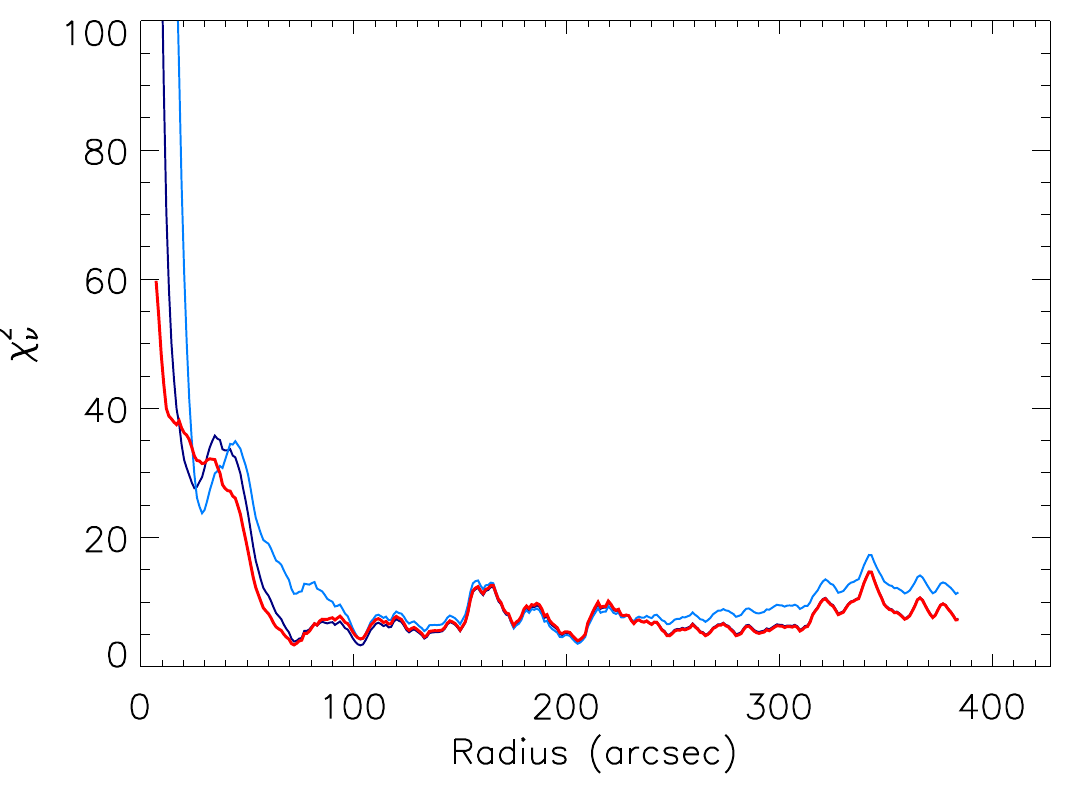}
}
}
\centerline{
\hbox{
\includegraphics[width=0.25\linewidth,angle=90]{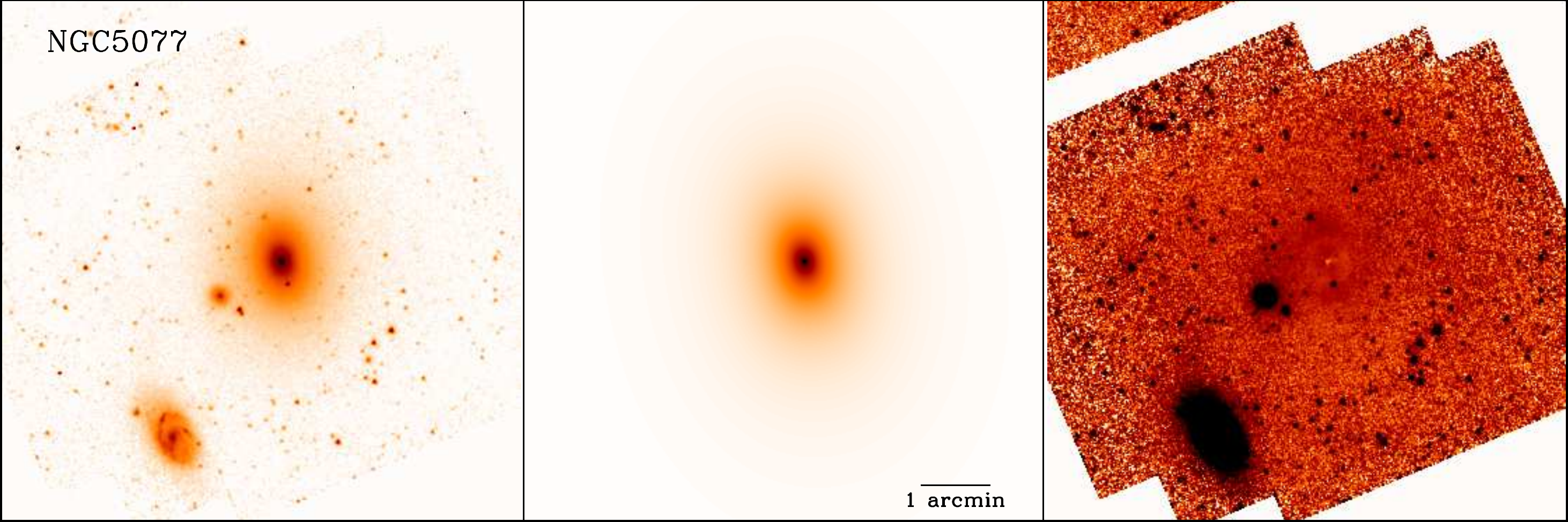}
\includegraphics[height=0.25\linewidth]{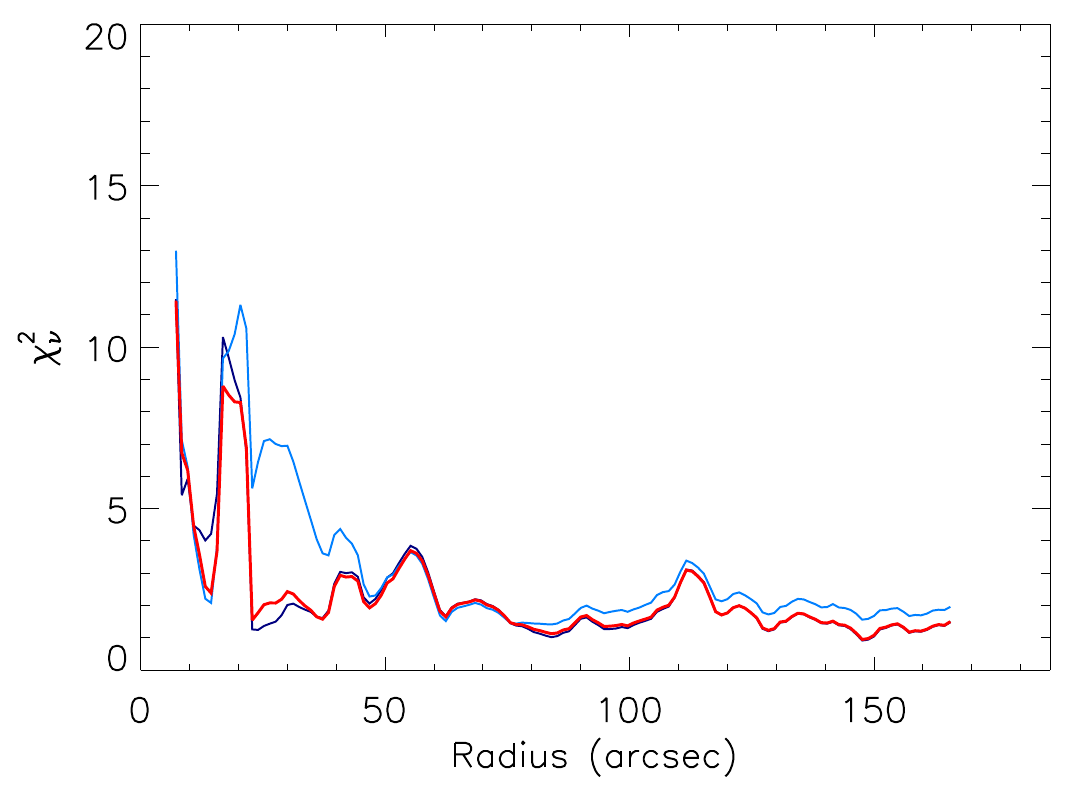}
}
}
\centerline{
\hbox{
\includegraphics[width=0.25\linewidth,angle=90]{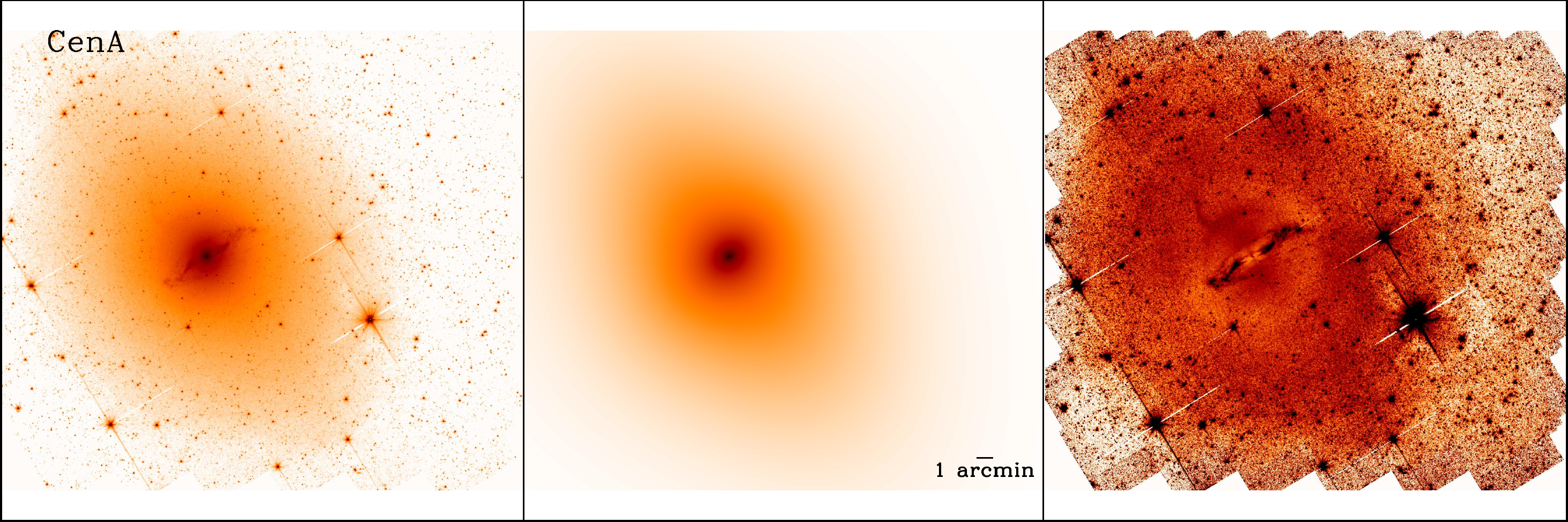}
\includegraphics[height=0.25\linewidth]{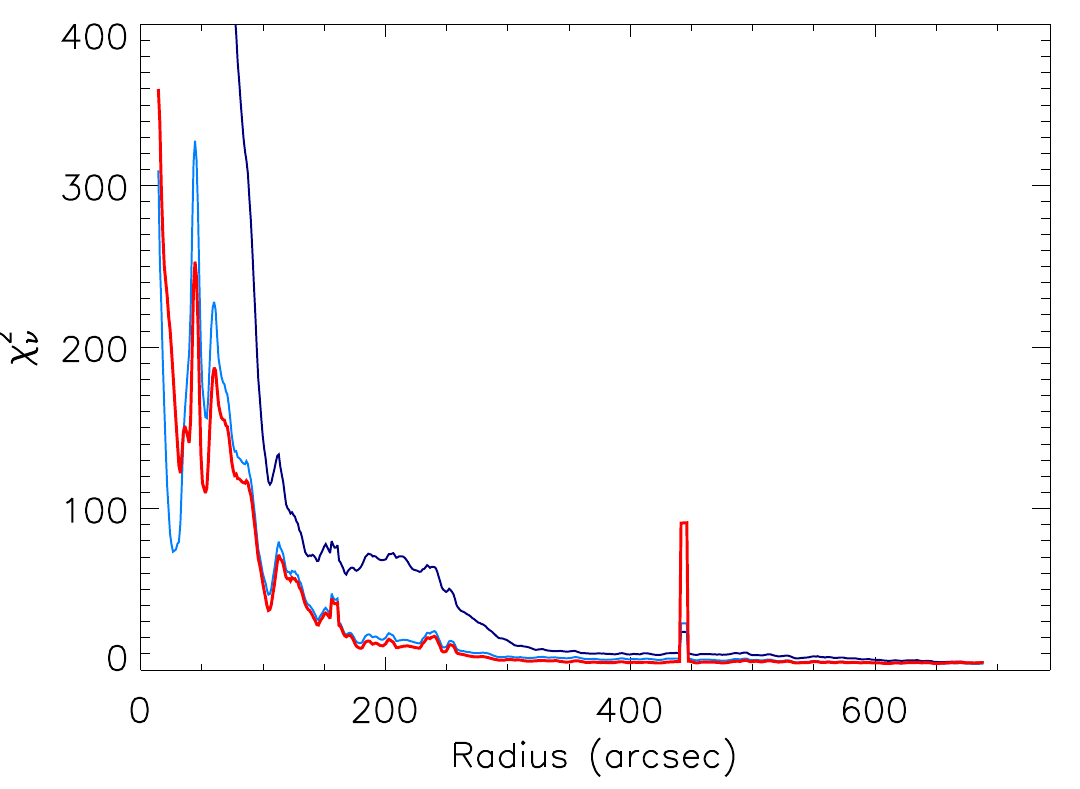}
}
}
\centerline{
\hbox{
\includegraphics[width=0.25\linewidth,angle=90]{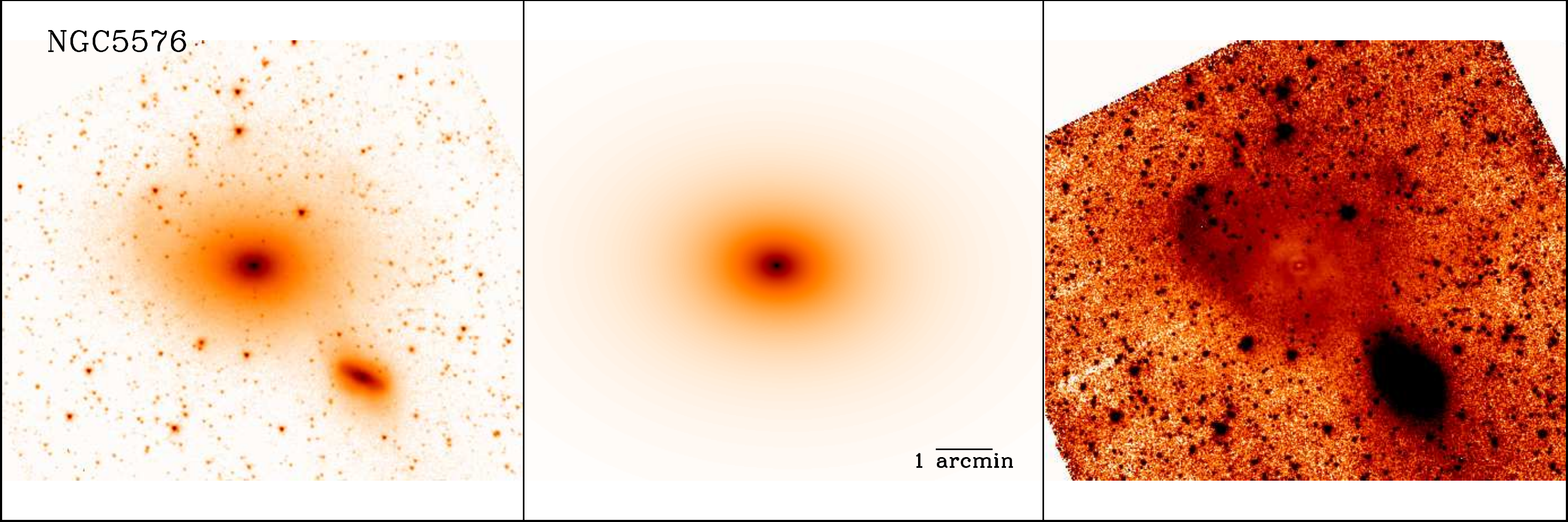}
\includegraphics[height=0.25\linewidth]{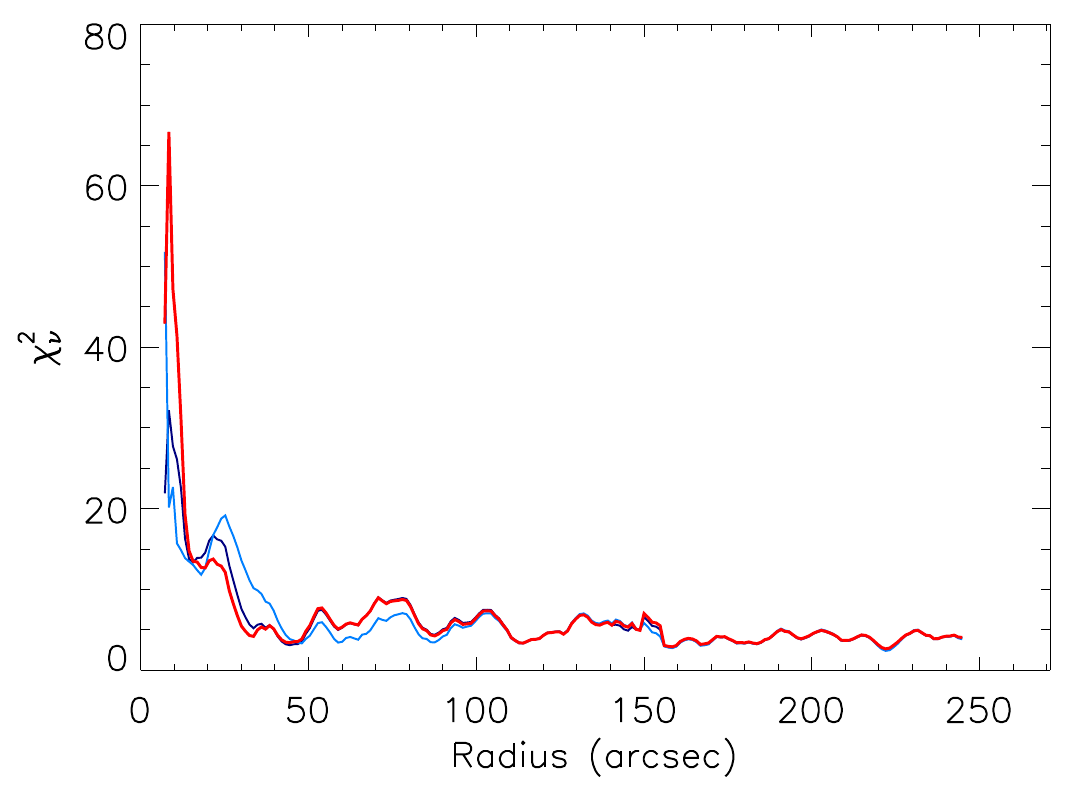}
}
}
\caption{Continued.}
\end{figure*}
\begin{figure*}
\centerline{
\hbox{
\includegraphics[width=0.25\linewidth,angle=90]{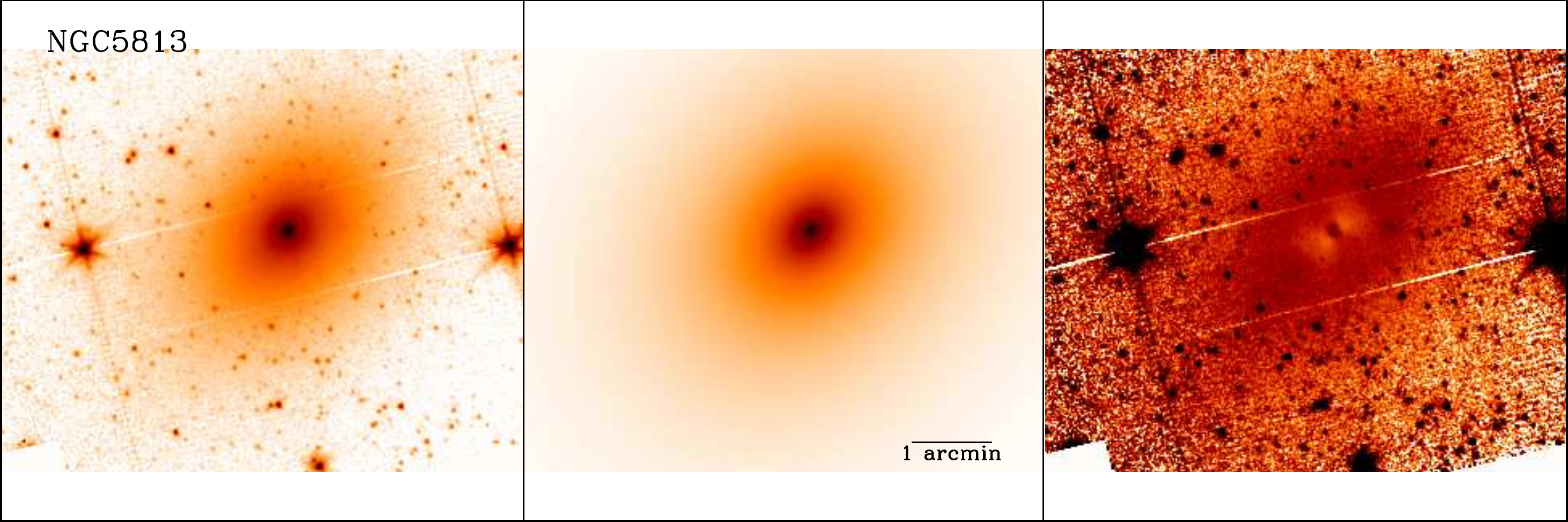}
\includegraphics[height=0.25\linewidth]{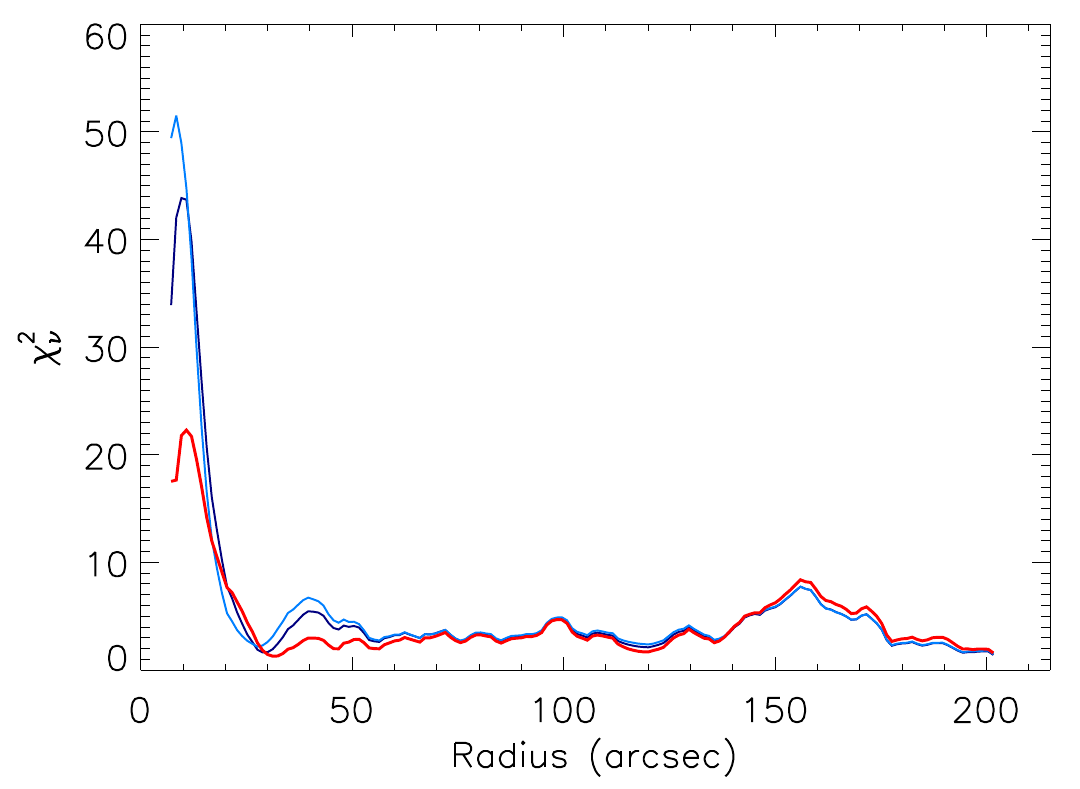}
}
}
\centerline{
\hbox{
\includegraphics[width=0.25\linewidth,angle=90]{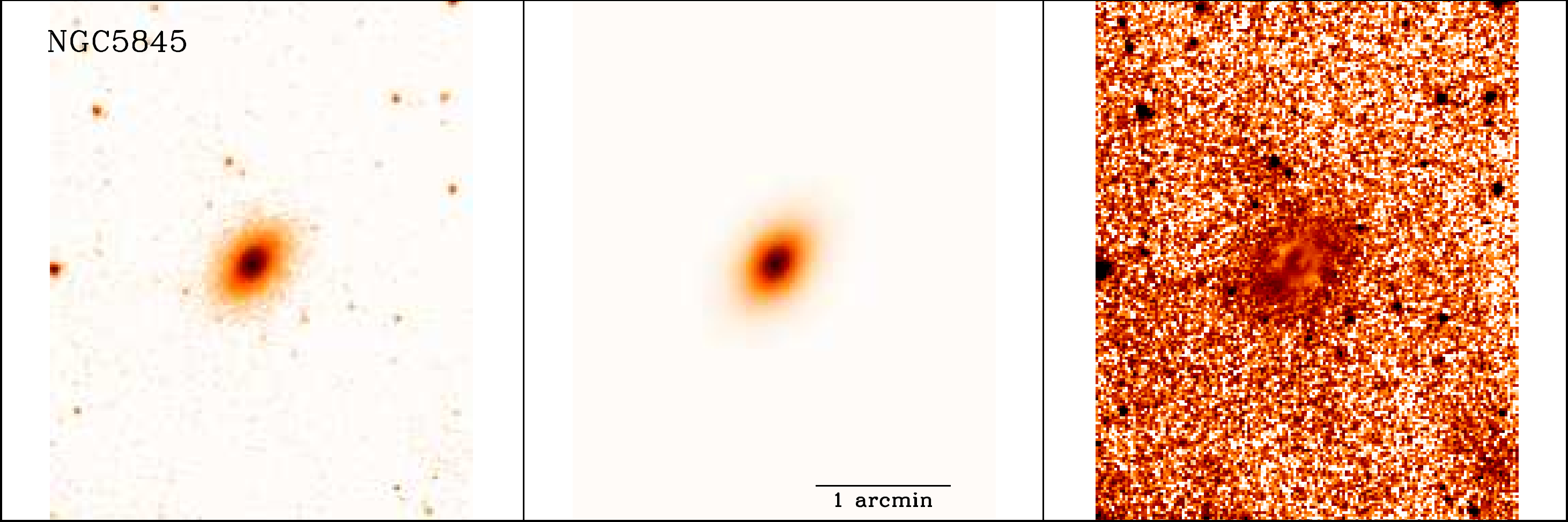}
\includegraphics[height=0.25\linewidth]{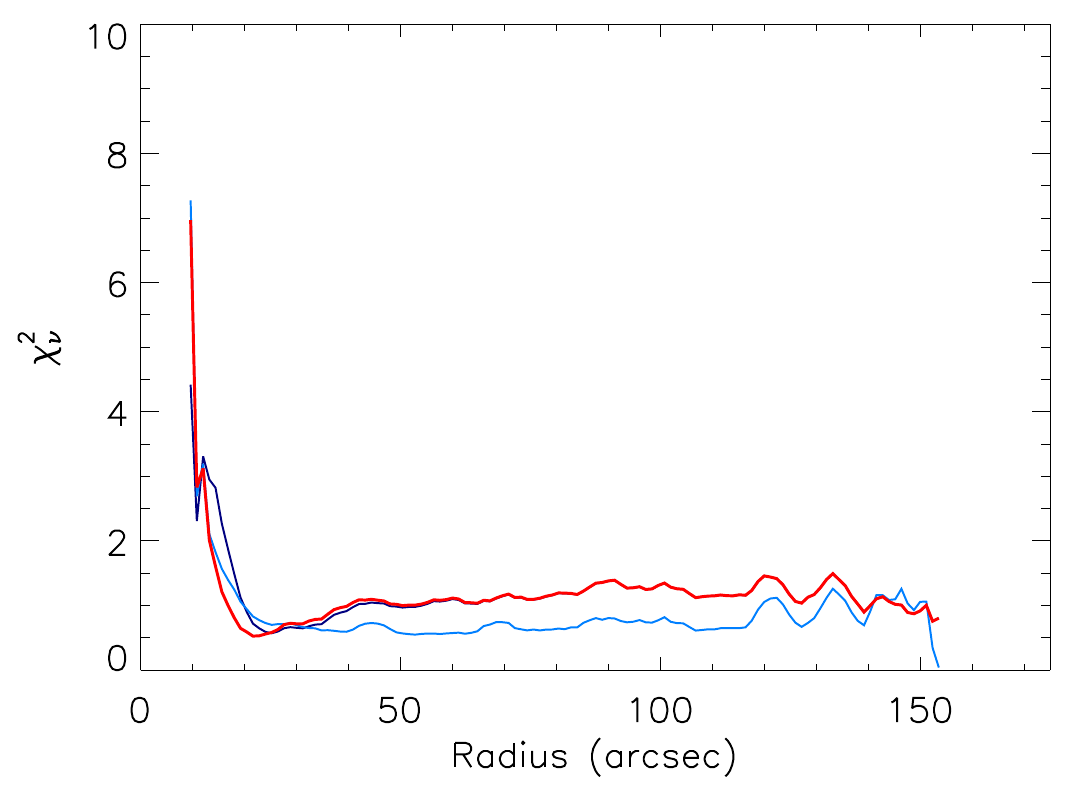}
}
}
\centerline{
\hbox{
\includegraphics[width=0.25\linewidth,angle=90]{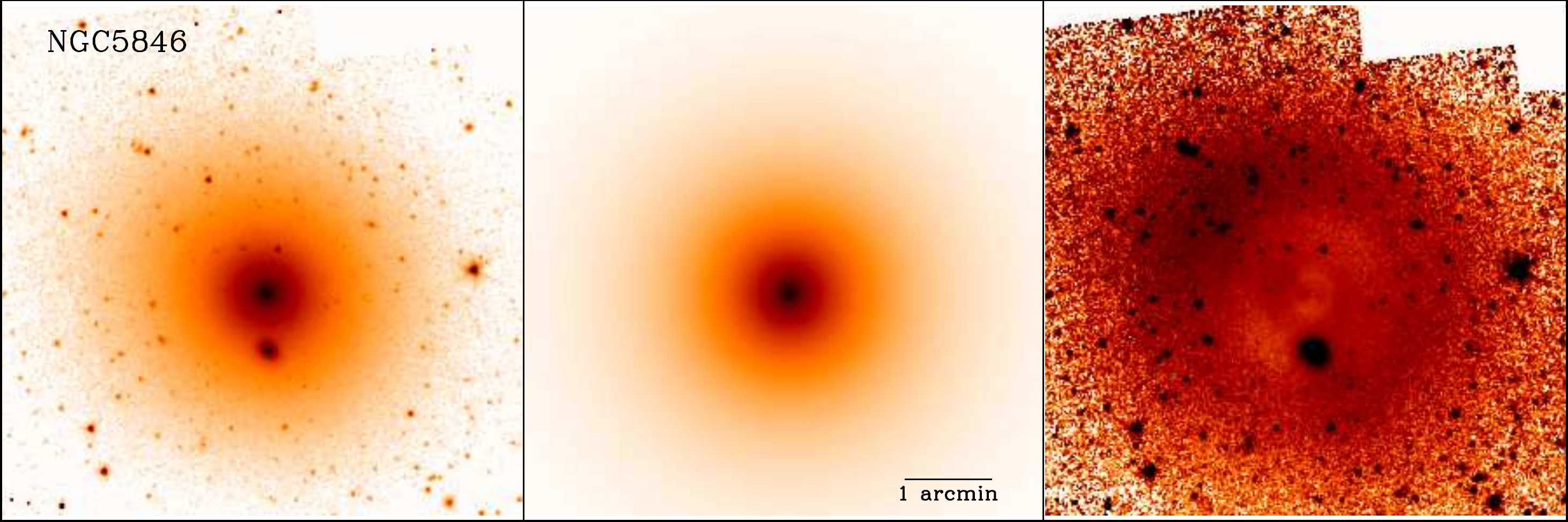}
\includegraphics[height=0.25\linewidth]{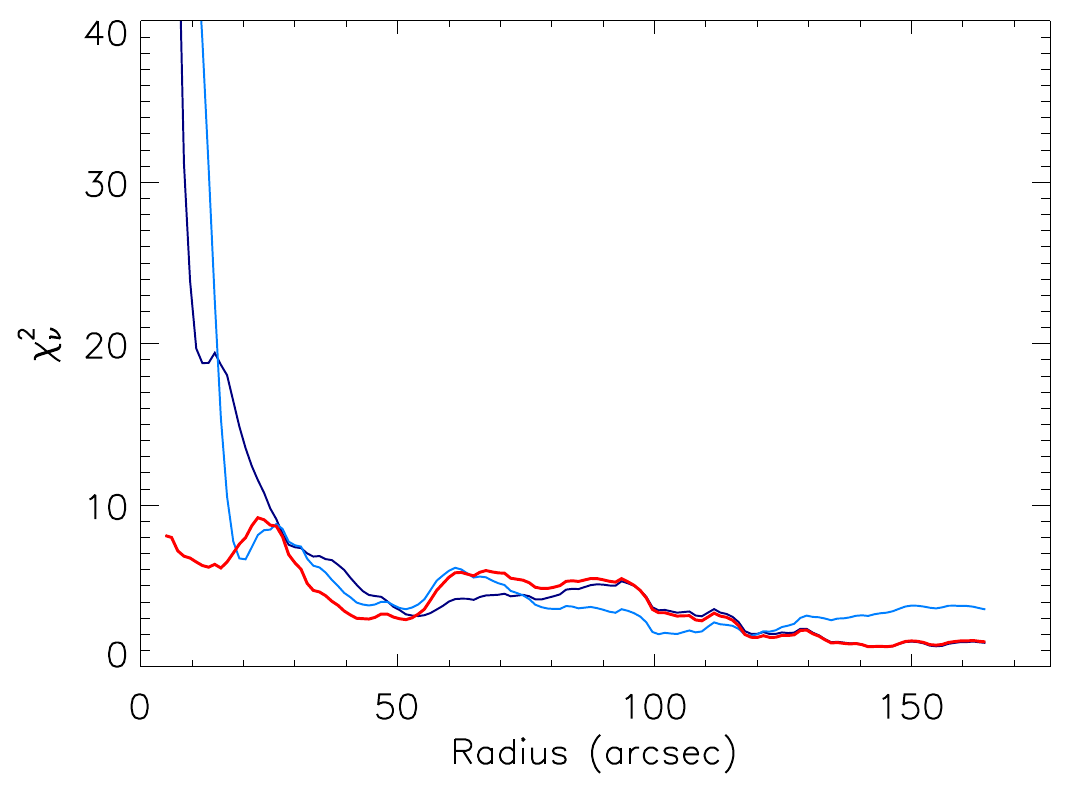}
}
}
\centerline{
\hbox{
\includegraphics[width=0.25\linewidth,angle=90]{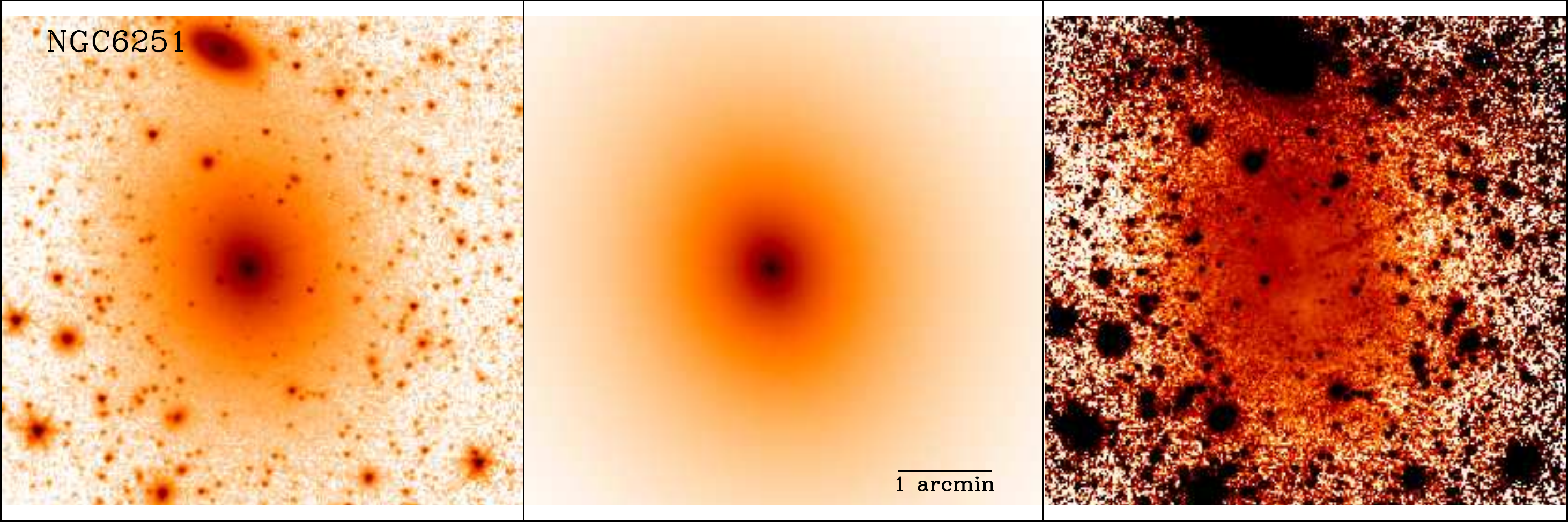}
\includegraphics[height=0.25\linewidth]{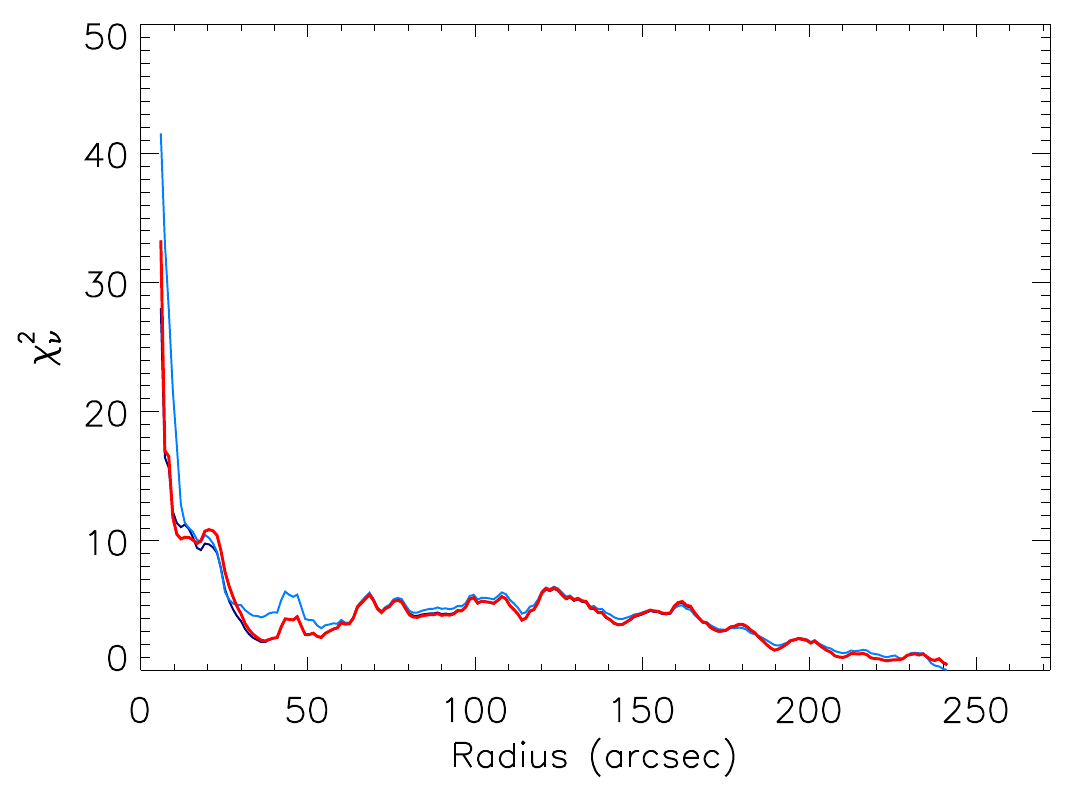}
}
}
\centerline{
\hbox{
\includegraphics[width=0.25\linewidth,angle=90]{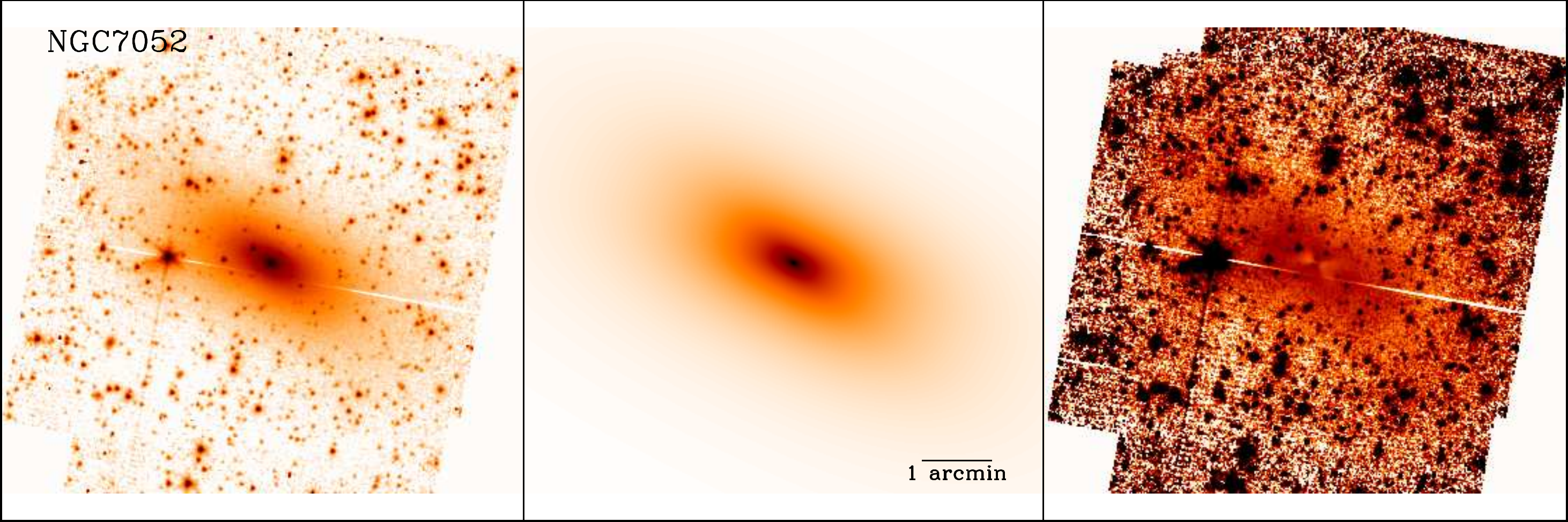}
\includegraphics[height=0.25\linewidth]{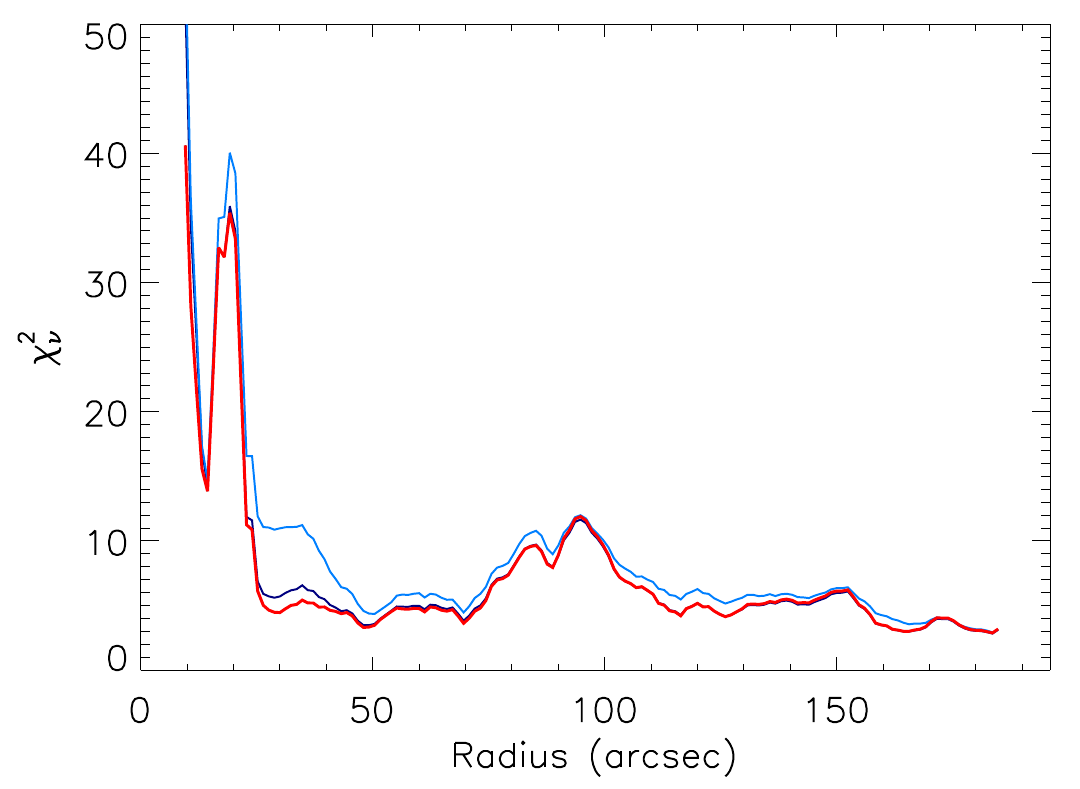}
}
}
\caption{Continued.}
\end{figure*}
\begin{figure*}
\centerline{
\hbox{
\includegraphics[width=0.25\linewidth,angle=90]{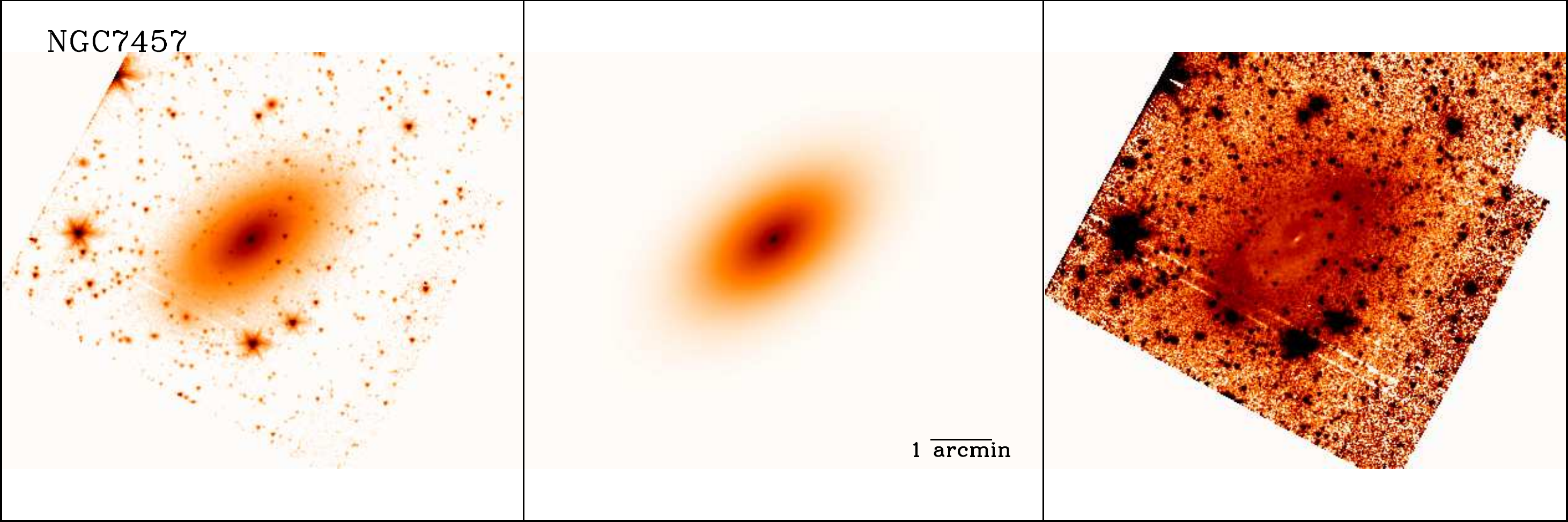}
\includegraphics[height=0.25\linewidth]{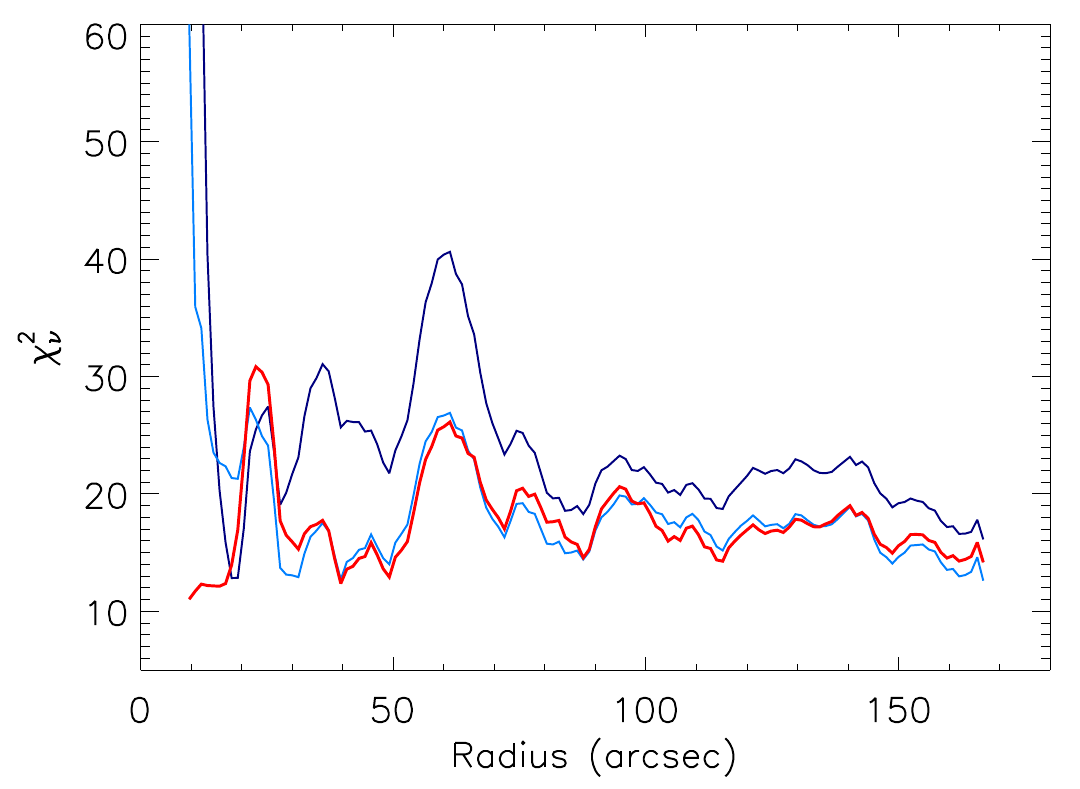}
}
}
\centerline{
\hbox{
\includegraphics[width=0.25\linewidth,angle=90]{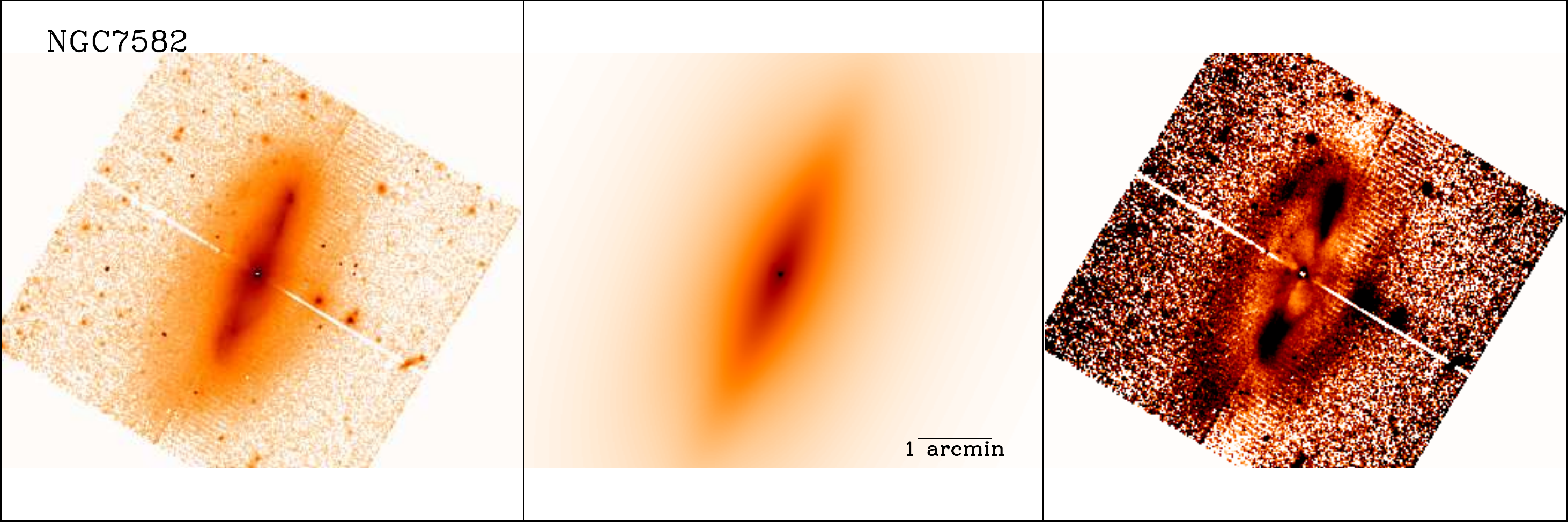}
\includegraphics[height=0.25\linewidth]{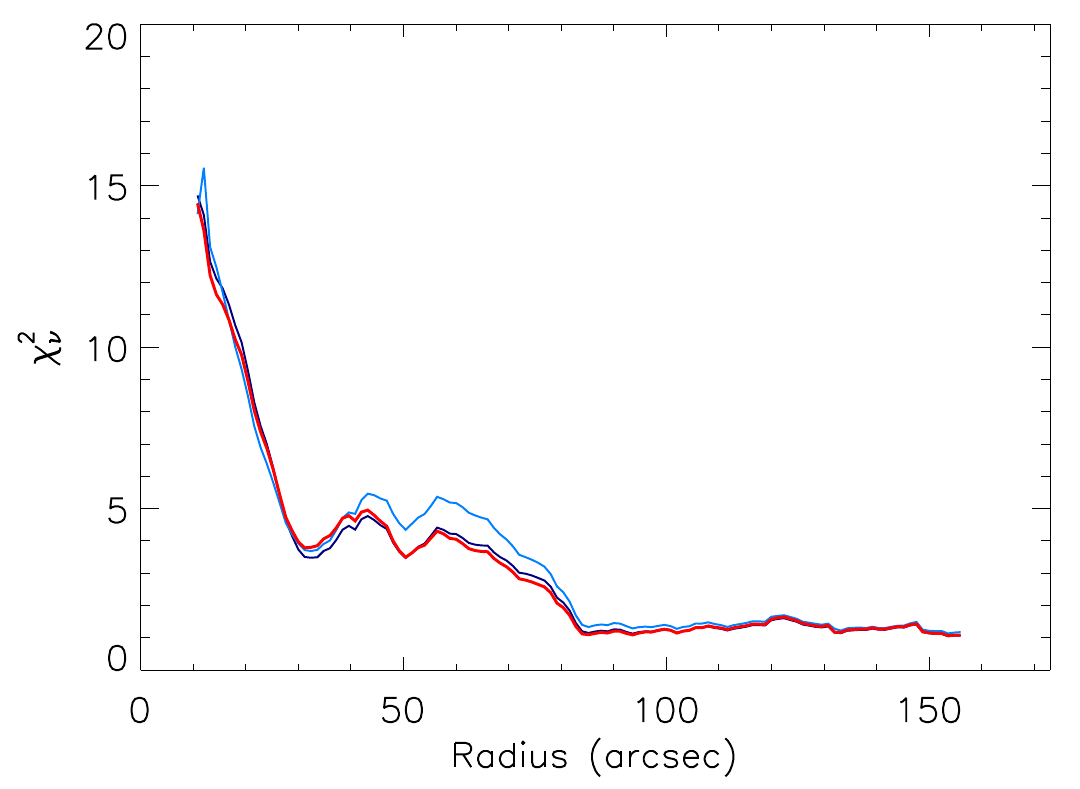}
}
}
\caption{Continued.}
\end{figure*}

\end{document}